\documentclass[aps,prd,showpacs,showkeys,preprint,amsmath,amssymb]{revtex4-1}
\usepackage[english]{babel}
\usepackage{epsfig,float,subfigure,amssymb,amsmath,amsthm}
\newcommand{\beq}{\begin{equation}}
\newcommand{\eeq}{\end{equation}}
\newcommand{\bea}{\begin{eqnarray}}
\newcommand{\eea}{\end{eqnarray}}
\newcommand{\meio}{{}^1\!/{}_{\!2}}
\newcommand{\tmeio}{{}^3\!/{}_{\!2}}
\newcommand{\tquarto}{{}^3\!/{}_{\!4}}

\newcommand{\terco}{{}^1\!/{}_{\!3}}
\newcommand{\dterco}{{}^2\!/{}_{\!3}}
\DeclareMathOperator{\sech}{sech}

\DeclareMathOperator{\arcsech}{arcsech}

\begin{document}
\title{Energy eigenfunctions for position-dependent mass particles in a new class of molecular hamiltonians}
\author{H. R. Christiansen$^\dag {}^*$ and M. S. Cunha$^\dag$}
\affiliation{$^\dag$ Grupo de F\'{\i}sica Te\'orica, State University of Ceara (UECE),
Av. Paranjana 1700, 60740-903 Fortaleza - CE, Brazil\\
$^*$ State University Vale do Acara\'{u}, Av. da Universidade 850, 62040-370 Sobral - CE, Brazil }
\begin{abstract}
Based on recent results on quasi-exactly solvable Schrodinger equations, we review a new phenomenological potential class lately reported. In the present paper we consider the quantum differential equations resulting from position-dependent mass (PDM) particles. We first focus on the PDM version of the hyperbolic potential $V(x) = {a}~{\sech^2x} + {b}~{\sech^4x}$, which we address analytically with no restrictions on the parameters and the energies. This is the celebrated Manning potential, a double-well widely used in molecular physics, until now not investigated for PDM. We also evaluate the PDM version of the sixth power hyperbolic potential $V(x) = {a}~{\sech^6x}+b~{\sech^4x}$ for which we could find exact expressions under some special settings. Finally, we address a triple-well case $V(x) = {a}~{\sech^6x}+b~{\sech^4x}+c~\sech^2x$ of particular interest for its connection to the new trends in atomtronics. The PDM Schrodinger equations studied in the present paper yield analytical eigenfunctions in terms of local Heun functions in its confluents forms. In all the cases PDM particles are more likely tunneling than ordinary ones. In addition, it is observed a merging of eigenstates  when the mass becomes nonuniform.
\end{abstract}
\pacs{3.65.Ge, 2.30 Hq, 2.30 Gp}
\keywords{Schrodinger equation, Position-dependent mass,  Heun equation, Molecular potentials}
\maketitle

\section{Introduction}
Over the years, the dynamics of quantum particles in every single substance
has been a target for analytical studies in order to have a full % complete, better
understanding of condensed matter systems. This is certainly an ambitious task but,
although generally frustrating, several phenomenologically relevant models have had
its differential equations analytically solved
\cite{morse,eckart,rosenmorse,manningrosen,poschlteller,manning,ma47,
bargmann,scarf,ff61,cimento62,cimento64,CPM66,natanzon79,nieto78,prl83,gino84}.

New models involving hyperbolic potentials, typically found in molecular physics,
have been reported very recently and in some cases have yield exact wavefunctions
for certain relations among the parameters \cite{sech246,cosh24g,downing}.
%We have lately addressed one of these problems \cite{downing} in connection with the issue
%of position-dependent mass and analytically found the corresponding eigenfunctions \cite{JPM2013}.
In the present paper it is our aim to deal with the family of potentials reported in \cite{sech246}
in connection with the issue of position-dependent mass (PDM) particles.

The number of solvable potentials in ordinary quantum mechanics is in fact limited, but
when we assume the particle mass has a nontrivial space distribution
the mathematical difficulties grow even more. For some relevant mass distributions
some phenomenological potentials have been solved in recent years
 \cite{JPM2013,CTP2013,ardasever2011,iran,severtezcan,midyaroy2009,bagchi,chinovalmir}.

The origin of the PDM approximation can be traced back in the domain of solid
state physics \cite{wannier,slater,luttingerkohn,BDD,bastard81,vonroos,einehem88,levy95}.
For instance, the dynamics of electrons in semiconductor heterostructures
has been tackled with an effective mass model related to the envelope-function approximation
\cite{BDD,zhukro83,weisvin91}. Besides this, position dependent mass particles have been used
to set to several important issues of low-energy physics related to the understanding of the electronic
properties of semiconductors, crystal-growth techniques \cite{gorawill,bastardetal,bastard88},
quantum wells and quantum dots \cite{dots}, Helium clusters \cite{helium}, graded crystals \cite{prl93},
quantum liquids \cite{qliq}, and  nanowire structures under size variations, impurities, dislocations,
and geometrical imperfections \cite{willatzenlassen2007}, among others.

In this paper, assuming a PDM distribution in the Schrodinger equation,
we take up on a new class of potentials \cite{sech246} recently reported,  viz.
\beq V(x) = -A\sech^6x-B\sech^4x-C\sech^2x. \label{ipotsech246}\eeq
For a large variety of constants, this family represents symmetric asymptotically flat double-well potentials,
related to problems of solid state and condensed matter physics.
Double-well potentials are emblematic since they allow studying typical quantal situations involving
bound states and particle tunneling through a barrier \cite{tunnel}.
For example, the case $A=0$ of Eq. (\ref{ipotsech246}) results in the renowned Manning potential \cite{manning}
originally used to address the vibrational normal modes of the NH$_3$ and ND$_3$ molecules
and found also appropriate for the understanding of the infrared spectra of organic compounds such as ammonia,
formamide and cyanamide. Coincidentally, the family given in \cite{cosh24g} also includes this potential
when the parameter $g>>1$. This class, also phenomenologically rich, was originally related to the double
sine-Gordon kink \cite{sinegordon} but is known for its interest in different physical
subjects \cite{oldcosh24g} such as the study of anti-ferromagnetic chains \cite{chains} and
 experimentally accessible systems like (CH3)$_4$ NMnC1$_3$ (TMMC) \cite{molecules}.
In both papers \cite{sech246,cosh24g} the ordinary constant-mass Schrodinger equations
have been found some analytical solutions proportional to Heun functions \cite{heun}. These special functions
are not very well-known but have been receiving increasing attention, particularly in the last decade
\cite{maier,maier192,conflu,ronveaux,hortacsu,hille,slavyanov,chebterrab,hounkonnou,fiziev,tolga}.

Interestingly, although not explored in \cite{sech246},
the class given by Eq. (\ref{ipotsech246}) also includes three parameter triple-well potentials
particularly interesting in atomtronics, associated with atomic diodes and transistors \cite{atomtronics}
and laser optics \cite{optics}.
Triple-well semiconductor structures have been used in experiments of light transfer in optical
waveguides \cite{waveguide,semiconductors} as well as in models of dipolar condensates with phase
transitions and metastable states \cite{condensates}.

In the present work we analyze the new potentials given by Eq. (\ref{ipotsech246})
with a physically significant input, namely a nonuniform mass. This phenomenological upgrade
of course induces highly nontrivial consequences in the associated differential equations and
yields new mathematical and physical results. Our goal is to
analytically handle the resulting equations and find their general solutions.
We succeed in some cases which we detail in what follows and find Heun functions in their confluent forms.
In the case of triple-wells we manage it numerically for its high analytical complexity.
In every case we compare the PDM results with the equivalent constant mass situations.

In the next section, \ref{sec:ordering}, we first address the problem of determining the correct
kinetic operator of the PDM Schrodinger equation and then, in Sec. \ref{sec:potential},
we obtain an effective potential in a convenient space. In Sec. \ref{sec:Manning}
we consider the \textit{PDM}-Manning potential, a particularly
important  member of the class (\ref{ipotsech246}), and find a complete set of eigenstates
built in terms of confluent Heun solutions. We plot all the six bound-eigenstates of the PDM differential
equation together with those of the constant mass problem to show their deviation from the ordinary ones.
Next, in section \ref{sec:sech64} we find exact expressions for the $E=0$ eigenfunctions
of the PDM version of the
sixth power potential $V(x) = {a}~{\sech^6x}+b~{\sech^4x}$ under some special settings.
In this case, the eigenstates are proportional to \textit{triconfluent} forms of the Heun functions.
Finally, we address the triple-well phase of the potential class, with and without PDM,
and discuss our results. The final remarks are drawn in Sec.\ref{sec:conclusion}.

%%%%%%%%%%%%%%%%%%%%%%%%%%%%%%%%%%%%%%%%%%%%%%%%%%%%%%%%%%%%%%%%%%%%%%%%%%%%%%%%%%%%%%%%%%%%
\section{The PDM kinetic operator \label{sec:ordering}}

The effective hamiltonian of a PDM nonrelativistic quantum particle
has received much attention along the years for both phenomenological and mathematical reasons \cite{bastard81,vonroos,einehem88,levy95,BDD,zhukro83,gorawill,shewell,morrowbron84,vonroofmavro,
morrow,galgeo,likhun,thomsenetc,young,einevol1,einevol2,tdlee,DA}.
The full expression for the kinetic-energy hermitian operator
for a position-dependent mass $m(x)$ reads
\beq
\hat T =\, \frac 14 \, \hat T_0\,+\frac 18\left\{\,\,\hat P^2\,m^{-1}(x)
+\,  m^{\alpha }(x)\,\hat{P}\ m^{\beta}(x)\,\hat{P}\ m^{\gamma }(x)\,\,+\,m^{\gamma }(x)\,\hat{P}
\ m^{\beta }(x)\,\hat{P}\ m^{\alpha }(x)\, \right\},\label{Top}
\eeq
where, for constant mass, $\hat T_0= \frac 12m^{-1}\, \hat P^2$ is the
standard quantum kinetic energy and $\hat P=-i\hbar\; d/dx$ is the momentum operator.
The above parameters have to fulfill the condition  $\alpha +\beta +\gamma =-1$  \cite{vonroos}.

Recalling the basic postulate $[\hat X,\hat P]=i\hbar$, we get
\begin{equation}
\hat T\,=\,\frac{1}{2\,m}\,\hat{P}^{2}\,+\,\frac{i\hbar }{2}\frac{1}{m^{2}}\frac{dm}{dx}
\,\,\hat{P}\,\,+\,U_{\rm K}\left( x\right),
\end{equation}
where
\beq
U_{\rm K}\left( x\right) =\frac{-\hbar ^{2}}{
4m^{3}(x)}\left[ \left( \alpha +\gamma -1\right) \frac{m(x)}{2}\left(\frac{
d^{2}m}{dx^{2}}\right)+\left(1-\alpha \gamma -\alpha -\gamma
\right) \left( \frac{dm}{dx}\right) ^{2}\right]\label{kinpot}
\end{equation}
is an effective potential of {kinematic} origin. This is of course
a source of ambiguity in the hamiltonian for it depends on the values of $\alpha, \beta, \gamma$.
In order to fix this issue, we can kill the kinematic potential by adding the constraint
$
\alpha \,+\,\gamma \,=\,1\,=\alpha \,\gamma \,+\,\alpha\,+\,\gamma.  \label{cond}
$
Its solution is $\alpha =0$ and $\gamma=1$, or  $\alpha=1 $ and $\gamma =0$, which corresponds to
the Ben-Daniel--Duke  $\hat T$ ordering \cite{BDD}.
Now, the hamiltonian is free of ambiguities but the resulting effective Schr\"odinger equation
still looks weird for it now includes a first order derivative term. For an arbitrary external
potential $V(x)$ the PDM-Schr\"odinger equation turns out
\beq
\frac{d^2\psi(x)}{dx^2} { -\left(\frac{1}{m(x)} \frac{d m(x)}{dx}\right)} { \frac{d\psi(x)}{dx}} +
\frac{2}{\hbar^2}\,{  m(x)}\, [E-V(x)] \psi(x) =0. \label{pdm}
\eeq
Noticeably, not only the last term has been strongly modified from the ordinary
Schr\"odinger equation but the differential operator turned out to be dramatically changed.
This will have of course deep consequences on the physical wave solutions of the system.

%%%%%%%%%%%%%%%%%%%%%%%%%%%%%%%%%%%%%%%%%%%%%%%%%%%%%%%%%%%%%%%%%%%%%%%%%%%%%%%%%%%%%%%%%%%%%%%%%%%%%%%%%
\section{Effective new PDM potential\label{sec:potential}}

Here we will adopt a solitonic smooth effective mass distribution
\beq m(x)=m_0\, {\sech}^2(x/d) \label{massa} \eeq
(see e.g. \cite{iran} and \cite{bagchi}). Besides its convenient analytical nature,
its shape is familiar
in effective models of condensed matter and low energy nuclear physics and depicts
a soft symmetric distribution.
%  btw, it looks appropriate for the aimed family of potentials.
The effective Schrodinger equation (\ref{pdm}) thus reads
\beq
\psi'' (x)+2 \tanh(x) \psi'(x)+ \frac{2m_0}{\hbar^2}(E-V(x))\, \sech^2(x) \psi(x)=0, \label{eqSch}
\eeq
but for the ansatz solution
\beq
\psi(x)=\cosh^{\nu}\!(x)\,\varphi(x), \label{transf} \eeq
becomes
\beq
\varphi''(x)+2 (\nu +1) \tanh(x)\varphi'(x)+\left[\nu(\nu+2) \tanh^2\!x
+ \left(\nu +\frac{2m_0}{\hbar^2}(E-V(x)) \right)\,{\sech}^2(x) \right] \varphi(x)=0.
\label{eqtransf}
\eeq
%
%Nota-se que  (\ref{transf}) preservou a paridade da Eq. (\ref{eq2}).
%
Now, a change of variables \beq\sech\,x = \cos\,z,  \label{transf2}\eeq
maps the domain $(-\infty,\infty) \rightarrow (-\frac{\pi}{2},\frac{\pi}{2})$ and provides
\beq
\varphi''(z)+(2\nu+1)\tan(z)\varphi'(z)+\left[
\nu+\nu(\nu+2)\tan^2(z)+\frac{2m_0}{\hbar^2}\left(E-{V}(z)\right) \right]\varphi(z)=0
\eeq
(we call $\varphi(x(z))=\varphi(z)$, etc.).
%\bea %%
%\varphi''(z)&&+(2\nu+1)\tan(z)\varphi'(z)\nonumber\\
%&&+\left [
%\nu+\nu(\nu+2)\tan^2(z)+\frac{2m_0}{a^2\hbar^2}(E-V(z)) \right ]
%\varphi(z)=0 %%
%\eea%%
The choice $\nu=-1/2$ allows the removal of the first derivative and grants an ordinary
Schrodinger equation
\beq
\left[-\frac{d^2}{dz^2}+ \mathcal{V}(z)\right]\varphi(z)=\mathcal{E}\varphi(z) \label{schrodinger-cons}
\eeq
where $\mathcal{E}=\frac{2m_0}{\hbar^2}E$.
% ATENTI, EN LOS GRAFICOS USAREMOS \frac{2m_0}{\hbar^2}=1
%
The potential class we are dealing with is
\beq V(x) = -A\sech^6x-B\sech^4x-C\sech^2x, \label{potsech246}\eeq
where the adjustable parameters
determine a large variety of possible shapes.
Thus, in Eq. (\ref{schrodinger-cons}) we shall employ
\beq
\mathcal{V}(z)=\frac{1}{2} +\frac{3}{4}\tan^2\!z -\mathcal{A}\cos^6 z
-\mathcal{B}\cos^4 z-\mathcal{C}\cos^2 z, \label{potcalII}
\eeq
where $\mathcal{A,B,C}$ incorporated the factor $\frac{2m_0}{\hbar^2}$, e.g. $\mathcal{A}=\frac{2m_0}{\hbar^2}A$.
The dynamics of the original PDM particle is therefore described by one of constant mass $m_0$
moving in z-space in a non-ambiguous effective potential with no kinematical contributions.
%(exception in the particular case  $\tilde{V}(z)=-\frac{3}{4}\tan^2(z)+$cons.)
%
Although restricted to within
$z =(-\frac{\pi}{2},\frac{\pi}{2})$, with $\varphi(z)=0$ at the borders,
we can eventually transform everything back to
the original x-variable and $\psi$ wave-function to obtain the real space solution.

%%%%%%%%%%%%%%%%%%%%%%%%%%%%%%%%%%%%%%%%%%%%%%%%%%%%%%%%%%%%%%%%%%%%%%%%%%%%%%%%%%%%%%%%%%%%%%%%%%%%%%%%%%%%

\subsection{The \textit{PDM}-Manning potential \label{sec:Manning}}

\begin{figure}[h]
\center
{\includegraphics[width=6.cm,height=6cm]{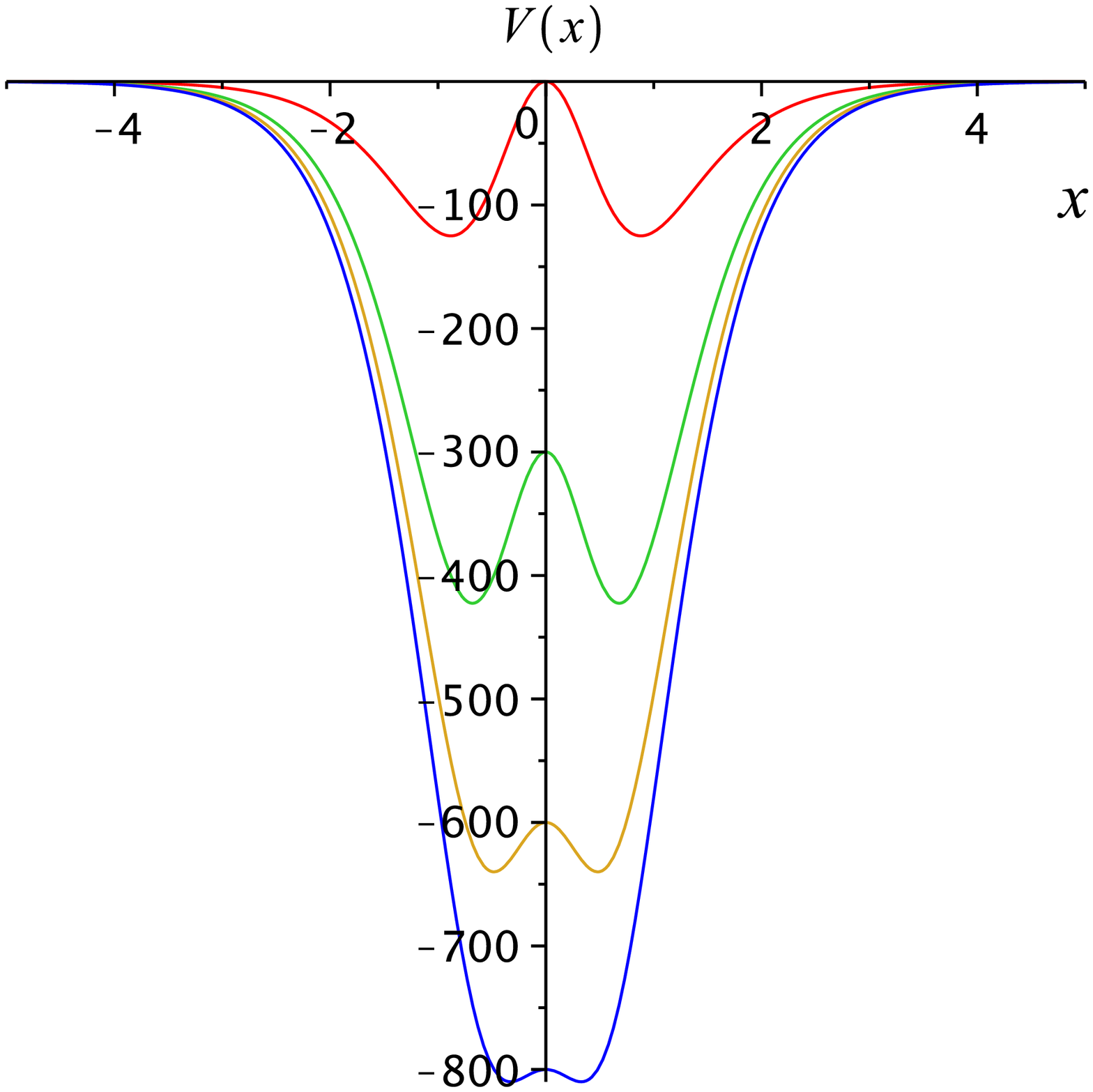}}\hspace{1cm}
{\includegraphics[width=6.cm,height=6cm]{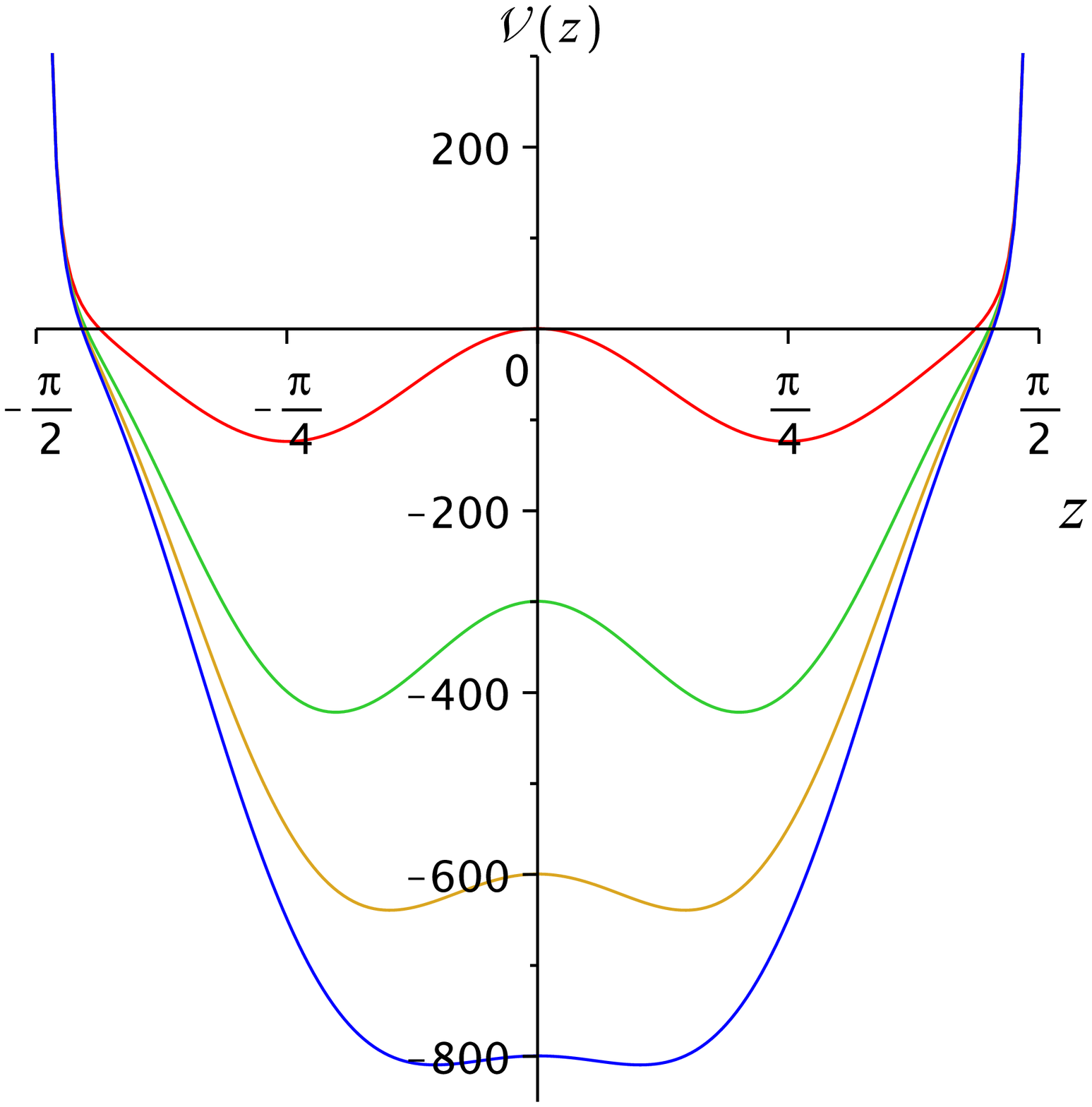}}
\caption{\label{graf_VA0} From top to bottom, plot of the Manning potential $V(x)$ (left) and
$\mathcal{V}(z)$ (right),
for $(\mathcal{B},\mathcal{C})$ = $(-500,500)$, $(-1000,1300)$, $(-1000,1600)$; and $(-1000,1800)$.}
\end{figure}

The celebrated Manning potential, $V(x) = a\sech^2x+b\sech^4x$,
very much used in molecular physics, %highly familiar
is here addressed in the very interesting situation of an effective spatially dependent mass.
This phenomenological potential corresponds to the case $A=0$ of  Eq. (\ref{potsech246}).
Note that when $B < 0, C > 0$ and  $-C/2B < 1$, the Manning potential is a double-well
potential with two minima at $x =\pm \arcsech(\sqrt{-C/2B})$. In Fig.\ref{graf_VA0}
we show this potential for several values of the free parameters.

In $z$-space we have
\beq
\mathcal{V}(z)=\frac{1}{2} +\frac{3}{4}\tan^2\!z - {\mathcal{B}}\,{\cos^4(z)}
- {\mathcal{C}}\,{\cos^2(z)},  \label{Vsech}
\eeq
to be considered in eq.(\ref{schrodinger-cons}). By means of the ansatz
\beq
\varphi(z)=\cos^\mu\!z \,\phi(z) \label{map}
\eeq
we obtain
\bea
\phi''(z)-2\mu\tan(z)\,\phi'(z)+\Big[(\mu^2-\mu-\tquarto)\tan^2\!z-\mu+\mathcal{E} -\meio
+ \mathcal{B}\cos^4\!z+\mathcal{C}\cos^2\!z\Big]\,\phi(z) = 0
\eea
which can be simplified by choosing $\mu^2-\mu-\tquarto=0,$ namely $\mu=3/2$ or $\mu=-1/2$.
If we now transform coordinates by $y=\sin^2\!z$, the above equation results in
\bea
y\,(1-y)\,\phi''(y)&+&\Big[\meio-(1+\mu)\,y\Big]\,\phi'(y)+\frac{1}{4}\\ && \Big[-\mu +\mathcal{E}-\frac{1}{2}+\mathcal{B}\,(1-y)^2+\mathcal{C}(1-y)\Big]\phi(y) = 0.
\eea
A further transformation
\beq\label{maph}\phi(y)=e^{\nu y}H(y),\eeq
puts in evidence its Heun nature
\bea
h''(y) &+& \left(2\nu+\frac{\meio}{y}+\frac{\mu+\meio}{y-1}\right)\,h'(y) + \frac{1}{4y\,(y-1)}
\Big[\mu+\frac{1}{2}-2\nu-\mathcal{E-B-C}\nonumber\\
&-&4\Big(\nu^2-\nu-\mu\nu-\frac{\mathcal{B}}{2}-\frac{\mathcal{C}}{4}\Big)\,y
+(\mathcal{B}-4\nu^2)\,y^2\Big] h(y) = 0.
\eea
Since $\mu=-1/2$ is misleading and $\nu$ is arbitrary we choose $\mu=\tmeio$ and $2\nu=\sqrt{\mathcal{B}}$.
This yields
\bea
h''(y)+\left(\sqrt{\mathcal{B}}+\frac{\meio}{y}+\frac{2}{y-1}\right) \,h'(y)+\nonumber\\
\frac{1}{y\,(y-1)} \left[\frac{1}{4}\Big(\mathcal{B} +\mathcal{C} +5\sqrt{\mathcal{B}}\Big)\,y +\frac{1}{2}-\frac{\sqrt{\mathcal{B}}+\mathcal{E}+\mathcal{B}+\mathcal{C}}{4}\right]h(y) = 0, \label{eqhA0}
\eea
which is a canonical non-symmetric confluent Heun equation \cite{ronveaux,chebterrab,hounkonnou} of the form
\bea
Hc''(y)+ \left(\alpha + \frac{\beta + 1}{y} + \frac{\gamma + 1}{y-1} \right)Hc'(y)+ \frac{1}{y(y-1)}\\
\left[\left(\delta + \frac{\alpha}{2}(\beta + \gamma +2)\right)y + \eta + \frac{\beta}{2}  + \frac{1}{2}
(\gamma - \alpha)(\beta+1) \right] Hc(y) = 0, \nonumber \label{heunc}
\eea%
with
\begin{eqnarray*}
\alpha &=& \sqrt{\mathcal{B}}\\
\beta &=& -\meio\\
\gamma &=& 1\\
\delta &=&\frac{1}{4}(\mathcal{B}+\mathcal{C})\\
\eta &=&\frac{1}{2}-\frac{\mathcal{E+\mathcal{B}+\mathcal{C}}}{4}.
\end{eqnarray*}

%%%%%%%%%%%%%%%% SOLUÇÕES MANNING

\begin{figure}[h]
\center
{\includegraphics[width=4.5cm,height=4.5cm]{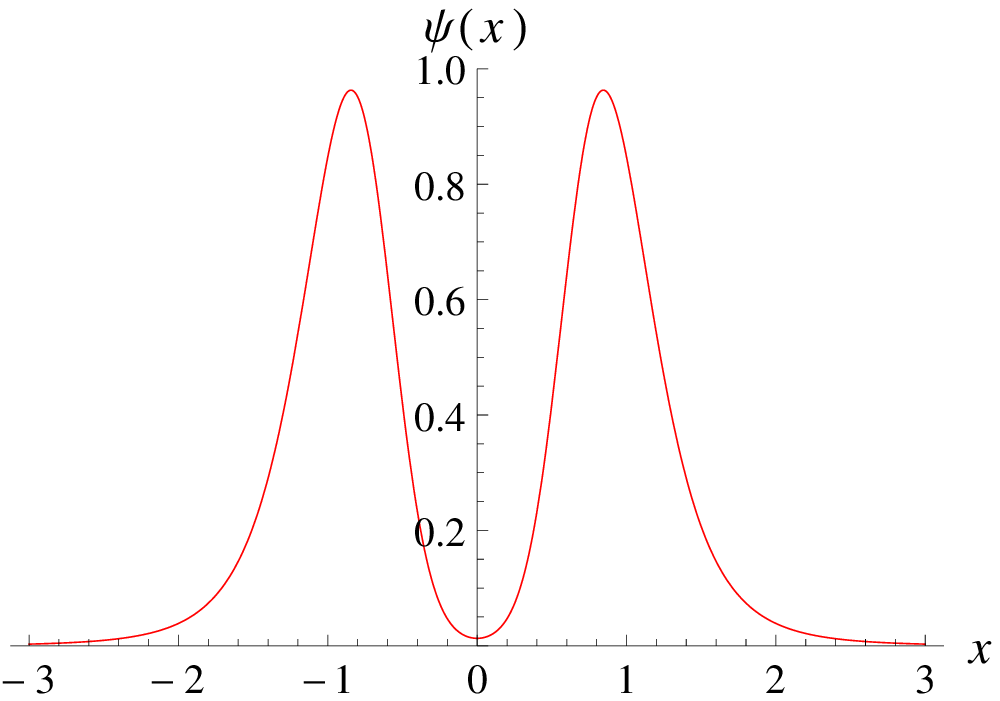}}
{\includegraphics[width=4.5cm,height=4.5cm]{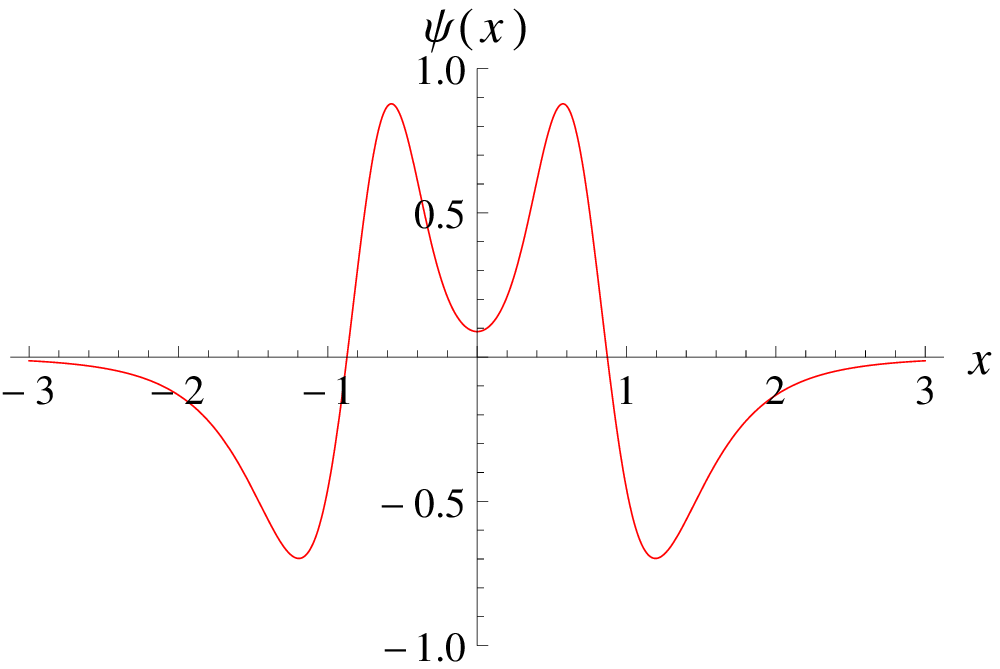}}
{\includegraphics[width=4.5cm,height=4.5cm]{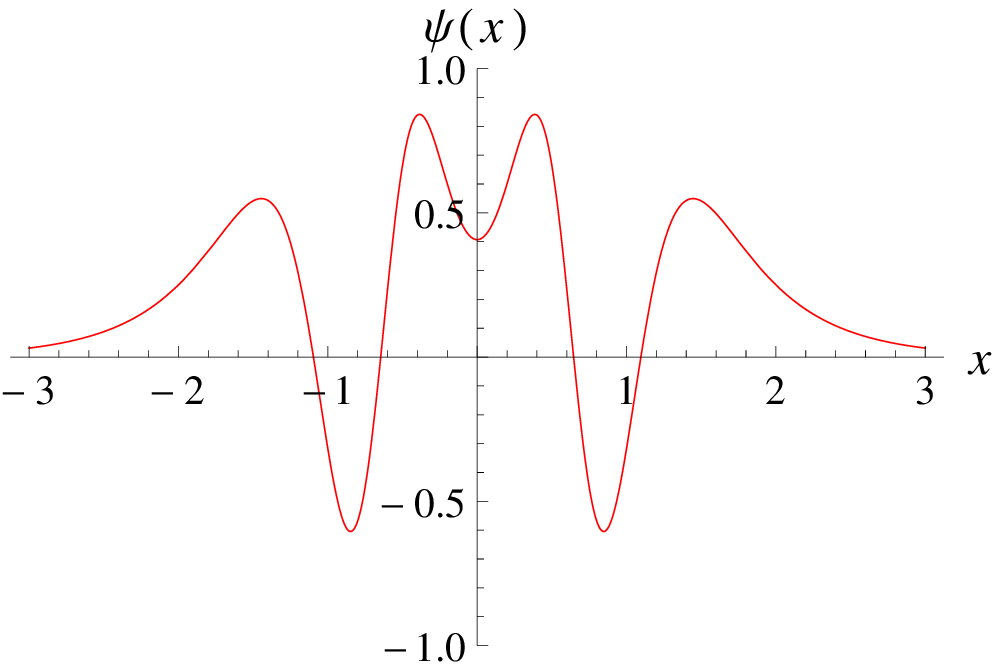}}\\
{\includegraphics[width=4.5cm,height=4.5cm]{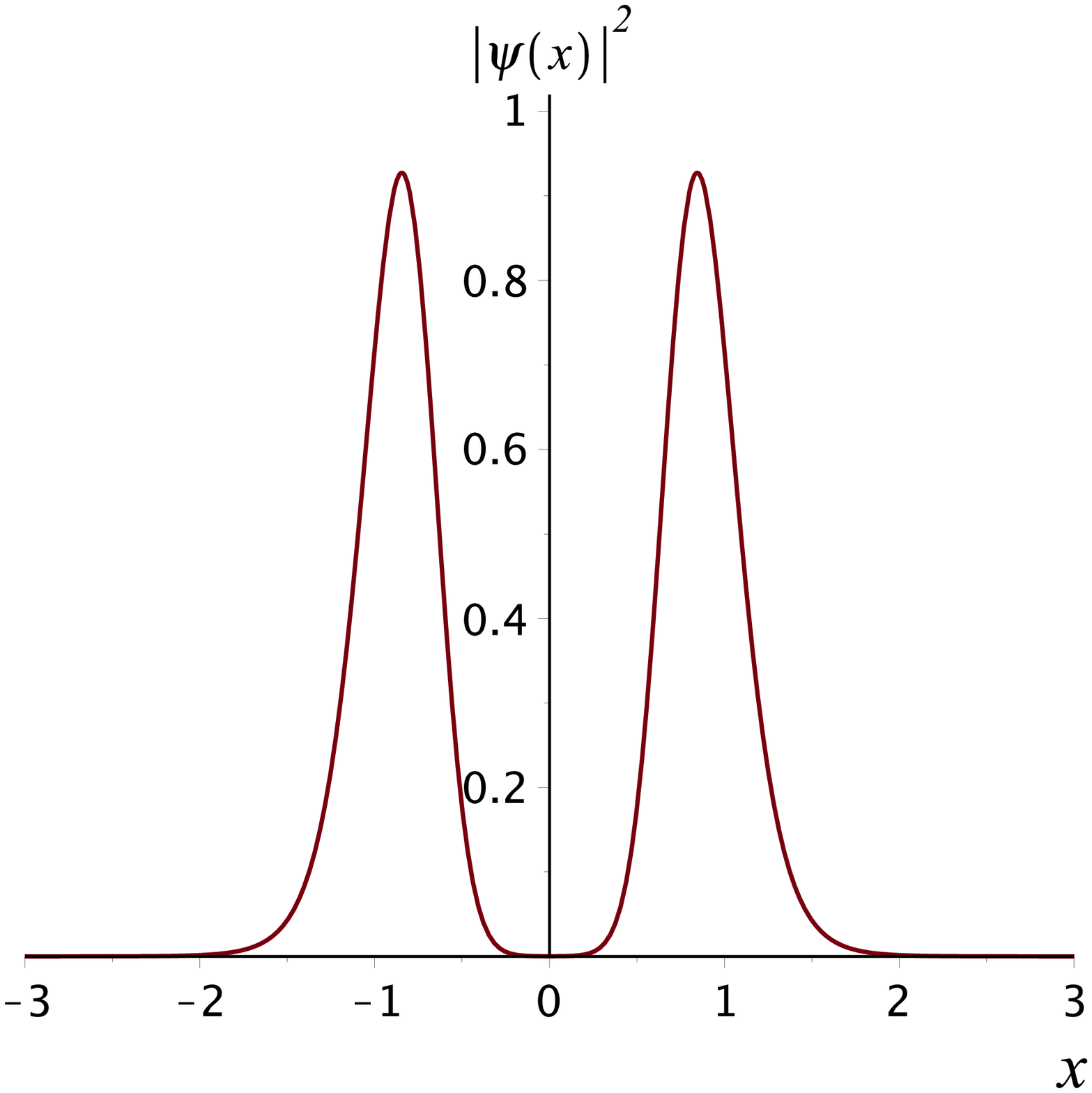}}
{\includegraphics[width=4.5cm,height=4.5cm]{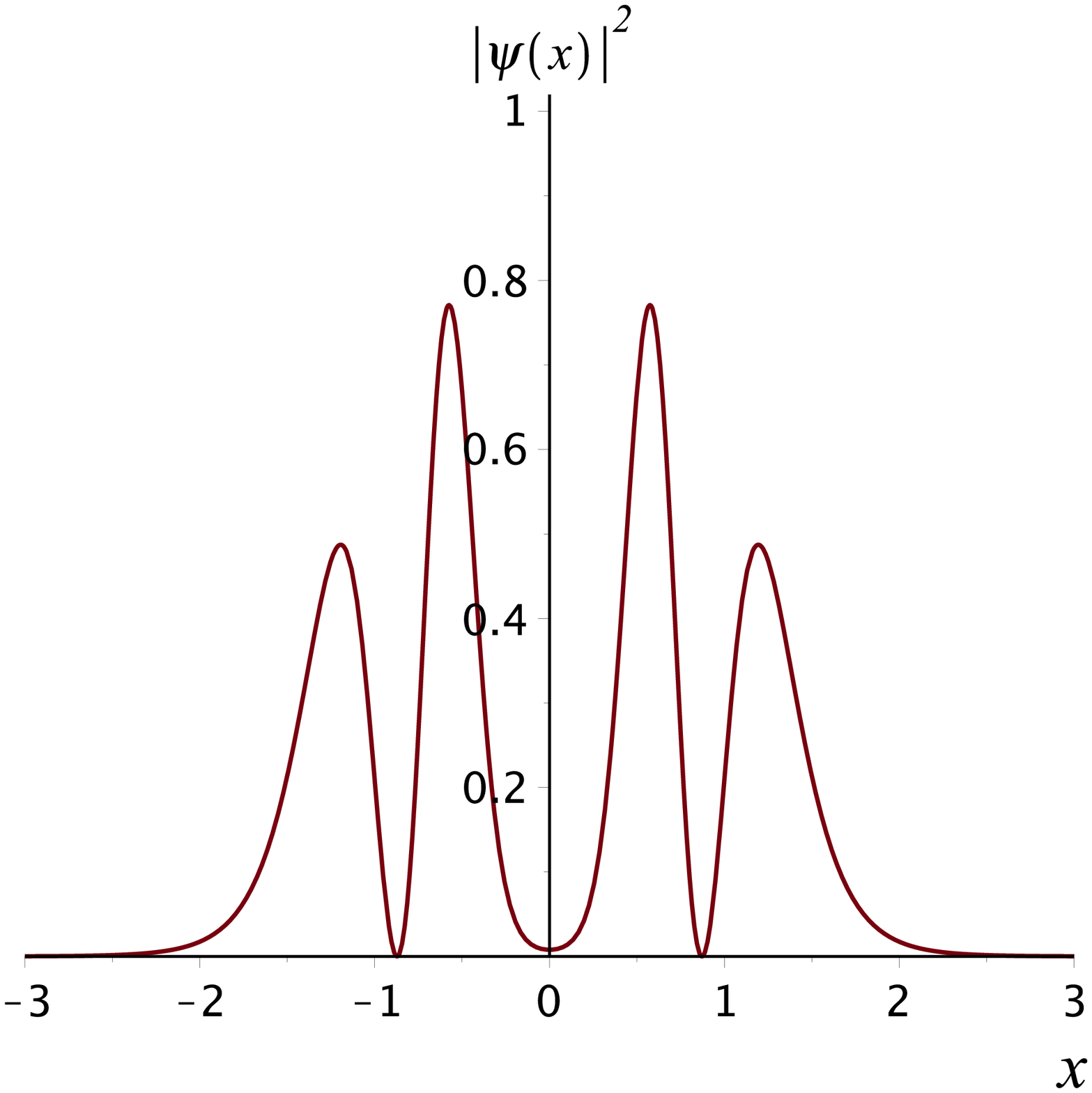}}
{\includegraphics[width=4.5cm,height=4.5cm]{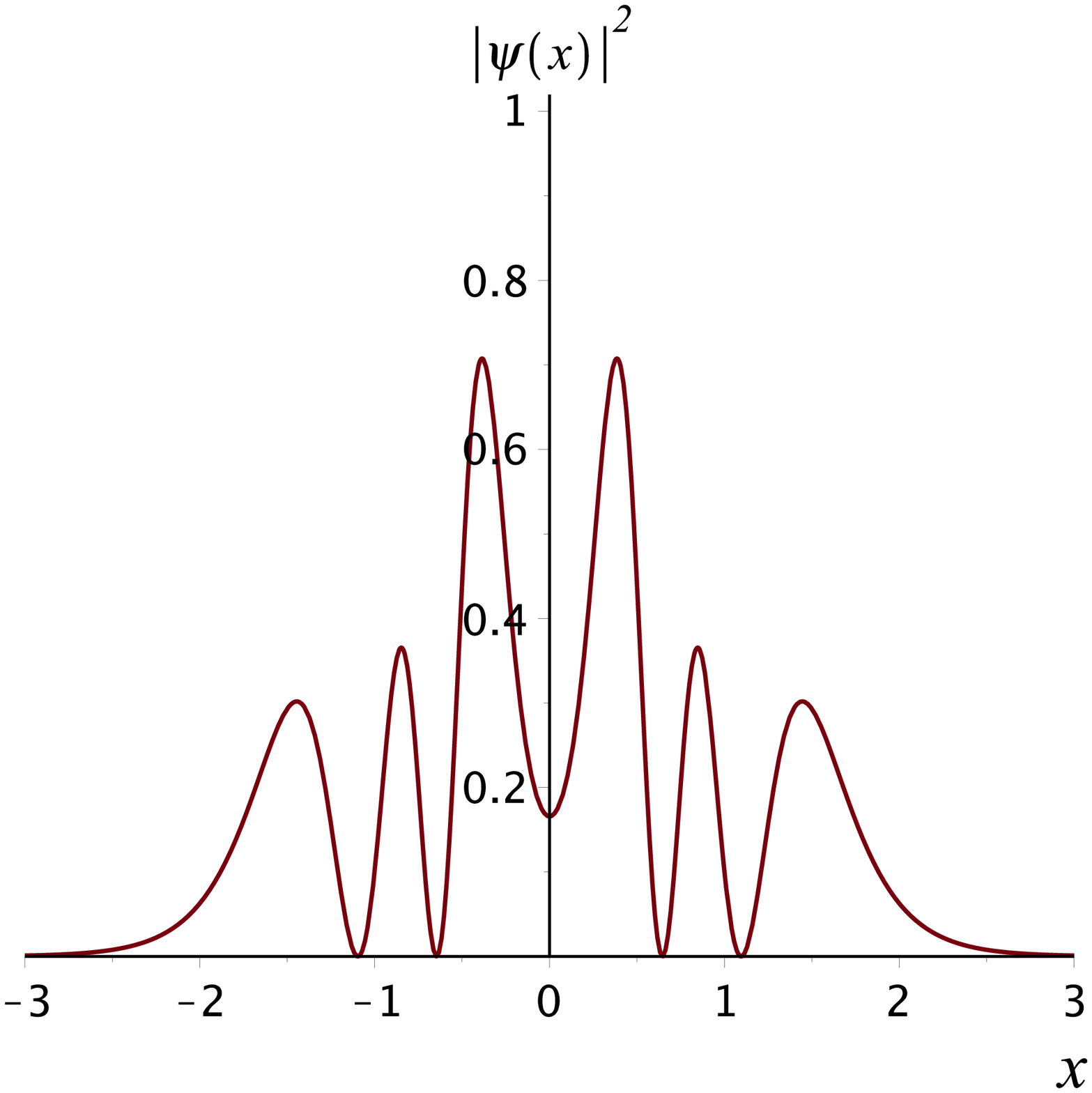}}
\caption{\label{figS_x_A0}  Plot of the normalized symmetric solutions
$\psi^{(1)}(x)$  [Eq. (\ref{SManningx})], when $B=-C=-500$
(see top (red) curve in Fig.\ref{graf_VA0}),
in their three symmetric bound states $\mathcal{E}_0=-102.25913969050$ (left),
$\mathcal{E}_2=-61.4581676270$ (center) and  $\mathcal{E}_4=-25.9419535530$ (right).
Below are shown the corresponding probability densities.}
\end{figure}

\begin{figure}[h]
\center
{\includegraphics[width=5cm,height=4.5cm]{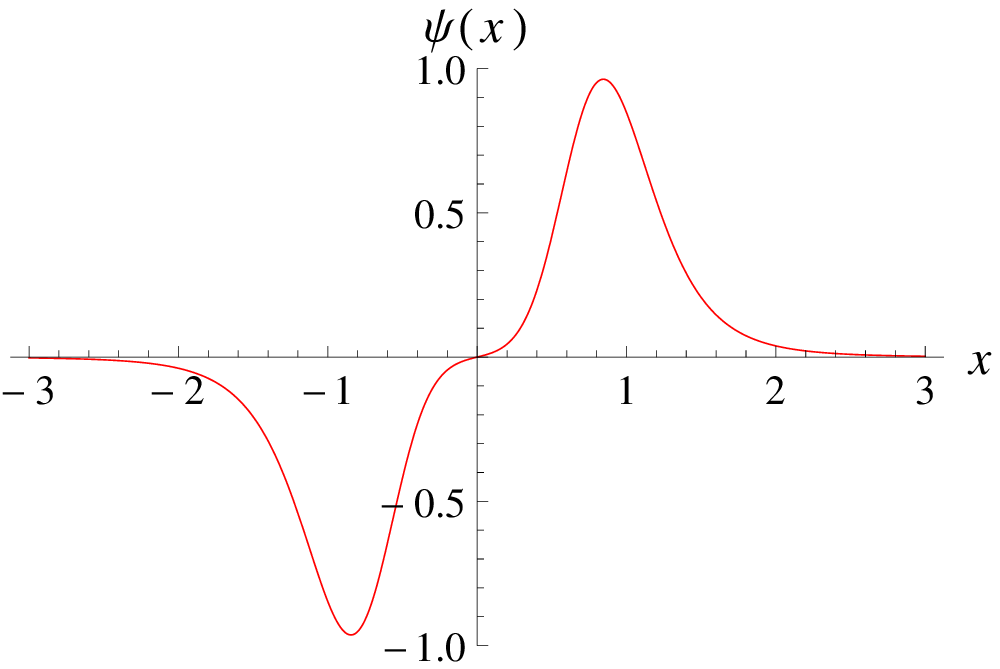}}
{\includegraphics[width=5cm,height=4.5cm]{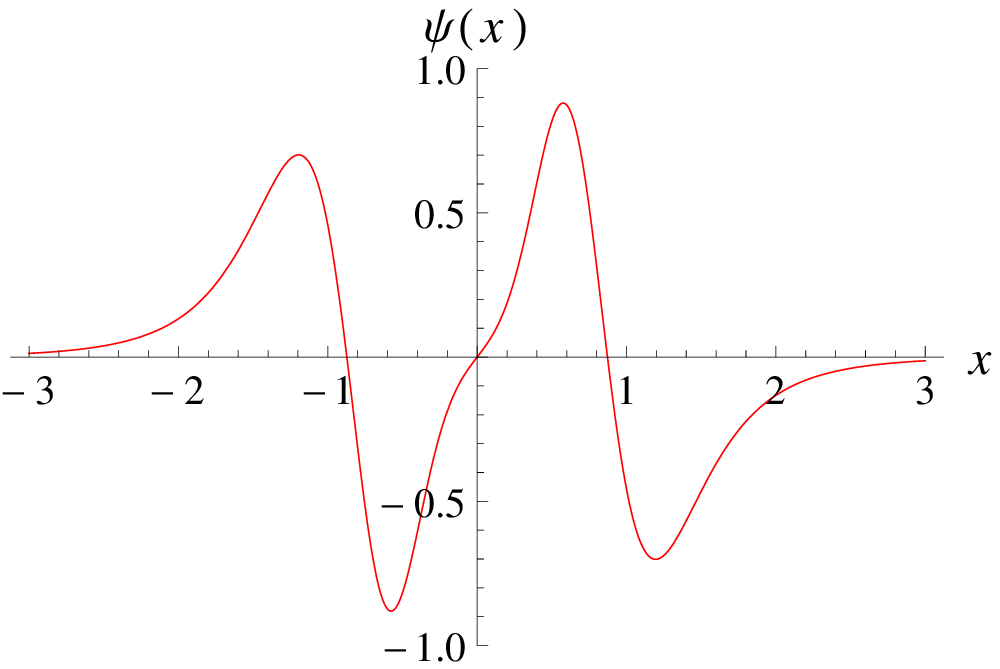}}
{\includegraphics[width=5cm,height=4.5cm]{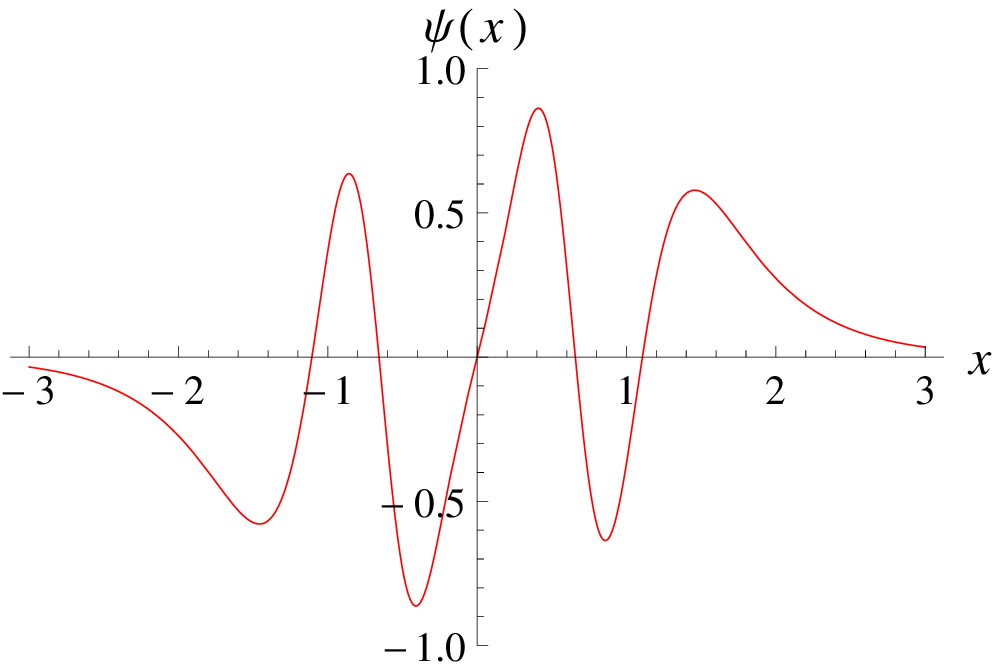}}\\
{\includegraphics[width=4.5cm,height=4.5cm]{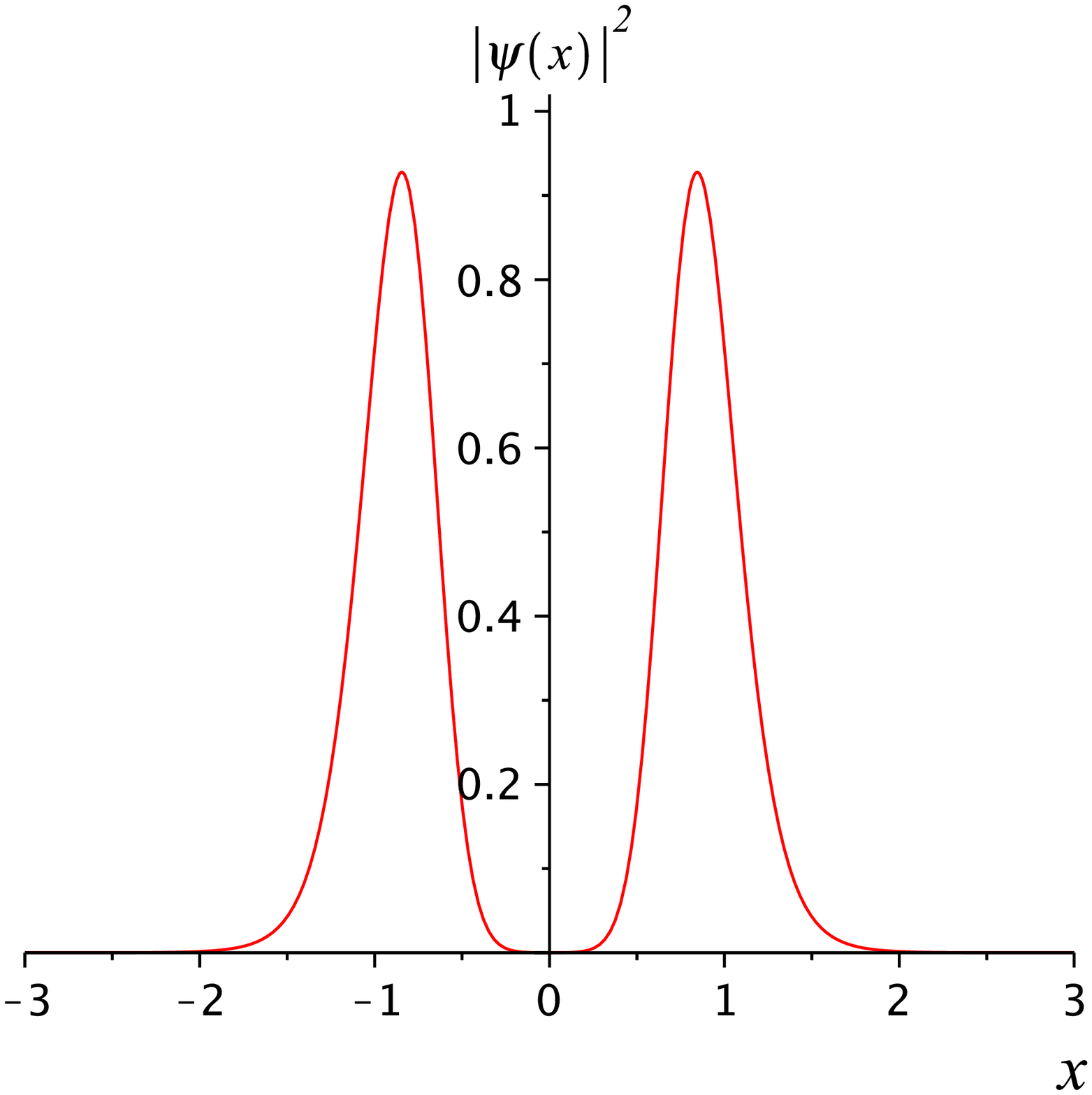}}
{\includegraphics[width=4.5cm,height=4.5cm]{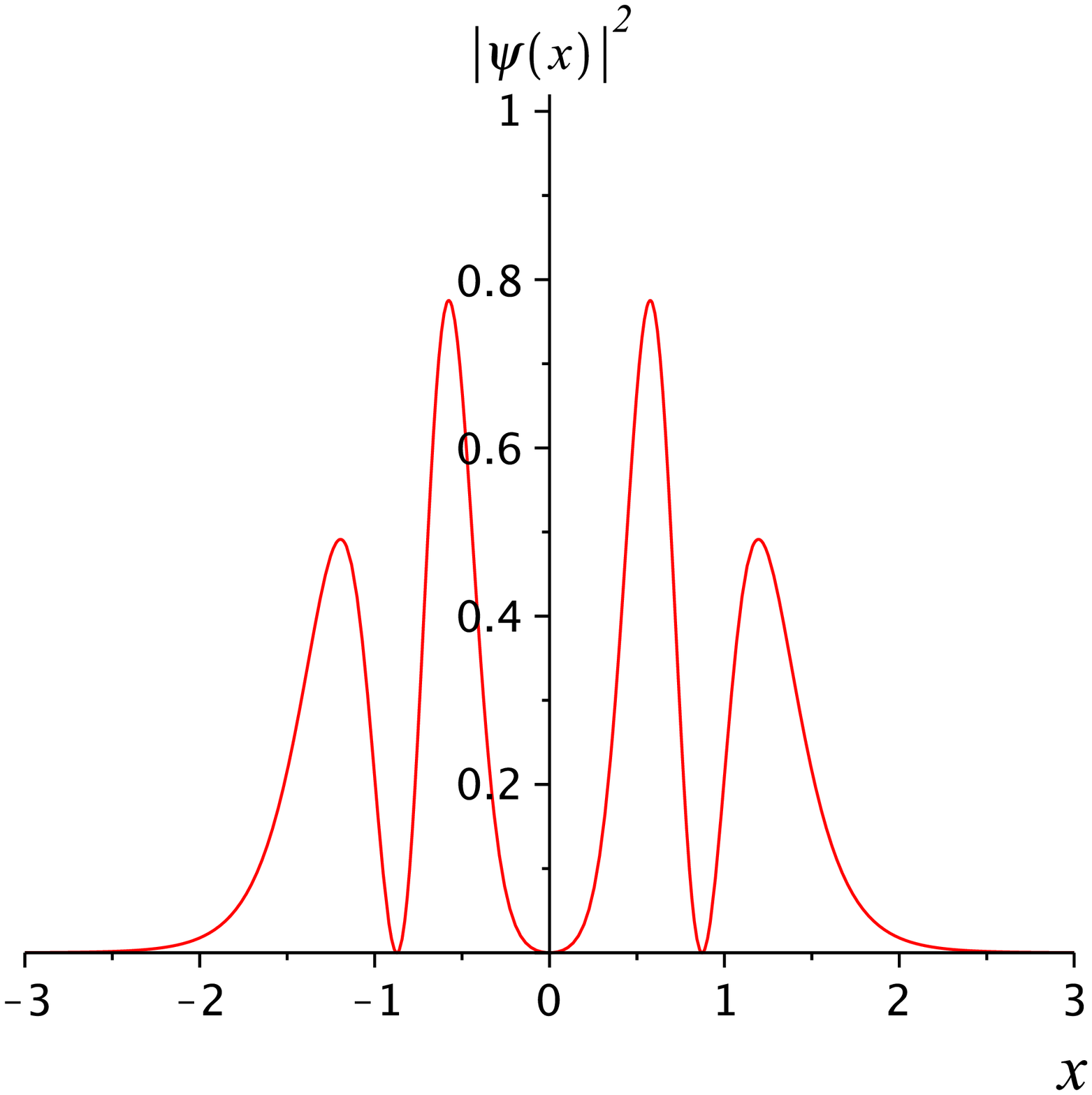}}
{\includegraphics[width=4.5cm,height=4.5cm]{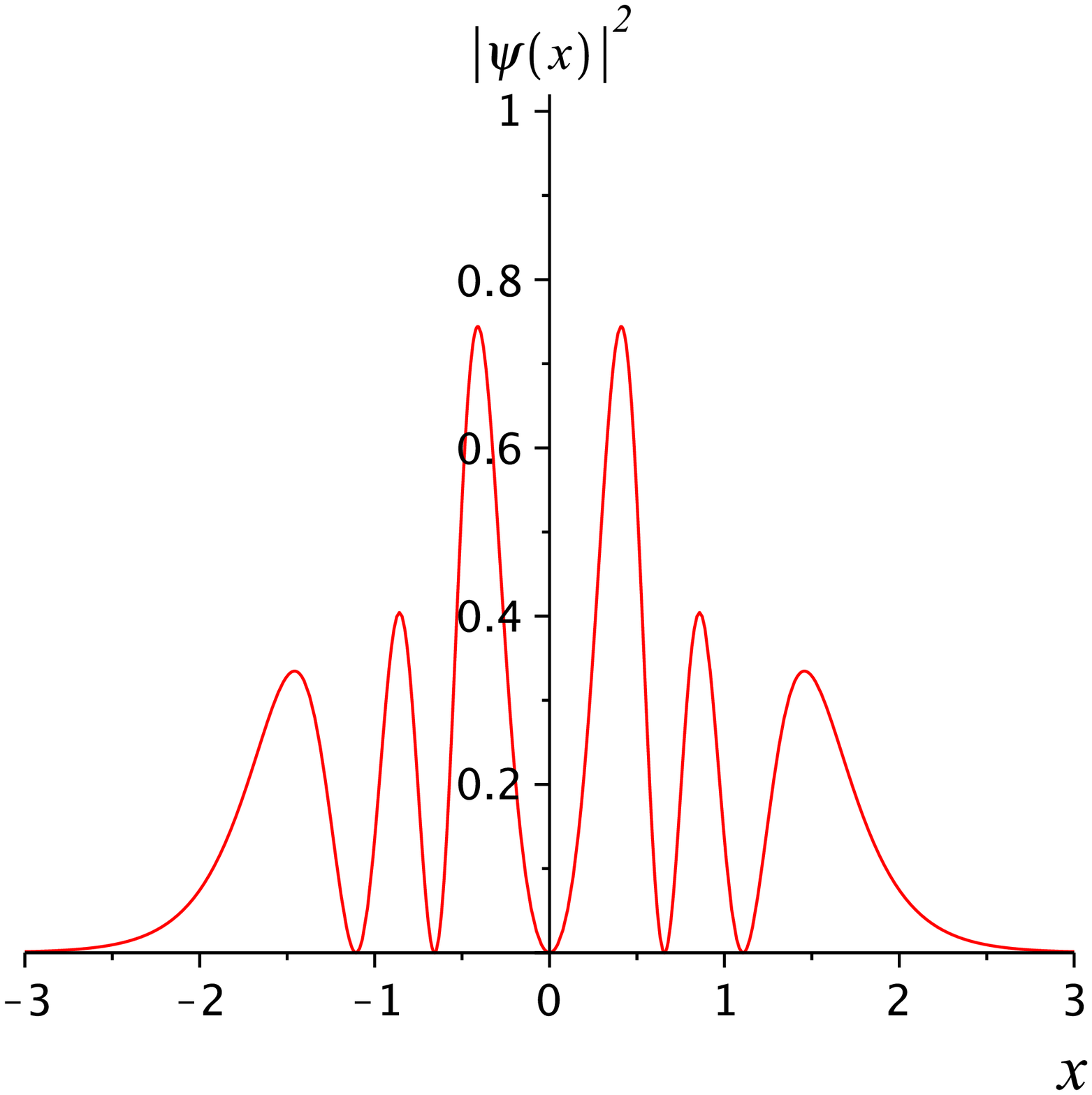}}
\caption{\label{figA_x_A0}
Plot of the normalized antisymmetric solutions
$\psi^{(2)}(x)$  [Eq. (\ref{AManningx})],
in their  three antisymmetric bound-states when $B=-500=-C$
(see top (red) curve in Fig.\ref{graf_VA0}), for $\mathcal{E}_1=-102.2558018532$ (left),
$\mathcal{E}_3=-61.3388827970$ (center) and $\mathcal{E}_5=-24.20206500$ (right).
Below are shown the corresponding probability densities.}
\end{figure}

\begin{figure}[h]
\center
{\includegraphics[width=4.5cm,height=4.5cm]{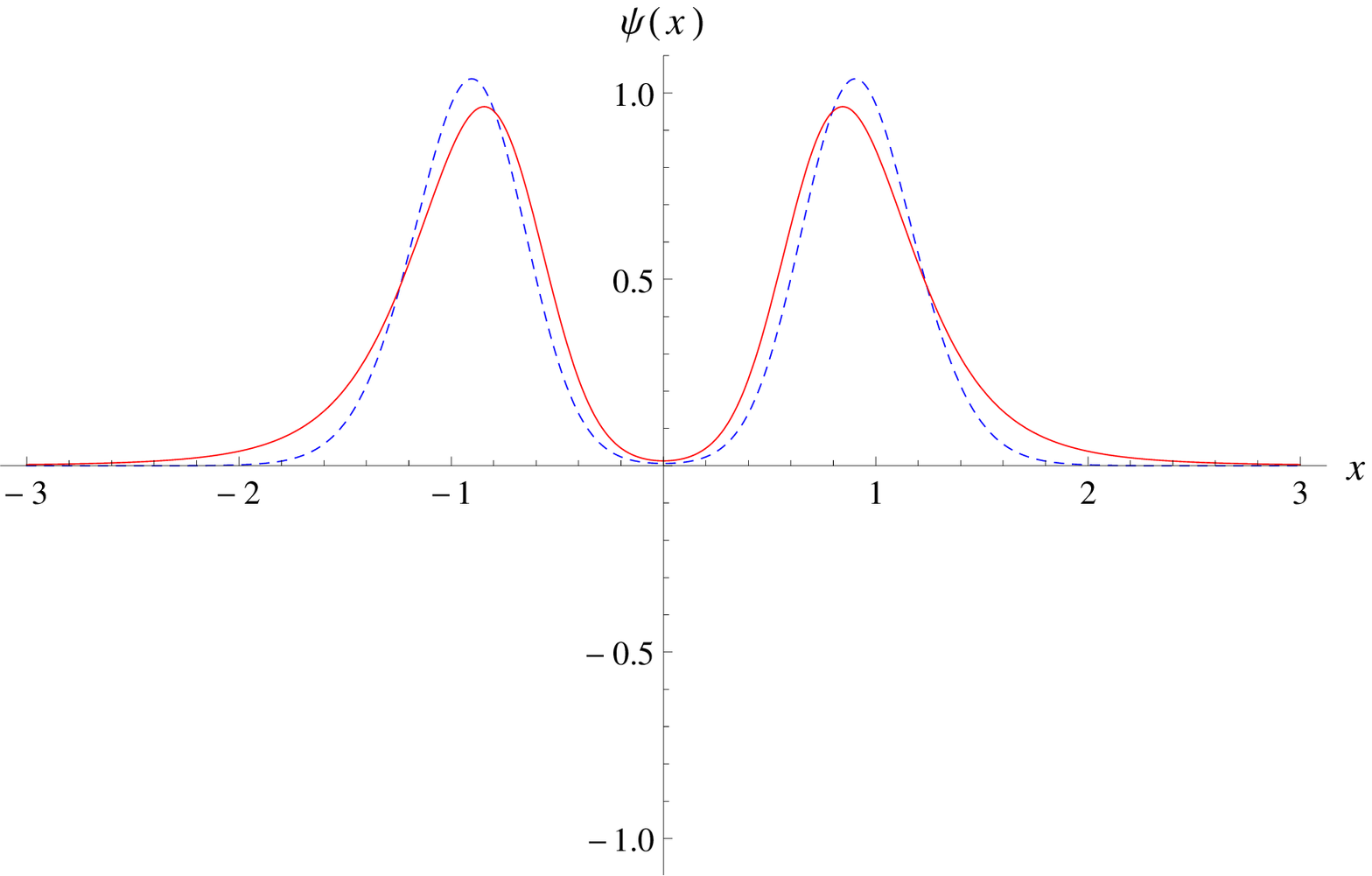}}
{\includegraphics[width=4.5cm,height=4.5cm]{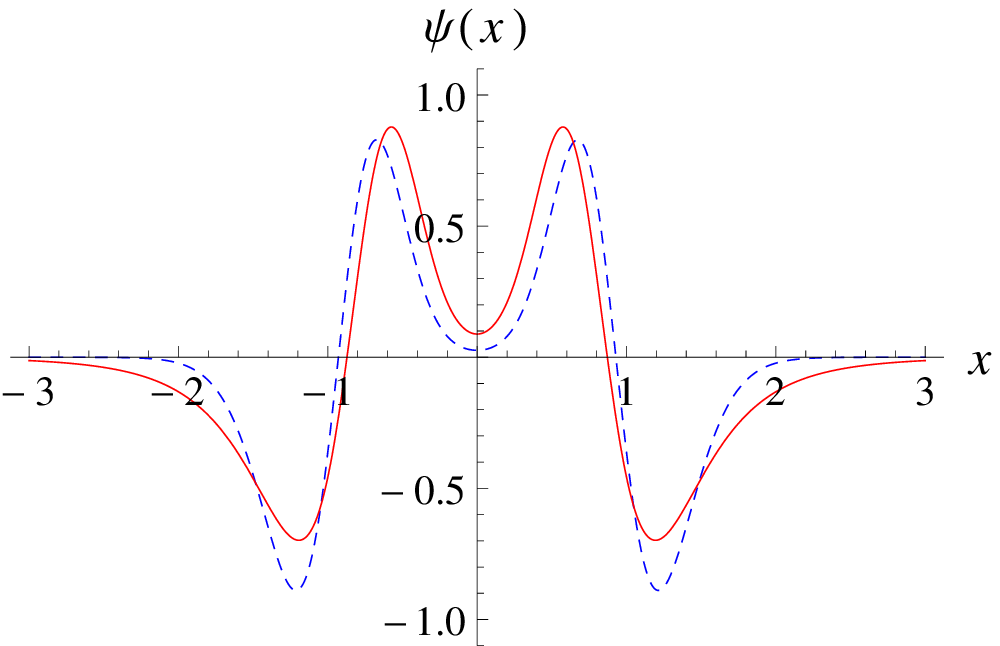}}
{\includegraphics[width=4.5cm,height=4.5cm]{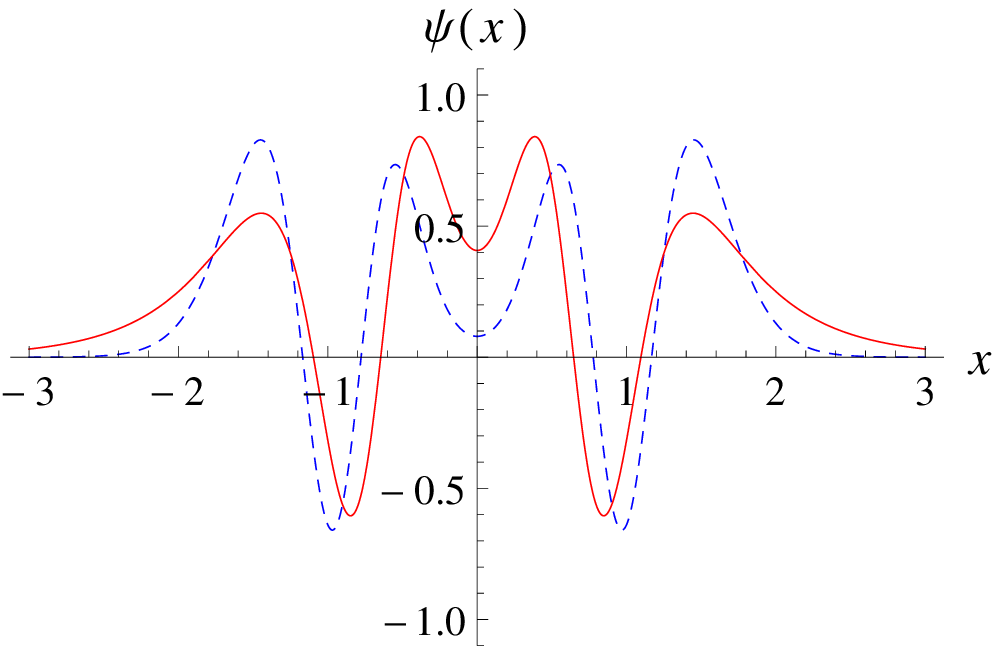}}\\
\caption{\label{figSimA0Compare}  In solid line, the normalized PDM  solutions
$\psi^{(1)}(x)$  [Eq. (\ref{SManningx})]
for the three symmetric bound states $\mathcal{E}_0=-102.25913969050$ (left),
$\mathcal{E}_2=-61.4581676270$ (center) and  $\mathcal{E}_4=-25.9419535530$ (right).
In dashed line the normalized ordinary solutions
$\chi^{(1)}(x)$  [Eq. (\ref{ordManning1})]
for the first three symmetric bound states $\mathcal{E}_0=-109.94122211881$ (left),
$\mathcal{E}_2=-81.887958347499$ (center) and  $\mathcal{E}_4=-57.567702358602$ (right).
Here $B=-C=-500$; see top (red) Manning potential curve in Fig.\ref{graf_VA0}.}
\end{figure}

\begin{figure}[h]
\center
{\includegraphics[width=4.5cm,height=4.5cm]{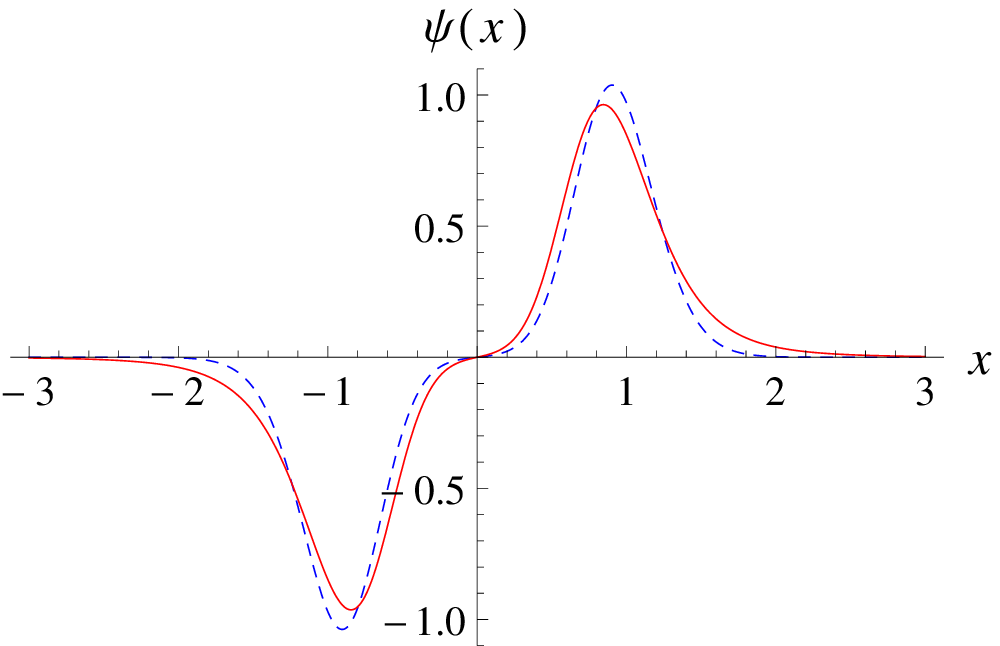}}
{\includegraphics[width=4.5cm,height=4.5cm]{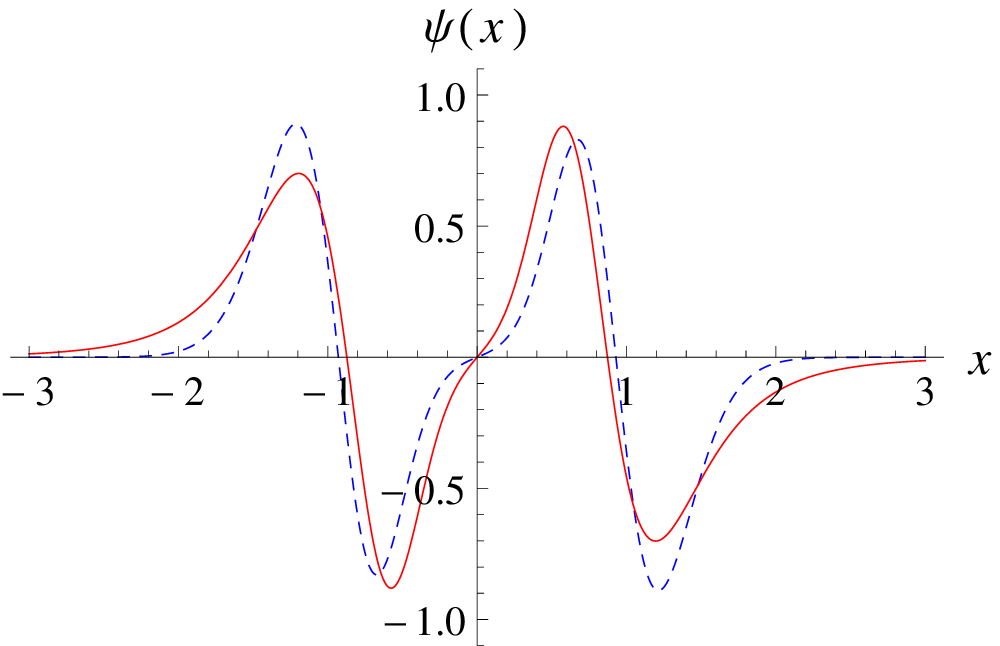}}
{\includegraphics[width=4.5cm,height=4.5cm]{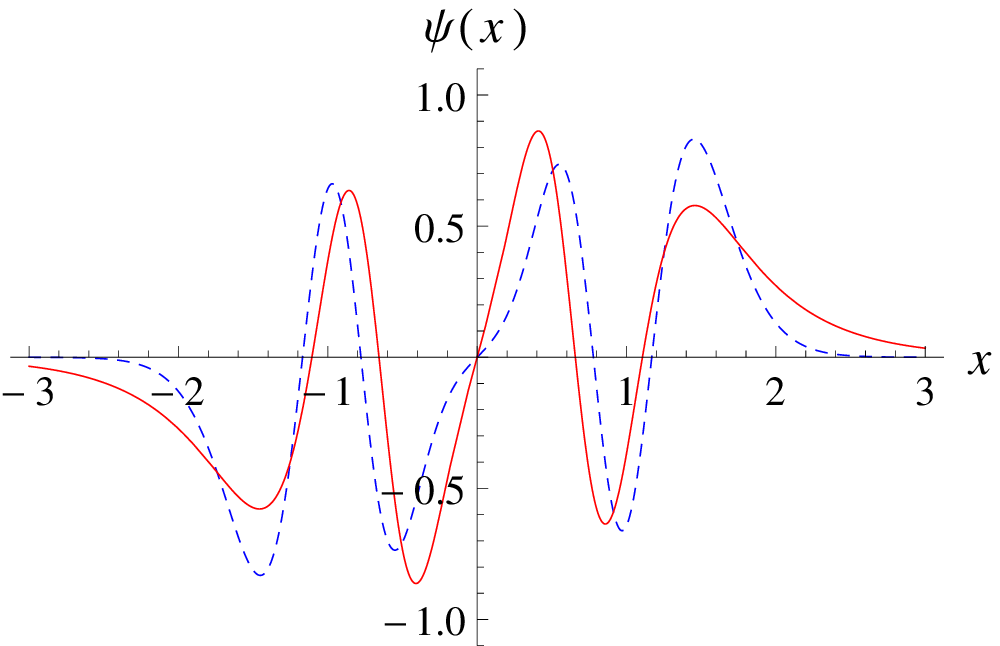}}\\
\caption{\label{figAsimA0Compare}  In solid line, the normalized PDM  solutions
$\psi^{(2)}(x)$  [Eq. (\ref{AManningx})]
for the  three antisymmetric bound states $\mathcal{E}_1=-102.25580185320$ (left),
$\mathcal{E}_3=-61.3388827970$ (center) and  $\mathcal{E}_5=-24.20206500$ (right).
In dashed line the normalized ordinary solutions
$\chi^{(2)}(x)$  [Eq. (\ref{ordManning2})]
for the first three antisymmetric bound states $\mathcal{E}_1=-109.9940489854443$ (left),
$\mathcal{E}_3=-81.87558412881$ (center) and  $\mathcal{E}_5=-57.474984727067$ (right).
Here $B=-C=-500$; see top (red) Manning potential curve in Fig.\ref{graf_VA0}.}
\end{figure}

\noindent
Therefore, the solutions of Eq. (\ref{eqhA0}) are
\bea
h^{(1)}(y) = Hc \left(\sqrt{\mathcal{B}},-\frac{1}{2}, 1, \frac{1}{4}(\mathcal{B}+\mathcal{C}),\frac{1}{2} -\frac{\mathcal{E+\mathcal{B}+\mathcal{C}}}{4};\, y\right)\\
h^{(2)}(y) = \sqrt{y} Hc \left(\sqrt{B},\frac{1}{2}, 1, \frac{1}{4}(\mathcal{B}+\mathcal{C}),\frac{1}{2} -\frac{\mathcal{E+\mathcal{B}+\mathcal{C}}}{4};\, y\right),
\eea
which in z-space result
\bea
\varphi^{(1)}(z) = \cos^\frac{3}{2}\!z\, e^{\frac{\sqrt{\mathcal{B}}}{2}\sin^2\!z}
Hc\! \left(\sqrt{\mathcal{B}},-\frac{1}{2}, 1, \frac{1}{4}(\mathcal{B}+\mathcal{C}),\frac{1}{2}
-\frac{\mathcal{E+\mathcal{B}+\mathcal{C}}}{4};\, \sin^2\!z \right)\\
\varphi^{(2)}(z) = \sin\!z\,\cos^\frac{3}{2}\!z\, e^{\frac{\sqrt{\mathcal{B}}}{2}\sin^2\!z}
Hc\! \left(\sqrt{\mathcal{B}},\frac{1}{2}, 1, \frac{1}{4}(\mathcal{B}+\mathcal{C}),\frac{1}{2}
-\frac{\mathcal{E+\mathcal{B}+\mathcal{C}}}{4};\, \sin^2\!z \right),
\eea
and in $x$-space read finally
\bea
&&\psi^{(1)}(x) = \sech^2\!x\, e^{\frac{\sqrt{\mathcal{B}}}{2}\tanh^2\!x}
Hc\! \left(\sqrt{\mathcal{B}},-\frac{1}{2}, 1, \frac{1}{4}(\mathcal{B}+\mathcal{C}),\frac{1}{2}
-\frac{\mathcal{E+\mathcal{B}+\mathcal{C}}}{4};\, \tanh^2\!x \right)\label{SManningx}\\
&&\psi^{(2)}(x) = \tanh x\,\sech^2\!x\, e^{\frac{\sqrt{\mathcal{B}}}{2}\tanh^2\!x} Hc\!
\left(\sqrt{\mathcal{B}},+\frac{1}{2}, 1, \frac{1}{4}(\mathcal{B}+\mathcal{C}),\frac{1}{2}
-\frac{\mathcal{E+\mathcal{\mathcal{B}}+\mathcal{C}}}{4};\, \tanh^2\!x \right).\label{AManningx}
\eea
In Figs.~\ref{figS_x_A0} and \ref{figA_x_A0} we plot the bound-state wavefunctions 
and probability densities of a PDM particle.  
We have numerically computed the energies of the eigenstates that satisfy vanishing boundary 
conditions and found just six bound states in this \textit{PDM}-Manning potential.
Although the three pairs of probability distributions are very close, the corresponding solutions
are certainly different (all the symmetric solutions are nonzero at the origin while
the antisymmetric ones are of course null). This is particularly apparent for the third pair of
eigenfunctions where the tunneling effect is highly manifest in the $\mathcal{E}_4$ eigenstate.
As we foreseen, the PDM analytic expressions are quite different from
the ordinary constant-mass solutions to the Manning potential found in \cite{sech246}
\bea
&&\chi^{(1)}(x) = (\sech x)^{\sqrt{-E}}\,
Hc\! \left(0,-\frac{1}{2}, \sqrt{-E}, \frac{1}{4} {B},\frac{1}{4}
-\frac{E+ {B}+ {C}}{4};\, \tanh^2\!x \right)  \label{ordManning1}\\
%= \\ && C_1\sech(x)^{\sqrt{-E}}\, \text{Hc}\left(0, \sqrt{-\mathcal{E}}, -\frac{1}{2}, %-\frac{\mathcal{B}}{4},\, \frac{1}{4}-\frac{\mathcal{E}+\mathcal{C}}{4};\,\sech^2(x)\right)\label{ordManning1}\\
&&\chi^{(2)}(x) = \tanh x\,(\sech x)^{\sqrt{-E}}\,
Hc\! \left(0,\frac{1}{2}, \sqrt{-E}, \frac{1}{4} {B},\frac{1}{4}
-\frac{E+ {B}+ {C}}{4};\, \tanh^2\!x \right).\label{ordManning2}
\eea
In Figs.~\ref{figSimA0Compare} and \ref{figAsimA0Compare} we show both pairs of curves for the
closest possible eigenenergies found for the two problems. We observe similar shapes in both sets
with a rapidly increasing deviation from the ordinary constant-mass case for the higher eigensates.
Note that
in all the eigenstates the PDM particle has more probability to be near the origin of coordinates and
thus keep on tunneling across the potential barrier.
Another remarkable point is that while in the  constant-mass case there exist fourteen bound-states
in the PDM case there are only six. This shows a kind of merging of eigensates
and a lower number of physical possibilities for growing energies assuming PDM (see Table \ref{table Manning}).

%%%%%%%%%%%%%%%%%%%%%%%%%%%%%%%%  TABLE  14x6  %%%%%%%%%%%%%%%%%%%%%%%%%%%%%
\begin{table}[hb]
\caption{\label{table Manning} Complete list of the energy eigenvalues of the
\textit{PDM} and constant-mass Manning hamiltonians for $A=0$ and $B=-C=-500$.
The $_S$ and $_A$ subindexes at left indicate symmetric and antisymmetric states.}
\vskip 0.2cm%
\begin{tabular}{c c c }
  \hline  \hline
            \,  \, & Constant mass                     \, & \, $PDM$\\ \hline
% \,  \, & \, $--$ 					 \, & \, $--$ \\
 \, $E_S^1$   \, & \,$-109.9412221188093$   \, & \, $-102.2591396905$\\
   \, $E_A^2$   \, & \,$-109.9940489854443$   \, & \, $-102.2558018532$\\
    \, $E_S^3$   \, & \, $-81.887958347499$    \, & \, $-61.458167627$ \\
     \, $E_A^4$   \, & \,$-81.875584128810$     \, & \, $-61.3388827970$  \\
      \, $E_S^5$   \, & \,$-57.567702358602$     \, & \, $-25.941953553$  \\
       \, $E_A^6$   \, & \,$-57.474984727067$     \, & \, $-24.202065000$  \\
        \, $E_S^7$   \, & \,$-37.240150270295$     \, & \,      $--$        \\
         \, $E_A^8$   \, & \,$-36.841246822072$     \, & \,      $--$        \\
          \, $E_S^9$   \, & \,$-21.195042009000$     \, & \,      $--$        \\
           \, $E_A^{10}$\, & \,$-20.147434878873$     \, & \,  	   $--$        \\
            \, $E_S^{11}$\, & \,\,\,$-9.457236339000$  \, & \,     $--$         \\
             \, $E_S^{12}$ \, & \,$-7.8621835775695$    \, & \,     $--$\\
              \, $E_S^{13}$ \, & \,$-2.02308205000\,\,\,$\, & \,    $--$          \\
               \, $E_S^{14}$ \, & \,$-0.961473079820$     \, & \,    $--$          \\
            \hline
             \hline
\end{tabular}
\end{table} %

\newpage
{~}
\newpage
%%%%%%%%%%%%%%%%%%%%%%%%%%%%%%%%%%%%%%%%%%%%%% sixth-order

\subsection{The \textit{PDM} sixth-order hyperbolic potentials \label{sec:sech64}}

\begin{figure}[h]
\center
{\includegraphics[width=6.cm,height=6cm]{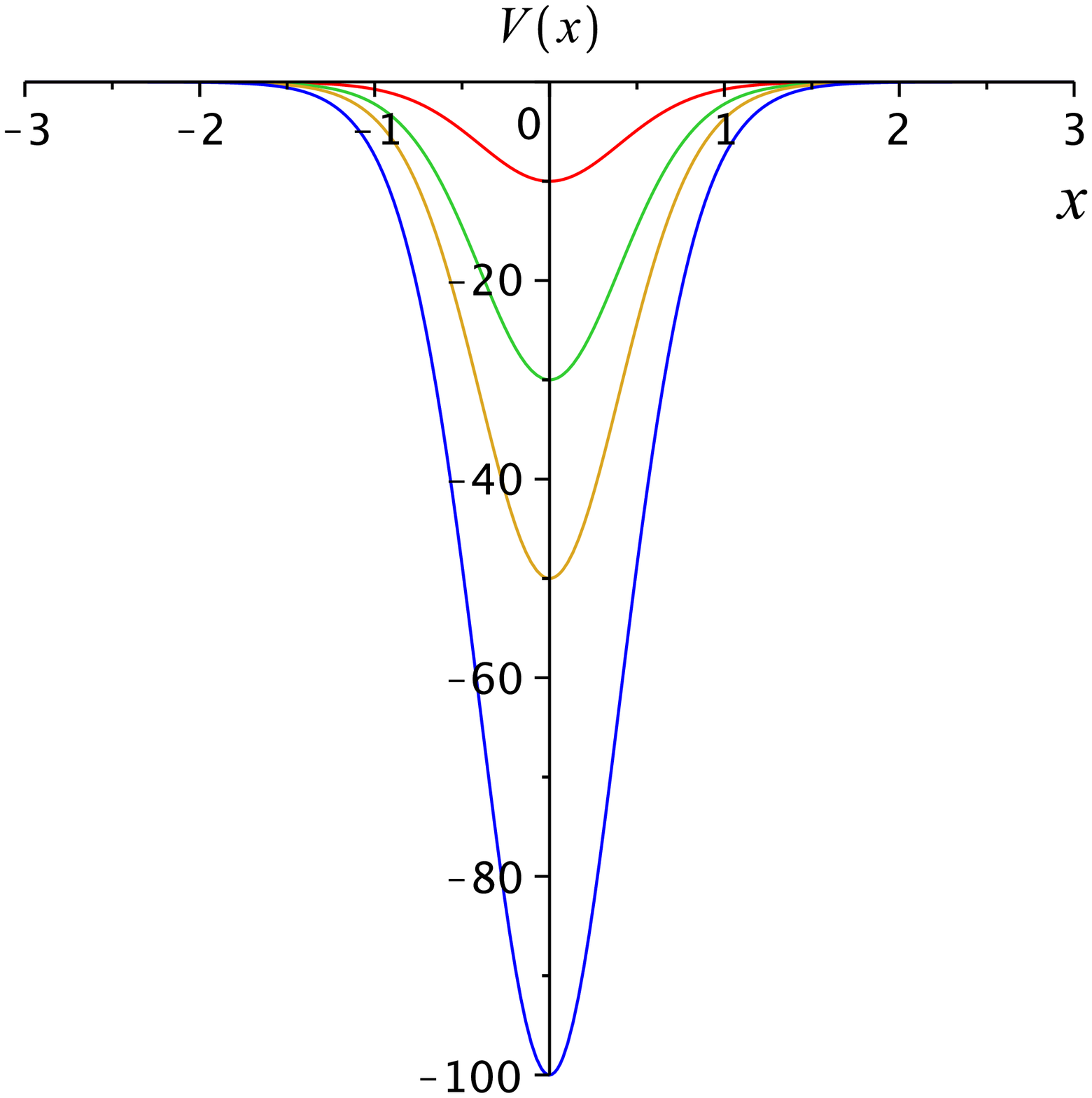}}\hspace{1cm}
{\includegraphics[width=6.cm,height=6cm]{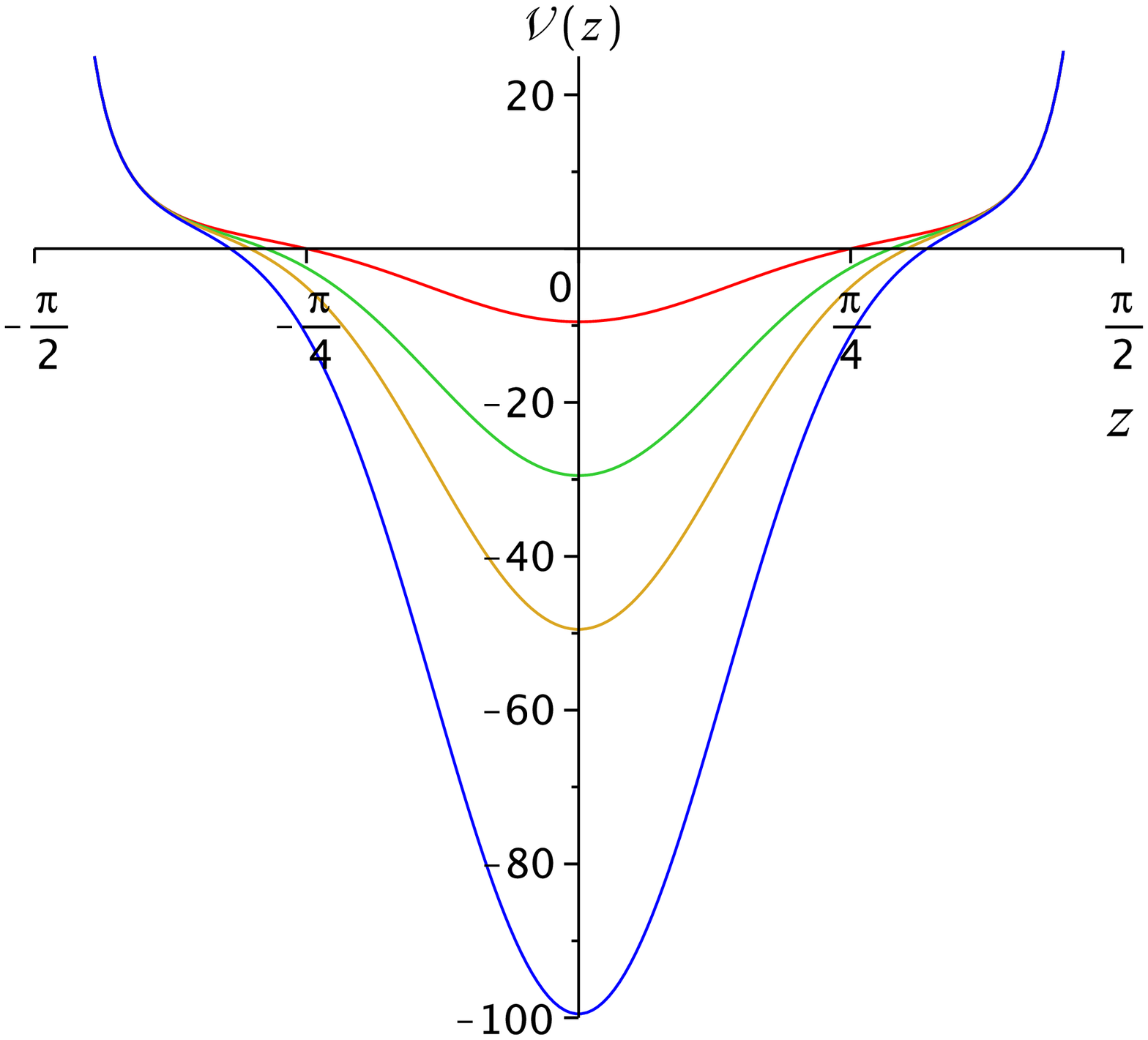}}
\caption{\label{graf_sech6x} From top to bottom, plot of well potentials $V(x)= A \sech^6(x)$ (left) and
the corresponding effective potentials $\mathcal{V}(z)$ (right) for ${A} = 10, 30, 50$ and 100.}
\end{figure}

Regarding the sixth-order members of family (\ref{ipotsech246}) we have tried
 to analytically disentangle the full problem but it seems too complex.
In any case,  we have been able to find the exact solution to
the \textit{PDM}-modified differential equation for one free parameter in two specific
cases:
%As a matter of fact, we have managed to analytically treat the two following situations:
$B = 0, C=0$, namely $V(x)= -A \sech^6(x)$, and $A = - B,  C=0$,
that is $V(x)=  -A (\sech^6(x) -  \sech^4(x))$,  for $E=0$, both yielding
\textit{triconfluent} Heun eigenfunctions \cite{ronveaux}  (see also e.g. \cite{scripta2002}).
In Fig. \ref{graf_sech6x} and Fig. \ref{graf_Vsech_C0}
we show these potentials for several values of the free parameter.

%%%%%%%%%%%%%%
\subsubsection{Case free $A$ and $ {B}=C=0$}
In the first case, the z-space eq. (\ref{schrodinger-cons}) to solve is
\beq
\varphi''(z)- \left(\frac{1}{2} + \frac{3}{4}\tan^2(z)-\mathcal{A}\cos^6(z)\right) \varphi(z)=0.
\eeq
We first factorize
$\varphi(z)=\cos^\sigma(z)\,\phi(z)$,
bearing
\beq
\phi''(z)-2\sigma\tan(z)\phi'(z)+\Big[(\sigma^2-\sigma-\tquarto)\tan(z)^2
-\sigma-\meio +\mathcal{A}\cos(z)^6\Big]\phi(z) = 0.
\eeq
We then cut down this equation by choosing $\sigma=-\meio$.
Now we transform the variable by means of $y=\sin^2z$ and get
\beq%
(-y^2+1)\,\phi''(y)+(A (-y^2+1)^3)\,\phi(y) = 0. \label{pre-tri}
\eeq%
It looks as the \textit{representative} form of
the triconfluent Heun equation and therefore we try with an exponential ansatz of the form
$\phi(y)=e^{ay^3+by} h(y)$ (see Proposition E.1.2.1 in \cite{ronveaux}),
so that Eq. (\ref{pre-tri}) results
\beq
h''(y)+(6 a y^2+2 b) h'(y)+\Big[6ay+(3ay^2+b)^2+\mathcal{A} (-y^2+1)^2\Big]\,h(y) = 0.
\eeq
This can be further shorten by means of $a=-\terco\sqrt{-\mathcal{A}}$ and $b=\sqrt{-\mathcal{A}}$
and if we next define the variable
$$\bar{y}=\left(\frac{2\sqrt{-\mathcal{A}}}{3}\right)^{\!\!\frac{1}{3}}\,y$$
 we obtain
\beq
h''(\bar{y})-\left[{3}\,\bar{y}^2 -(-12 \mathcal{A})^{\!\!\terco}
\right]h'(\bar{y}) -3\,\bar{y}\,h(\bar{y}) = 0.
\eeq
Now, this can be readily compared with
\beq
H''(u)- (\gamma+3u^2)H'(u)+[\alpha+(\beta-3) u]H(u)=0
\eeq
which is known as the \textit{canonical} triconfluent form of the Heun equation.
Its L.I. solutions are
\bea
H^{(1)}(u)&=&Ht(\alpha, \beta, \gamma; u)\\
H^{(2)}(u)&=&e^{u^3+\gamma u}Ht(\alpha, -\beta, \gamma; -u).
\eea

The triconfluent Heun equation is obtained from the \textit{biconfluent}
form through a process in which two singularities coalesce by redefining parameters
and taking the appropriate limits. See \cite{ronveaux}
for a detailed discussion about the confluence procedure
in the case of the Heun equation and its different forms.
%Proposition 1.2.1 of Part E \cite{ronveaux} relates precisely by a factor $e^{ay^3+by}$
%the solutions to the canonical and the representative forms of the triconfluent Heun equation.
The function $Ht(\alpha,\beta,\gamma;u)$  is a local solution around the origin,
which is a regular point.
Because the single singularity is located at infinity, this series
converges in the whole complex plane and consequently its solutions
can be related to the Airy functions \cite{abramowitz}.

\begin{figure}[h]
\center
{\includegraphics[width=4.5cm,height=4cm]{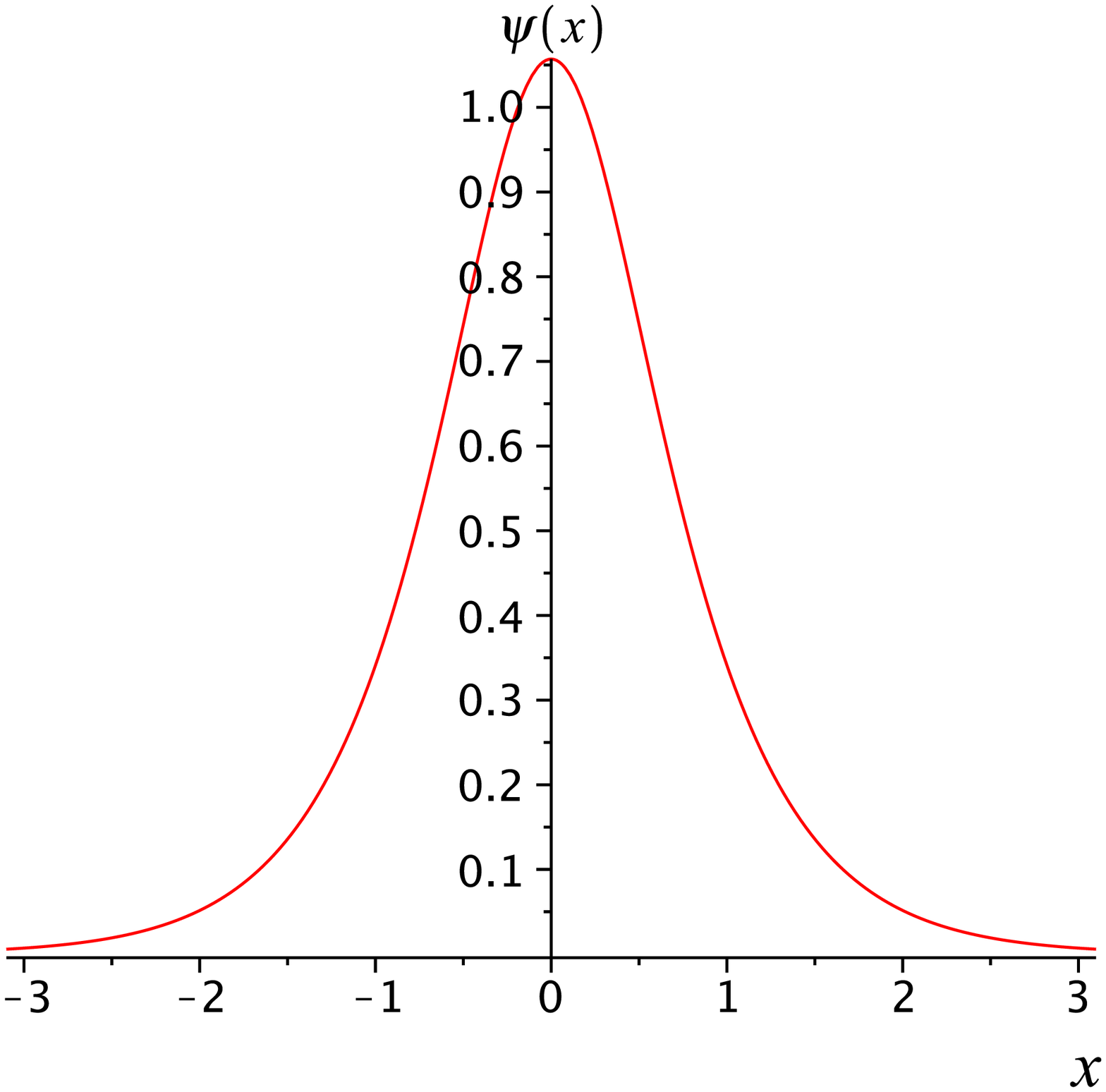}}
{\includegraphics[width=4.5cm,height=4cm]{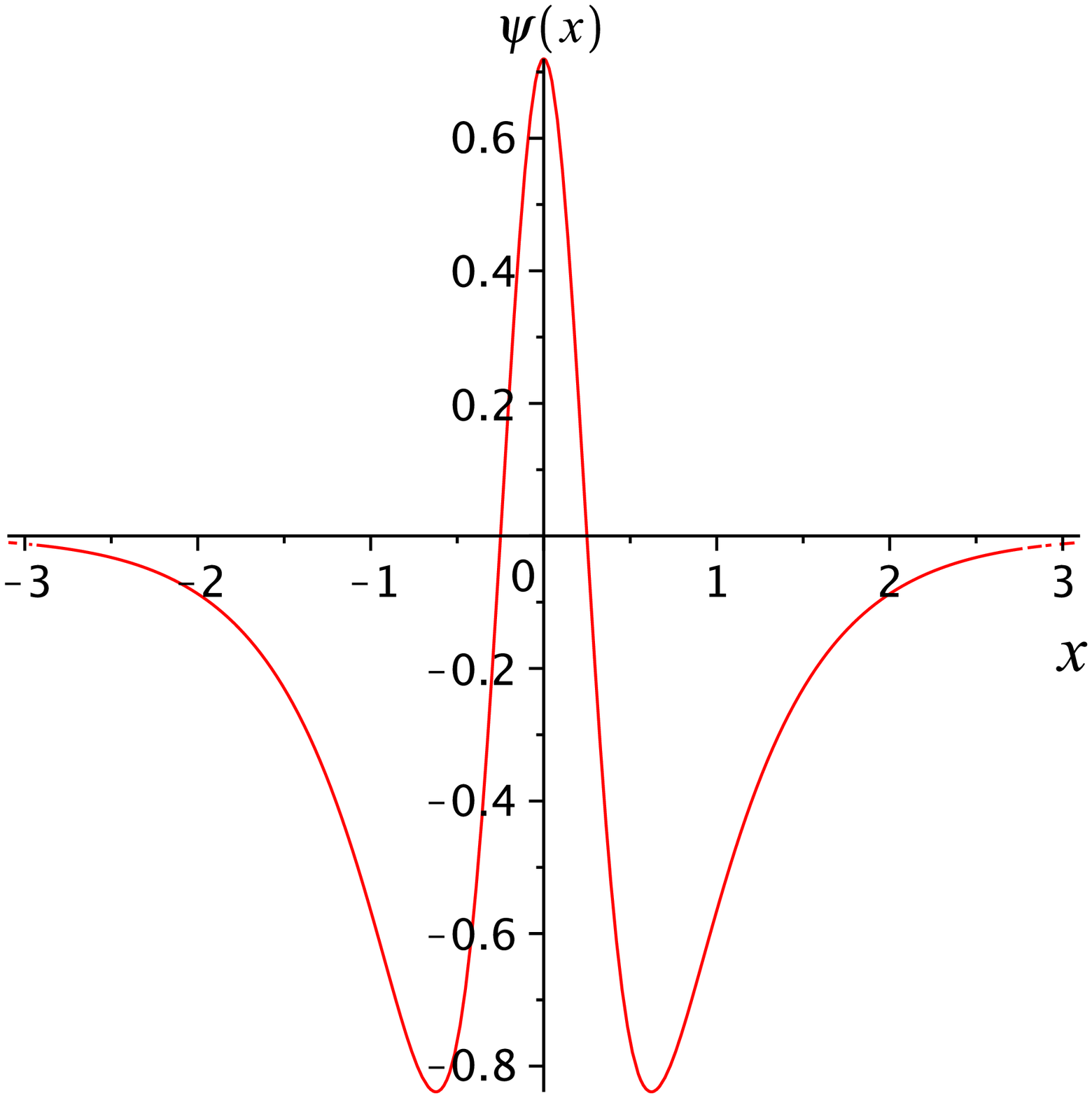}}
{\includegraphics[width=4.5cm,height=4cm]{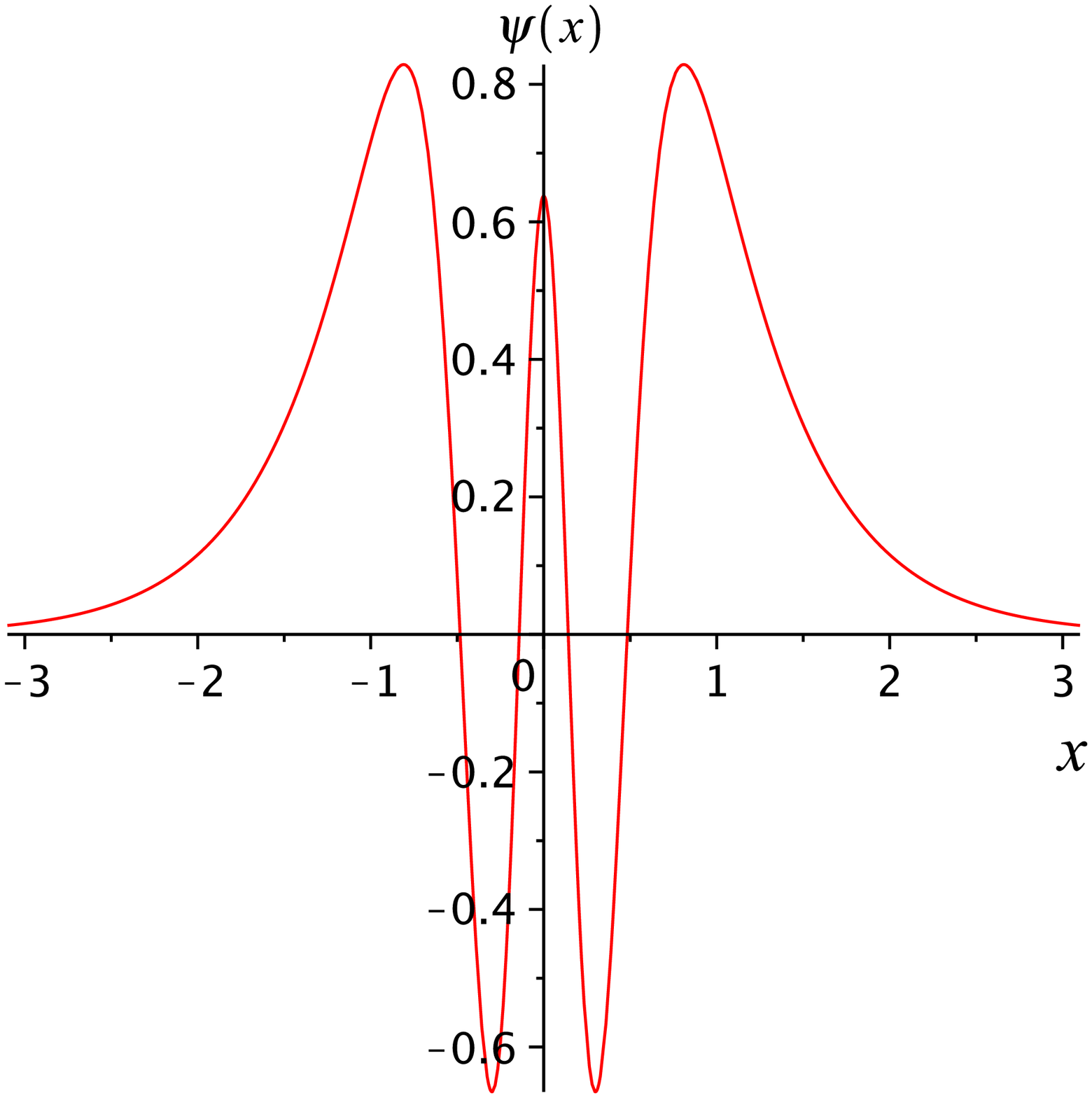}}\\
{\includegraphics[width=4.5cm,height=4cm]{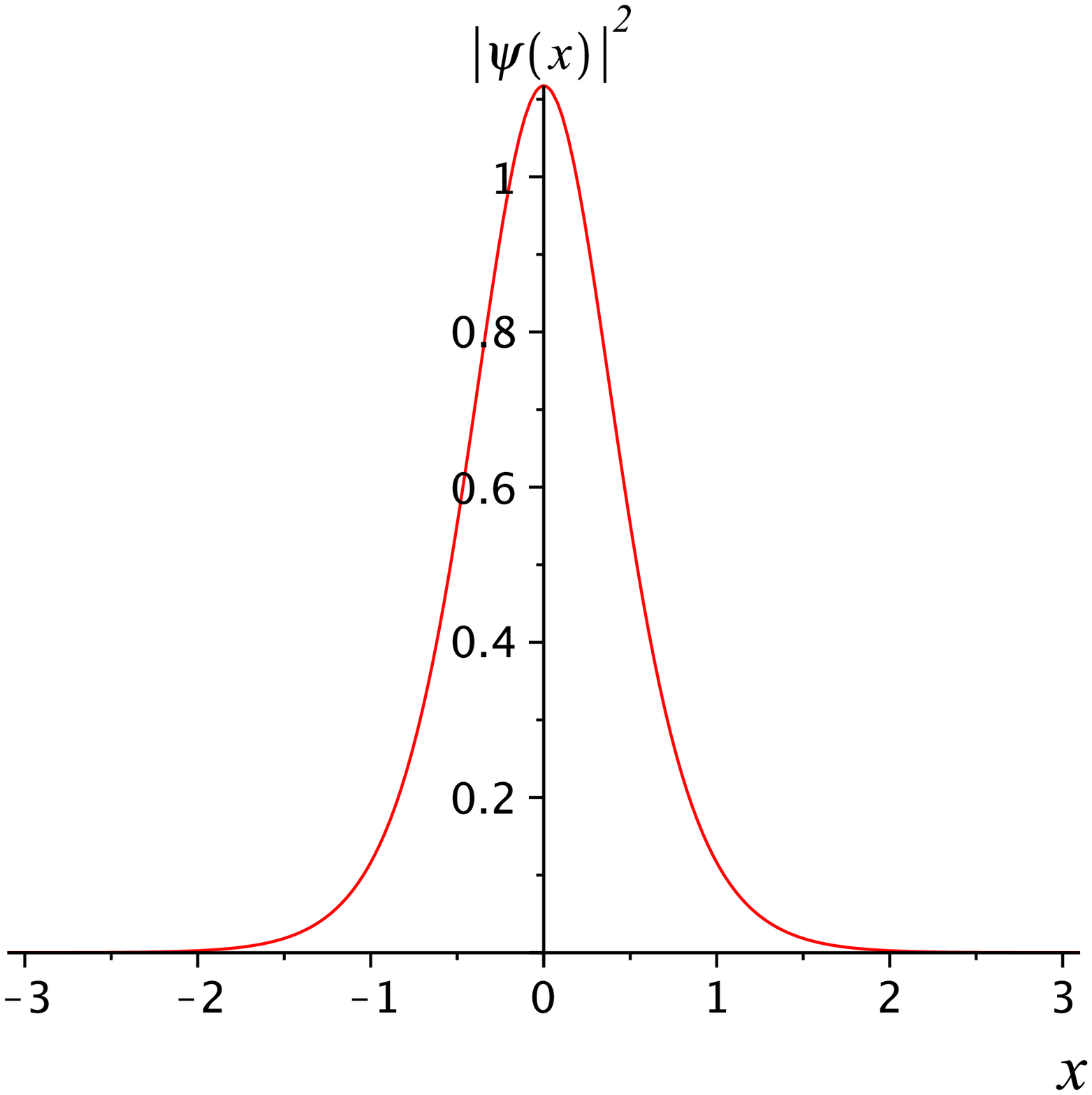}}
{\includegraphics[width=4.5cm,height=4cm]{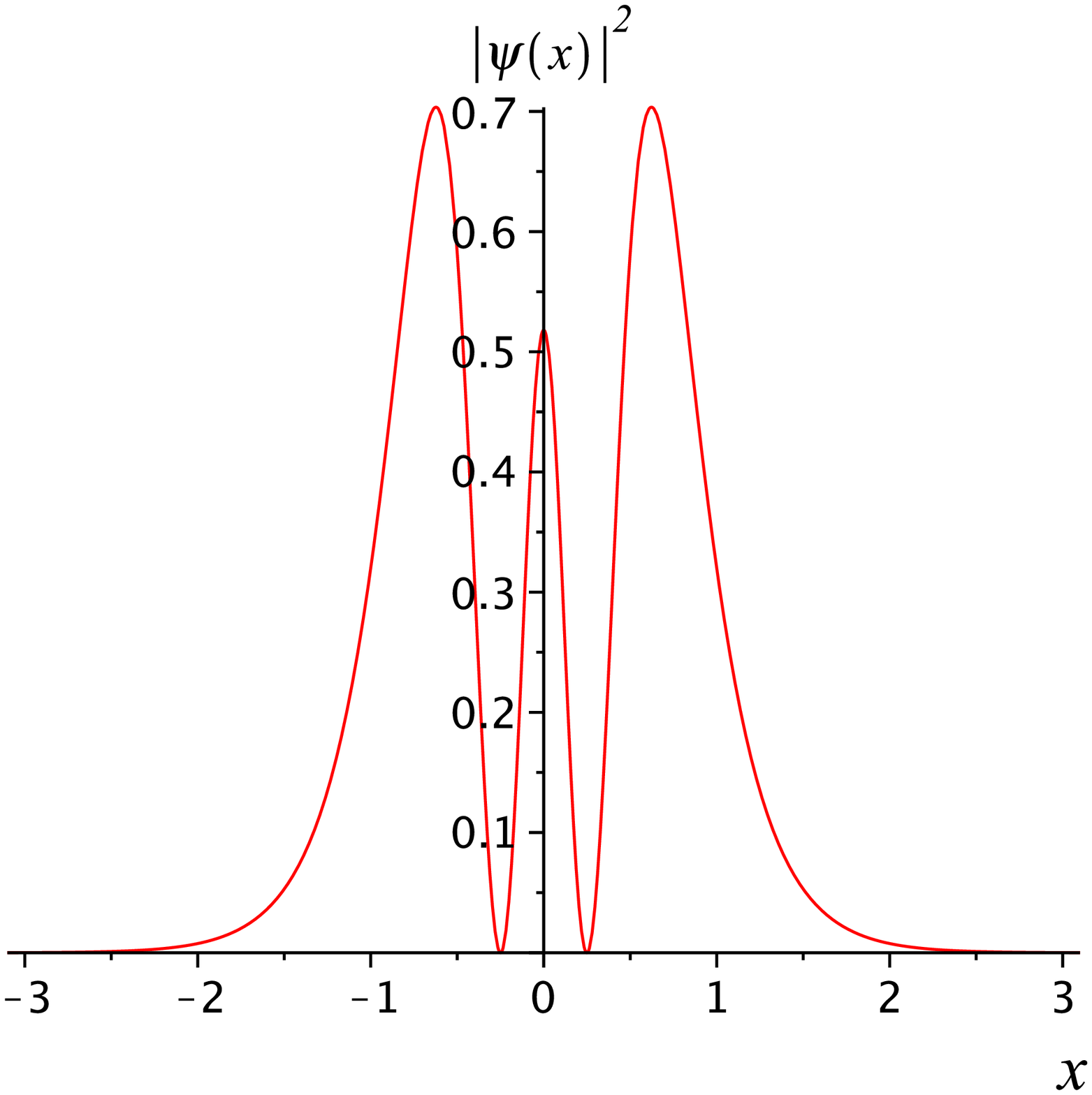}}
{\includegraphics[width=4.5cm,height=4cm]{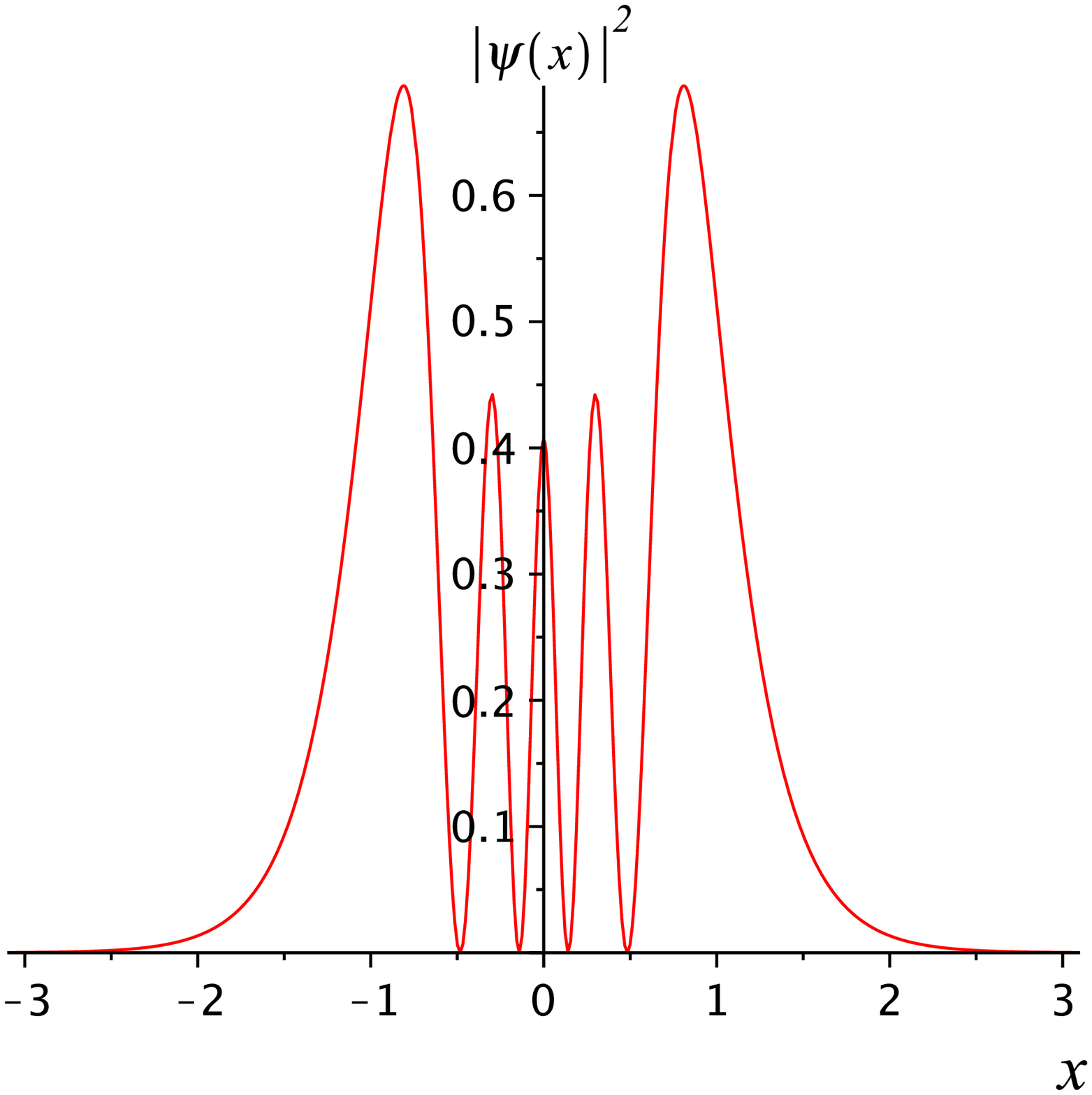}}
\caption{\label{figsech6_prob_x_Sim}
Plot of symmetric solutions $\psi_s(x)$ [up] and the
corresponding probability densities $|\psi_s(x)|^2$ [down]
given $\mathcal{B}=\mathcal{C}=0$  and $\mathcal{E}=0$,
for  $\mathcal{A} = 3.131784324$ (left);
$\mathcal{A} = 41.919051$ (center); and  $\mathcal{A} = 125.162981$ (right).}
\end{figure}

\begin{figure}[h]
\center
{\includegraphics[width=4.5cm,height=4cm]{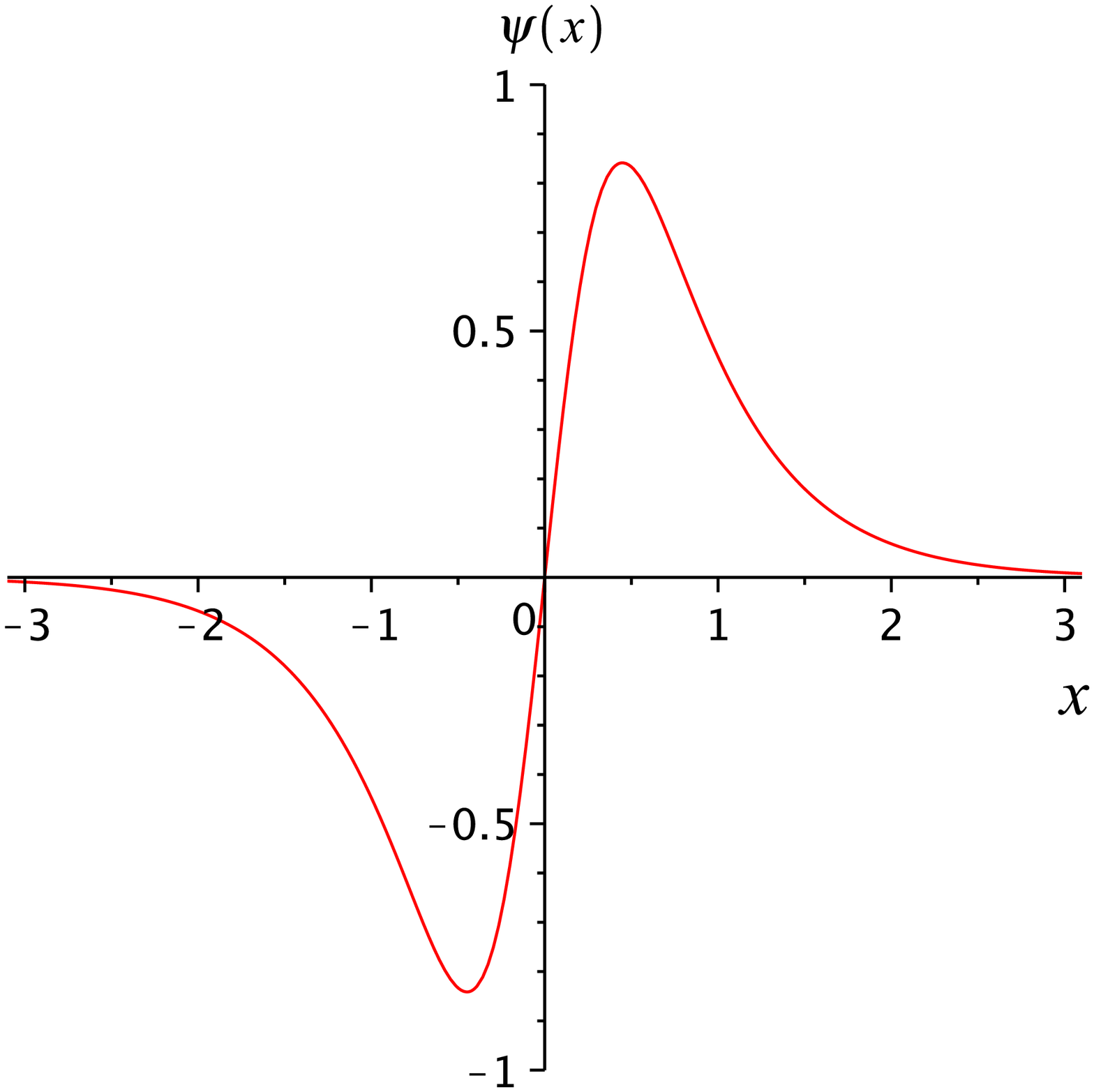}}
{\includegraphics[width=4.5cm,height=4cm]{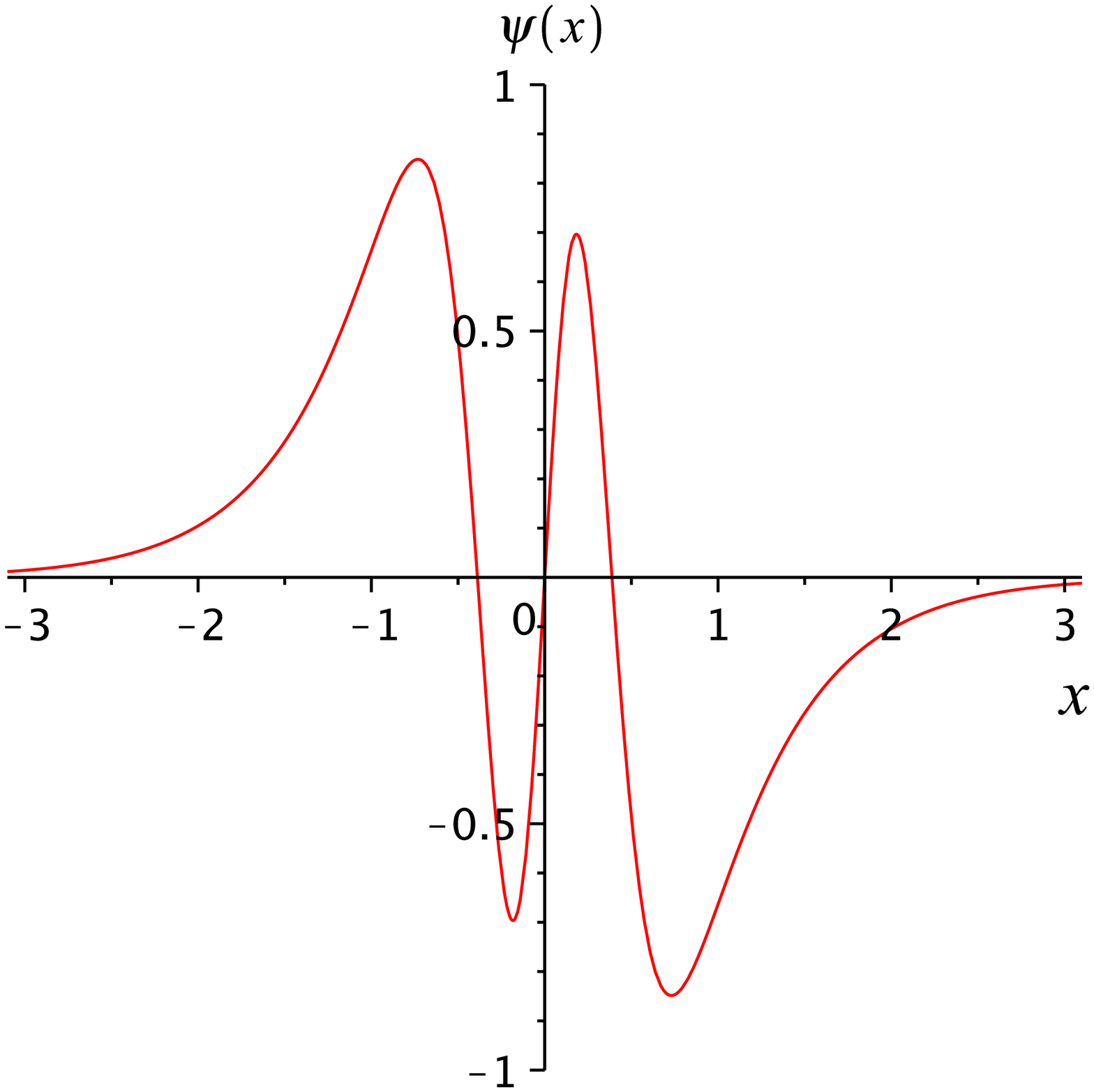}}
{\includegraphics[width=4.5cm,height=4cm]{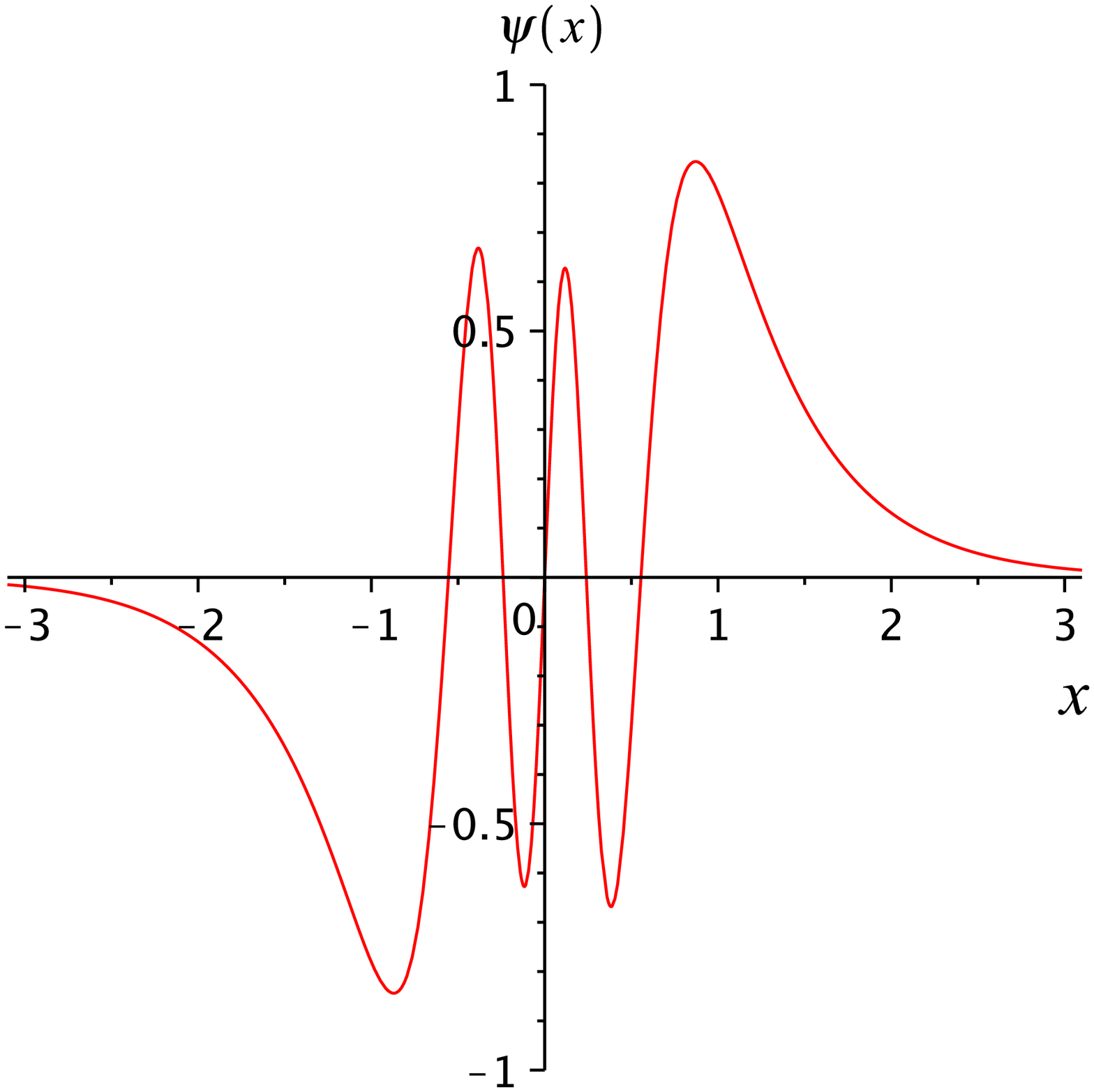}}\\
{\includegraphics[width=4.5cm,height=4cm]{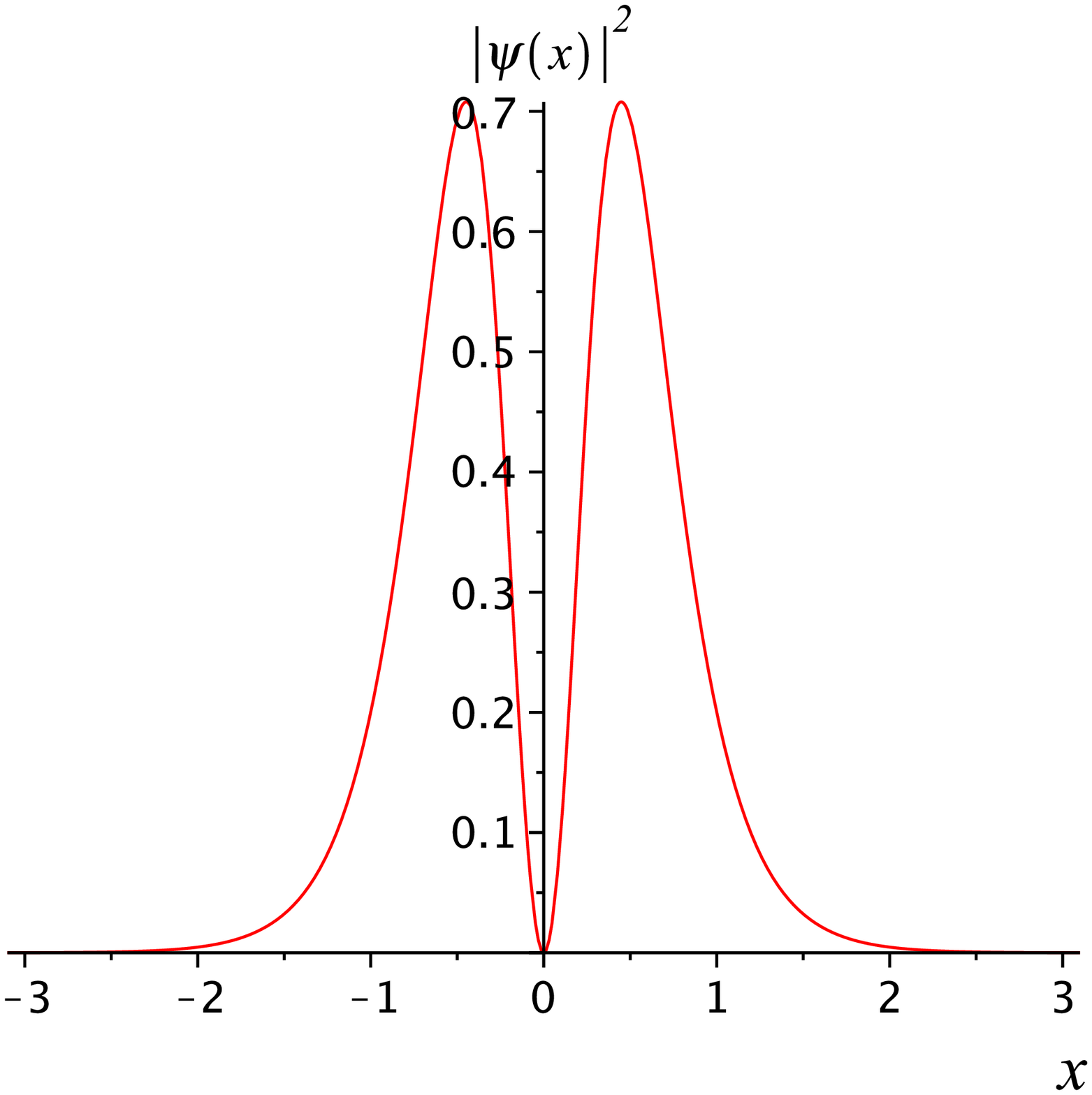}}
{\includegraphics[width=4.5cm,height=4cm]{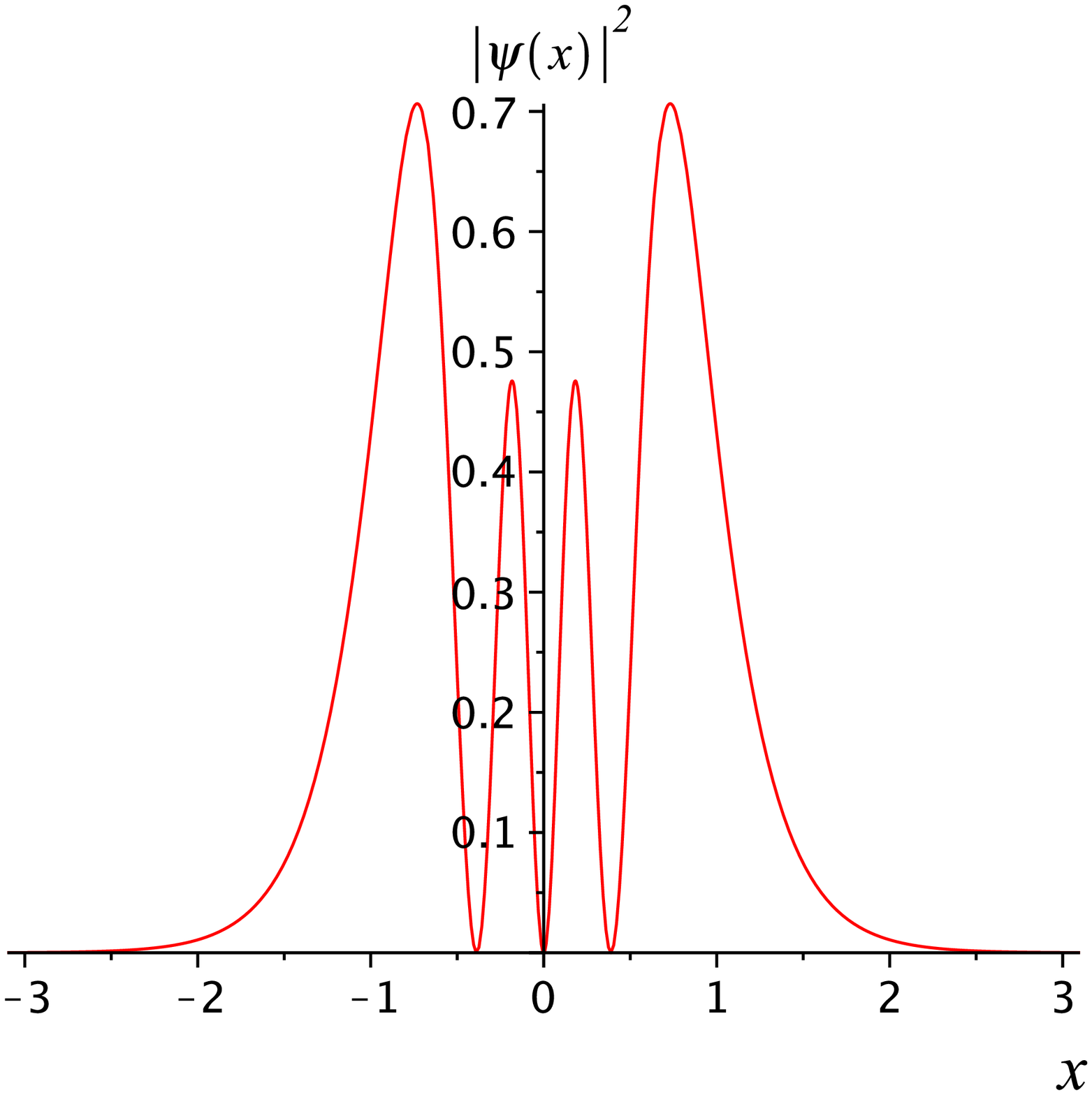}}
{\includegraphics[width=4.5cm,height=4cm]{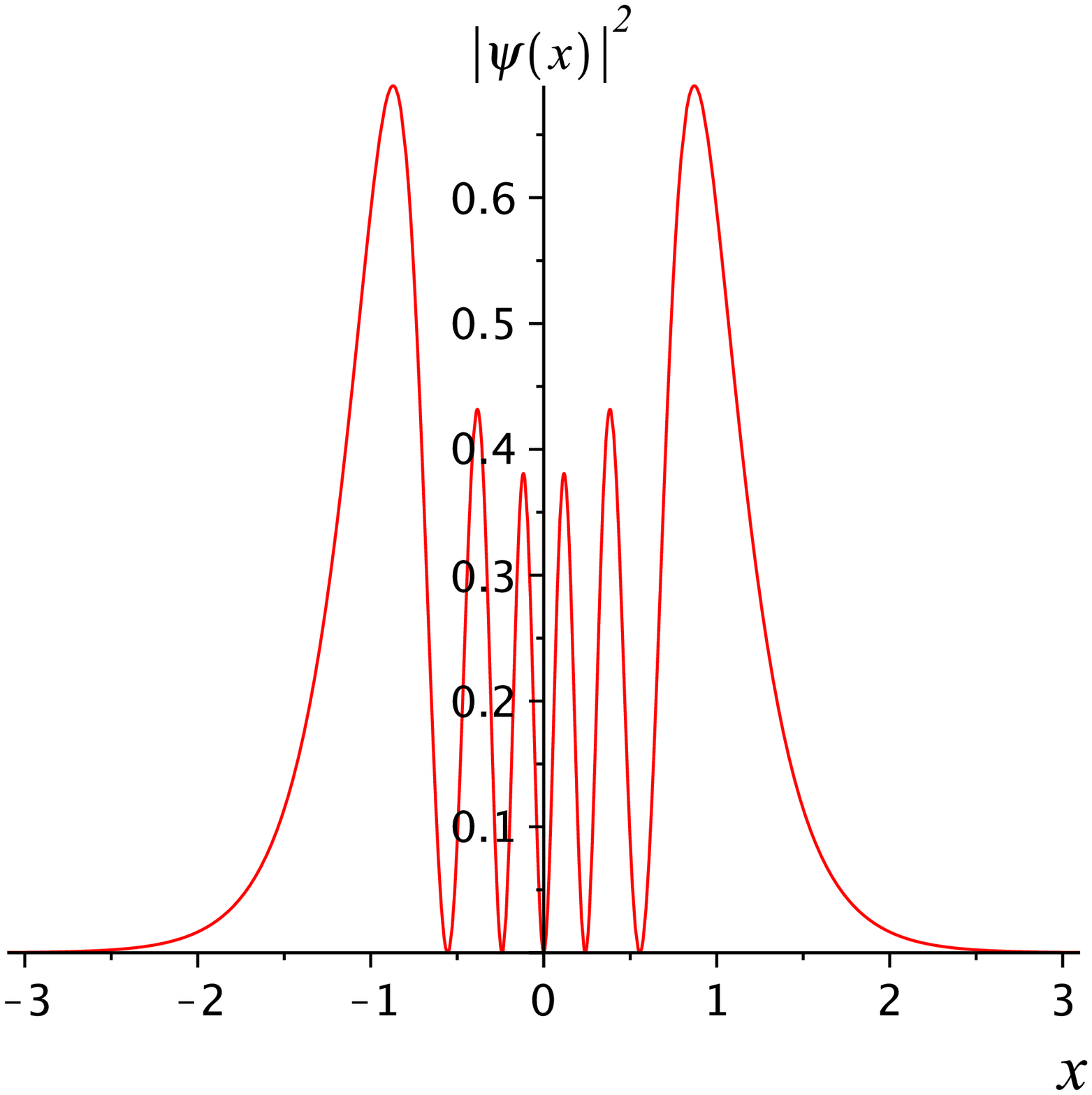}}
\caption{\label{figsech6_prob_x_Asim} Plot of antisymmetric solutions $\psi_a(x)$
[up] and the corresponding probability densities $|\psi_a(x)|^2$ [down] given
$\mathcal{B}=\mathcal{C}=0$ and $\mathcal{E}=0$,  for (left)
$\mathcal{A} = 16.962907791$; (center) $\mathcal{A} = 77.987131$;
and (right) $\mathcal{A} = 183.4448166$.}
\end{figure}

Our solutions are thus
\bea
h^{(1)}(y)&=&Ht\left(0, 0, -(-12 \mathcal{A})^{\!\terco};
({\dterco \sqrt{-\mathcal{A}}})^{\!\terco}\,y\right)\\
h^{(2)}(y)&=&\exp\!\left[\dterco\sqrt{-\mathcal{A}}\,\, y\!\left(y^2-3\right)\right]
Ht\left(0, 0, -(-12 \mathcal{A})^{\!\terco}; - ({\dterco\sqrt{-\mathcal{A}}})^{\!\terco}\,y\right),
\eea
namely
\bea
\phi^{(1)}(y)&=&\exp\!\Big[\!-\!\terco\sqrt{-\mathcal{A}}\,y\, (y^2-3)\Big]
Ht\left(0, 0, -(-12 \mathcal{A})^{\!\terco}; ({\dterco \sqrt{-\mathcal{A}}})^{\!\terco}\,y\right)\\
\phi^{(2)}(y)&=&\exp\!\Big[\,\terco\sqrt{-\mathcal{A}}\,y\,(y^2-3)\Big]
Ht\left(0, 0, -(-12 \mathcal{A})^{\!\terco}; -({\dterco \sqrt{-\mathcal{A}}})^{\!\terco}\,y\right)\!\!.
\eea
%In variable $z$ these read
%\bea
%\phi^{(1)}(z)&=&\exp\!\Big[\!-\!\terco\sqrt{-\mathcal{A}}\,\cos\!z\,
%(\cos^2\!z-3)\Big] Ht\left(0, 0, -(-12 %\mathcal{A})^{\!\terco};
%({\dterco \sqrt{-\mathcal{A}}})^{\!\terco}\!\cos\!z\right)\\
%\phi^{(2)}(z)&=&\exp\!\Big[\,\terco\sqrt{-\mathcal{A}}\,\cos\!z\,(\cos^2\!z-3)\Big]
%Ht\left(0, 0, -(-12 %\mathcal{A})^{\!\terco}; -({\dterco \sqrt{-\mathcal{A}}})^{\!\terco}\,\cos\!z\right)
%\eea
%
In variable $z$, recalling that $\varphi(z)=\cos\!z\,\phi(z)$, they result
\bea
\varphi^{(1)}(z)&=&\cos\!z\exp{\!\Big[\!-\!\terco\sqrt{-\mathcal{A}}\,\cos\!z\,
(\cos^2\!z-3)\Big]}Ht\left(0, 0, -(-12 \mathcal{A})^{\!\terco};
({\dterco \sqrt{-\mathcal{A}}})^{\!\terco}\!\cos z\!\right)\\
\varphi^{(2)}(z)&=&\cos\!z\exp\!\Big[\,\terco\sqrt{-\mathcal{A}}\,\cos\!z\,(\cos^2z-3)\Big]
Ht\left(0, 0, -(-12 \mathcal{A})^{\!\terco}; -({\dterco \sqrt{-\mathcal{A}}})^{\!\terco}\!\cos z\!\right)\!,
\eea
and finally, for $\psi(x)=\sech^{\frac{1}{2}}\!x\,\varphi(x)$ we have
\bea
\psi^{(1)}(x)&=&\sech^{\frac{3}{2}}\!xe^{\Big[\!-\!\terco\sqrt{-\mathcal{A}}\,\sech\!x\, (\sech^2\!x-3)\Big]}
 Ht\left(0, 0, -(-12 \mathcal{A})^{\!\terco}; ({\dterco \sqrt{-\mathcal{A}}})^{\!\terco}\sech\!x\right)
 \label{eqsech61}\\
\psi^{(2)}(x)&=&\sech^{\frac{3}{2}}\!xe^{\Big[\,\terco\sqrt{-\mathcal{A}}\,\sech\!x\, (\sech^2\!x-3)\Big]}
Ht\left(0, 0, -(-12 \mathcal{A})^{\!\terco}; -({\dterco \sqrt{-\mathcal{A}}})^{\!\terco}\,\sech\!x\right)\!.
\label{eqsech62}
\eea

Note that although these solutions, Eqs. (\ref{eqsech61}) and  (\ref{eqsech62}),
are both complex functions, their symmetric and antisymmetric combinations are real.
Furthermore, only the symmetric and antisymmetric solutions
satisfy the border conditions and make physical sense.
There is a discrete number of potential wells with an $E=0$ eigenstate.
We show these eigenfunctios for the first six values of ${A}$,
see Fig. \ref{figsech6_prob_x_Sim} and \ref{figsech6_prob_x_Asim}.
It can be seen that the number of nodes of the zero-modes depends
directly on the depths of the wells.

%As solucoes simetricas sao obtidas somando as HeunT: $\psi_S(x)= HeunT1+HenT2$
%As solucoes antisimetricas sao obtidas subtraindo as HeunT: $\psi_A(x)= HeunT1-HenT2$

\newpage
%%%%%%%%%%%%%%%%%%%%%%%%%%%%%%%%  Caso C=0; A=-B; E=0   %%%%%%%%%%%%%%%%%%%%%%%%%%%%%%%%%%%
\subsubsection{Case free ${B}=-{A}$, and $C=0$ \label{sec:sech6-4}}

\begin{figure}[h]
\center
{\includegraphics[width=6.cm,height=6cm]{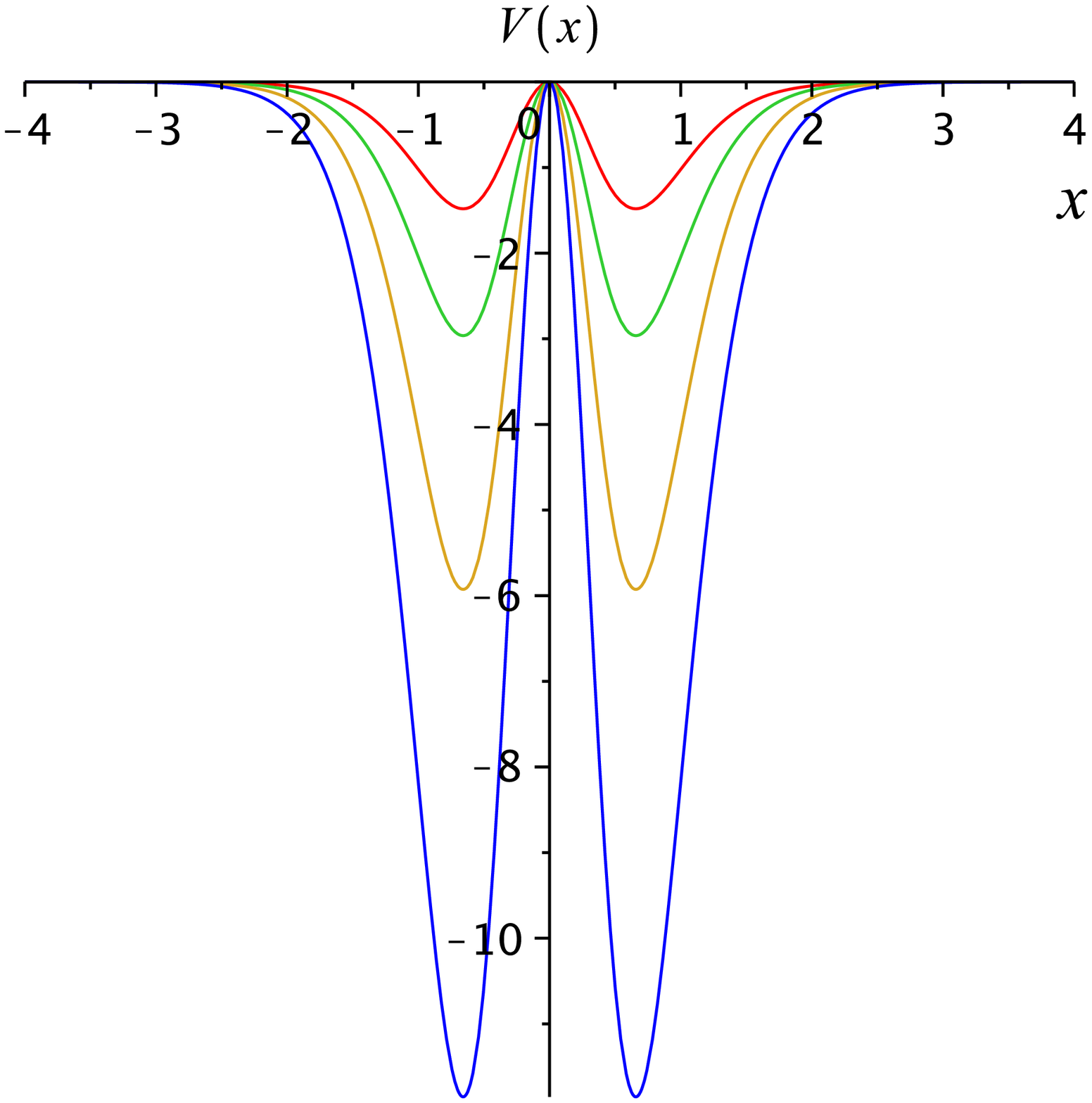}}\hspace{1cm}
{\includegraphics[width=6.cm,height=6cm]{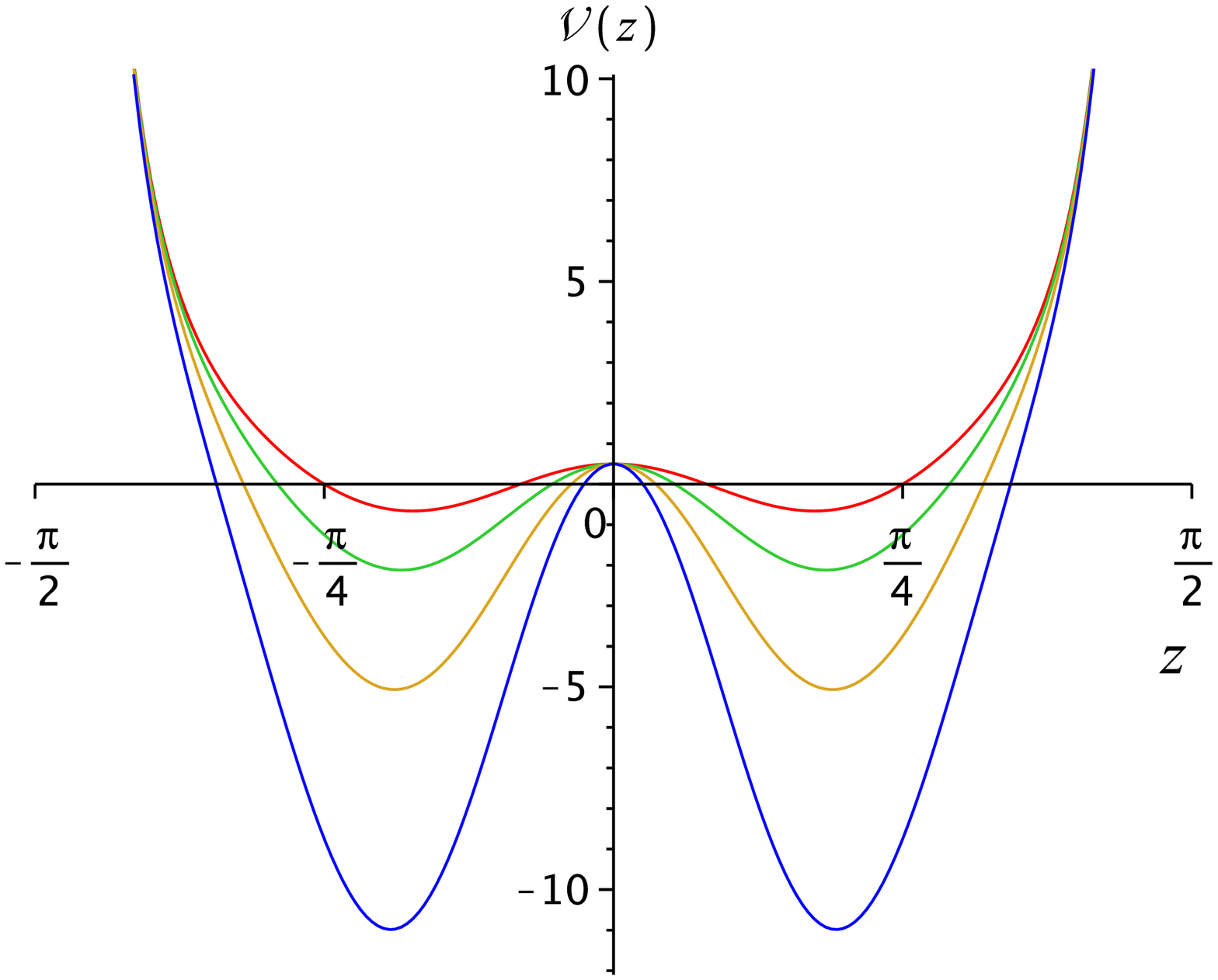}}
\caption{\label{graf_Vsech_C0} From top to bottom, plot of double-well
potentials $V(x)=  A (\sech^6(x) -  \sech^4(x))$ (left)
and the corresponding effective potentials $\mathcal{V}(z)$ (right)
for $-\mathcal{A}=10, 20, 40$ and $80$.}
\end{figure}

The second case we managed to analytically solve corresponds to the hyperbolic sixth-order
double-well plotted in Fig. \ref{graf_Vsech_C0}.

In order to show this we have to analyze the following instance of the
modified Schr\"odinger eq. (\ref{schrodinger-cons})
\beq
\varphi''(z)- \left(\frac{1}{2} +
\frac{3}{4}\tan^2(z)-\mathcal{A}\cos^6(z)+\mathcal{A}\cos^4(z)\right) \varphi(z)=0.
\eeq
After transformation $\varphi(z)=\cos^\mu(z)\,\phi(z)$ we get
\beq
\phi''(z)-2\mu\tan(z)\phi'(z)+\Big[(\mu^2-\mu-3/4)\tan^2z-\mu-1/2+\mathcal{A}\cos^6z
-\mathcal{A}\cos^4z\Big]\phi(z) = 0
\eeq
which shortens to
\beq
\phi''(z)+\tan(z)\phi'(z)+\Big[\mathcal{A}\cos(z)^6 -\mathcal{A}\cos(z)^4\Big]\phi(z) = 0,
\eeq
provided one chooses $\mu=-\meio$.
Now, we change coordinates by $y=\sin(z)$ yielding
\beq
\phi''(y)+\Big[\mathcal{A} (1-y^2)^2-\mathcal{A}(1-y^2)\Big]\phi(y) = 0.
\eeq
As in the previous case, we try the ansatz  $\phi(y)=e^{ay^3+by} h(y)$ and get
\beq
h''(y)+(6ay^2+2b)h'(y)+\Big[6ay+b^2+(\mathcal{A}+9a^2)\,y^4+(-\mathcal{A}+6ab)\,y^2\Big]h(y) = 0,
\eeq
which simplifies conveniently when we adopt $3a=-\sqrt{-\mathcal{A}}$
and $2b=\sqrt{-\mathcal{A}}$. We thus obtain
\beq
h''(y)+\sqrt{-\mathcal{A}}\Big(\!-2 y^2+1\Big)\,h'(y)+
\Big(-2\sqrt{-\mathcal{A}}\,y-\mathcal{A}/4\Big)h(y) = 0
\eeq
which, by redefining $\bar{y}=\left(\frac{2\sqrt{-\mathcal{A}}}{3}\right)^{\!\!\!\terco}\!\!y$, gives
\beq
h''(\bar{y})-\left[{3}\,\bar{y}^2 -\left(-\frac{3\mathcal{A}}{2} \right)^{\!\!\terco} \right]
h'(\bar{y})+\left[-3\,\bar{y} +\left(\frac{-3\mathcal{A}}{16}\right)^{\!\!\dterco}\right]\,h(\bar{y}) = 0.
\label{pretrisech64}\eeq
This is the canonical triconfluent Heun equation
\beq
H''(u)- (3u^2+\gamma)H'(u)+[(\beta-3) u+\alpha]H(u)=0,
\eeq
as soon as we identify
\bea
\alpha &=&\left(-\frac{3\mathcal{A}}{16}  \right)^{\!\!\dterco}\nonumber\\
\beta &=&0\nonumber\\
\gamma &=&-\left(-\frac{3\mathcal{A}}{2} \right)^{\!\!\terco}.\nonumber
\eea
The solutions of eq. (\ref{pretrisech64}) are  then
\bea
h^{(1)}(y)&=&
Ht\!\left(\left(-\frac{3\mathcal{A}}{16}\right)^{\!\!\dterco}\!\!\!\!\!,\,\,\,0,
-\!\left(-\frac{3\mathcal{A}}{2} \right)^{\!\!\terco},
\left(\frac{2\sqrt{-\mathcal{A}}}{3}\right)^{\!\!\!\terco}\!\!y\right)\\
h^{(2)}(y)&=&\exp\!\left[\dterco\sqrt{-\mathcal{A}}\,\, y\!\left(y^2-\tmeio\right)\right]
Ht\!\left(\left(-\frac{3\mathcal{A}}{16}  \right)^{\!\!\dterco}\!\!\!\!\!,\,\,\,0, -\!
\left(-\frac{3\mathcal{A}}{2} \right)^{\!\!\terco},
-\left(\frac{2\sqrt{-\mathcal{A}}}{3}\right)^{\!\!\!\terco}\!\!y\!\!\right)\!\!.
\eea
%
%\bea
%\phi^{(1)}(z)&=&\exp\!\left[\sqrt{-\mathcal{A}}\,\,
%\sin(z)\!\left(\meio-\terco \sin^2(z)\right)\right]
%Ht\!\left(\left(-\frac{3\mathcal{A}}{16}  \right)^{\!\!\dterco}\!\!\!\!\!,\,\,\,0,
%-\!\left(-\frac{3\mathcal{A}}{2} %\right)^{\!\!\terco},
%\left(\frac{2\sqrt{-\mathcal{A}}}{3}\right)^{\!\!\!\terco}\!\!\sin(z)\right)\\
%\phi^{(2)}(z)&=&\exp\!\left[\sqrt{-\mathcal{A}}\,\, \sin(z)\!
%\left(\terco \sin^2(z)-\meio\right)\right] Ht\!\left(\left(-\frac{3\mathcal{A}}{16}
%\right)^{\!\!\dterco}\!\!\!\!\!,\,\,\,0, -\!\left(-\frac{3\mathcal{A}}{2} %\right)^{\!\!\terco}, -\left(\frac{2\sqrt{-\mathcal{A}}}{3}\right)^{\!\!\!\terco}\!\!\sin(z)\right)
%\eea
%%
Finally, by transforming everything back to the original $x$-space,
we have the starting eigenfunctions of the PDM hamiltonian
\bea
&&\psi^{(1)}\!(x)={\rm e}^{\sqrt{-\mathcal{A}}\ \tanh(x)
\!\left(\meio-\terco \tanh^2(x)\right)}\ Ht\!\left(\!\!\left(-\frac{3\mathcal{A}}{16}
\right)^{\!\!\dterco}\!\!\!\!\!,\,\,\,0, -\!\left(-\frac{3\mathcal{A}}{2}
\right)^{\!\!\terco}\!\!\!\!,\,
\left(\frac{2\sqrt{-\mathcal{A}}}{3}\right)^{\!\!\!\terco}\!\!\tanh x\!\!\right)
\label{eqsech461}\\
&&\psi^{(2)}\!(x)={\rm e}^{\sqrt{-\mathcal{A}}\, \tanh(x)\!\left(\terco \tanh^2(x)-\meio\right)}
Ht\!\left(\!\!\left(-\frac{3\mathcal{A}}{16}  \right)^{\!\!\dterco}\!\!\!\!\!,\,\,0,
-\!\left(-\frac{3\mathcal{A}}{2} \right)^{\!\!\terco}\!\!\!\!,
-\left(\frac{2\sqrt{-\mathcal{A}}}{3}\right)^{\!\!\!\terco}
\!\!\!\tanh x\!\!\right)\!\!.
\label{eqsech462}\eea
%

% vide falta de $\varphi(z)=\cos^\mu(z)\,\phi(z)$ !

\begin{figure}[h]
\center
{\includegraphics[width=4.5cm,height=4cm]{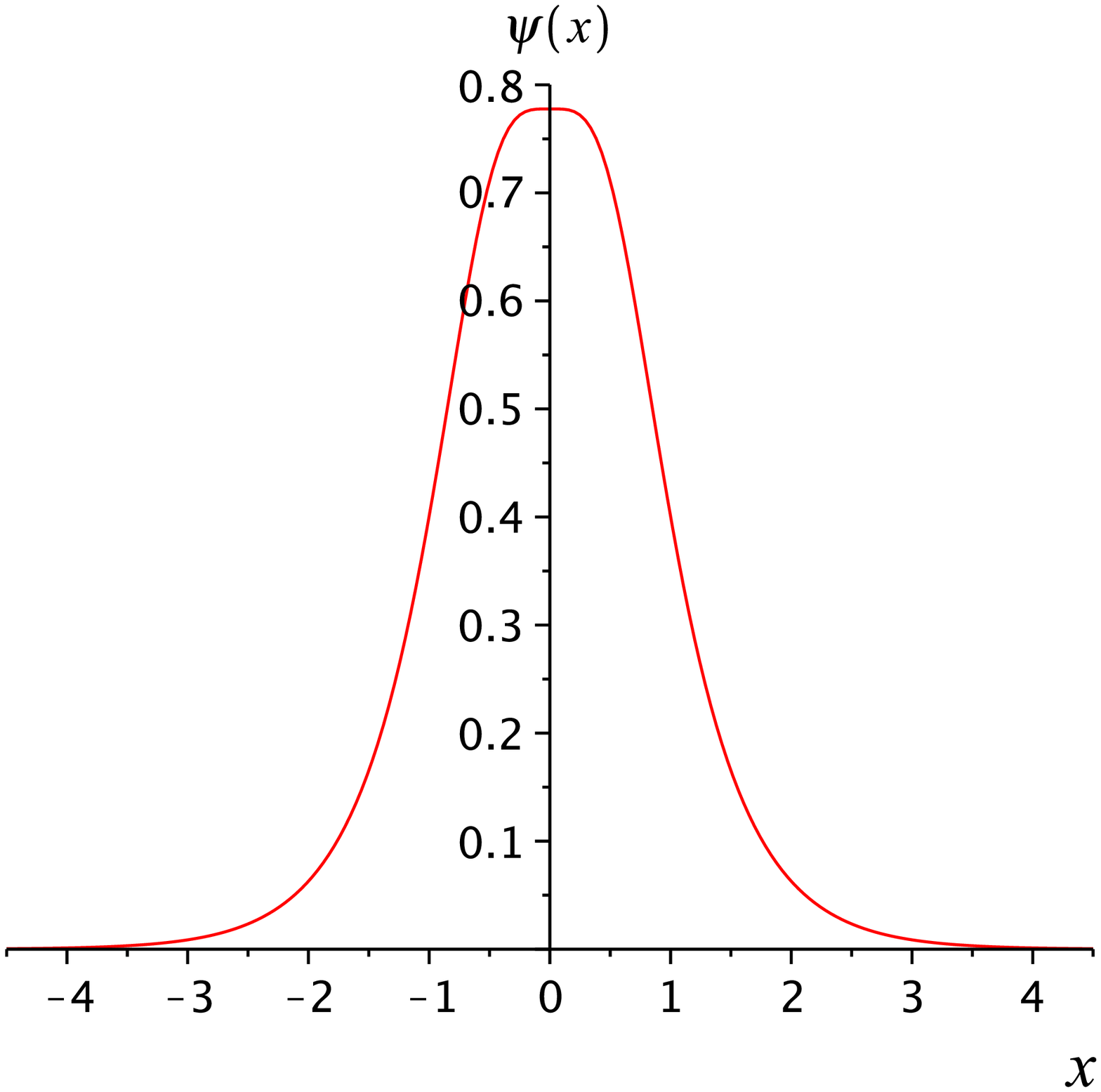}}
{\includegraphics[width=4.5cm,height=4cm]{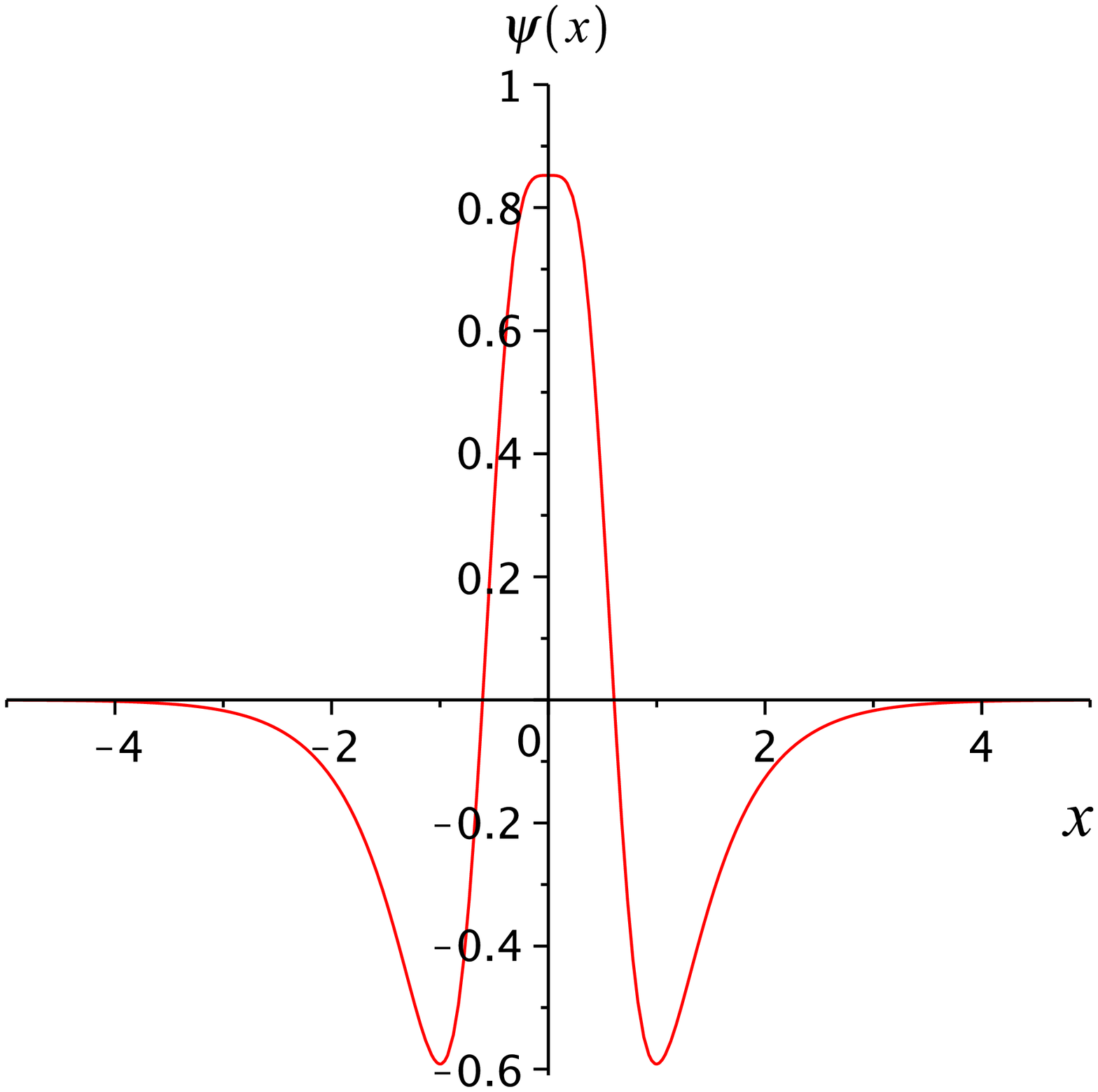}}
{\includegraphics[width=4.5cm,height=4cm]{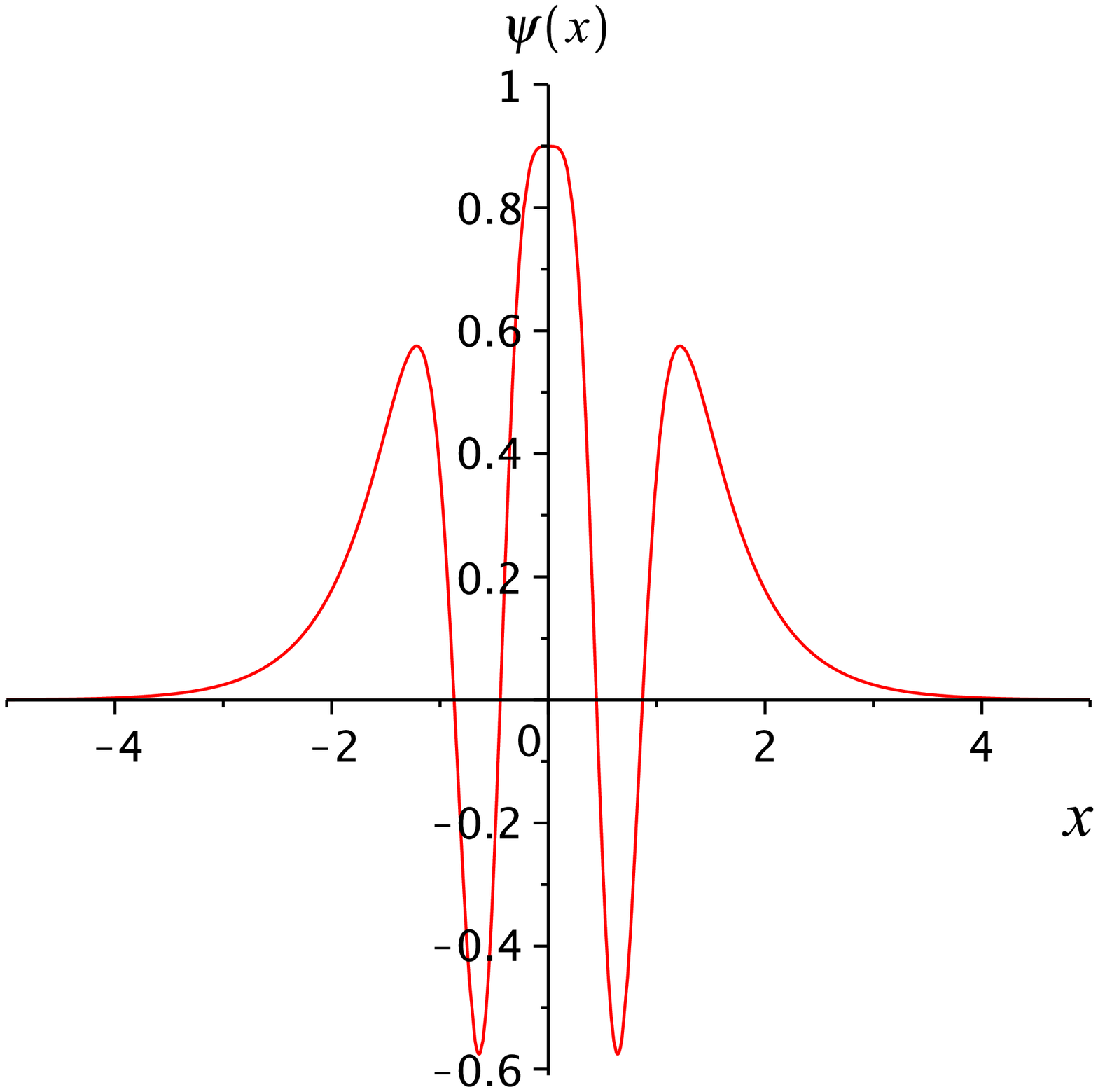}}\\
{\includegraphics[width=4.5cm,height=4cm]{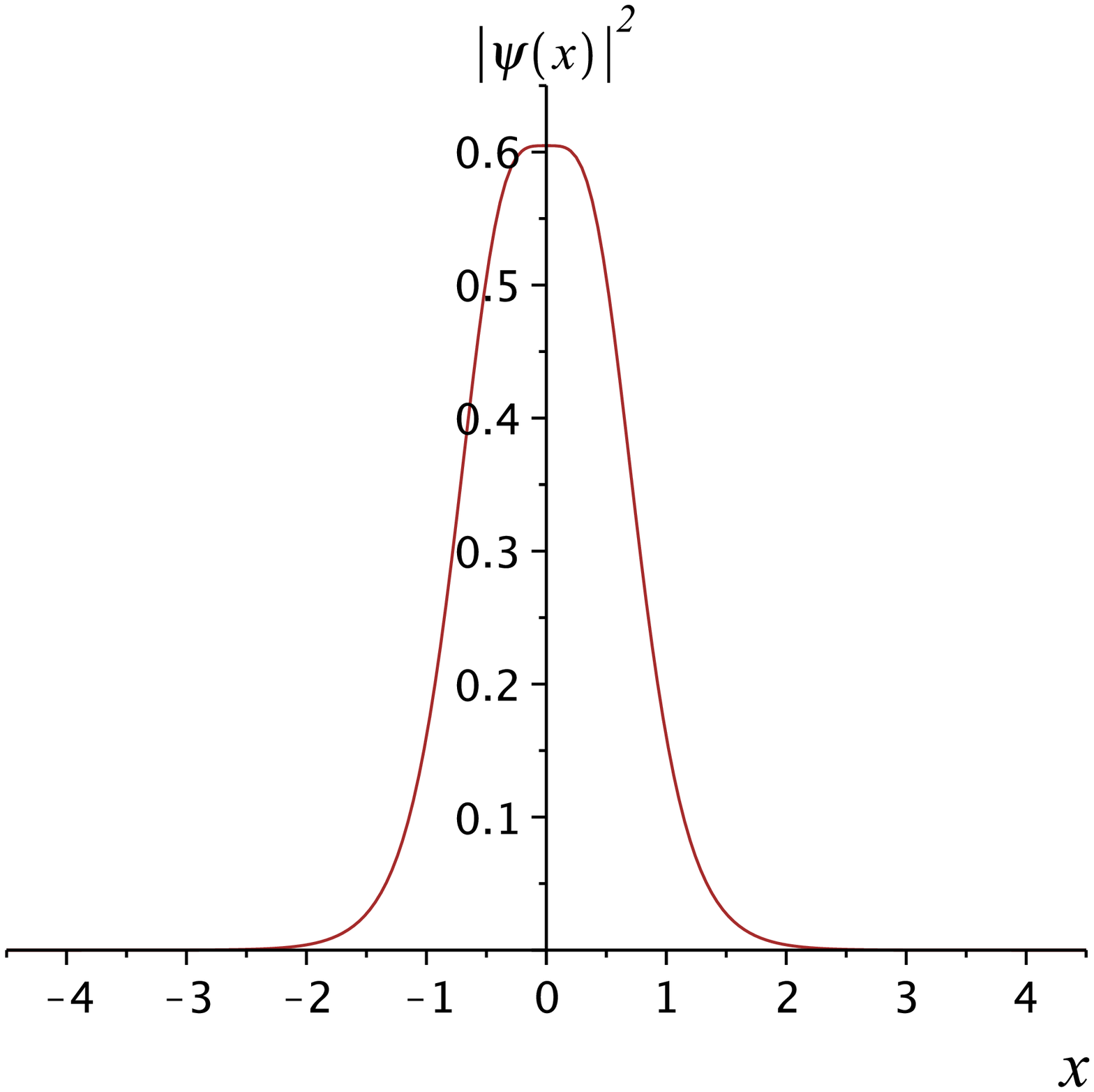}}
{\includegraphics[width=4.5cm,height=4cm]{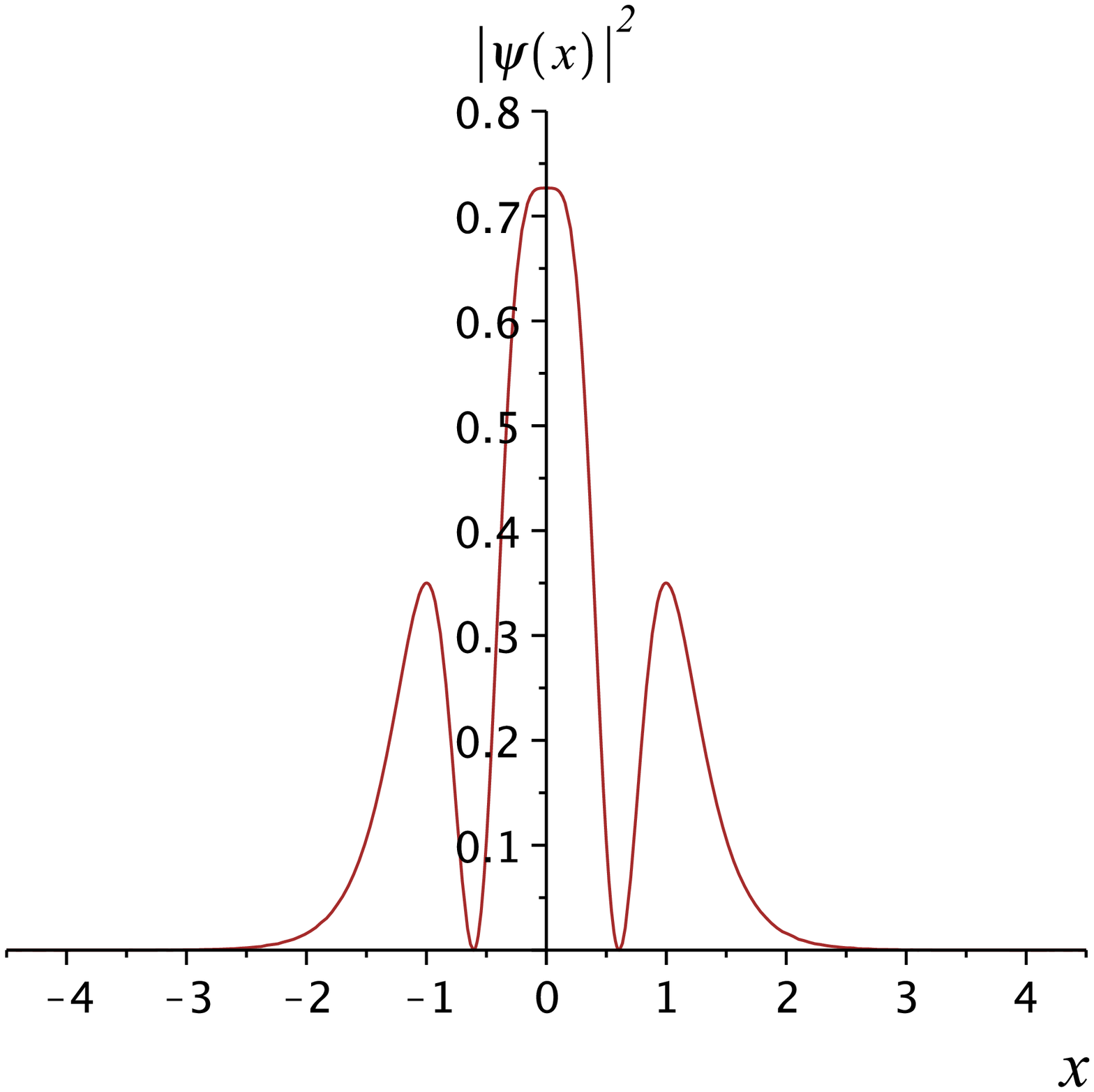}}
{\includegraphics[width=4.5cm,height=4cm]{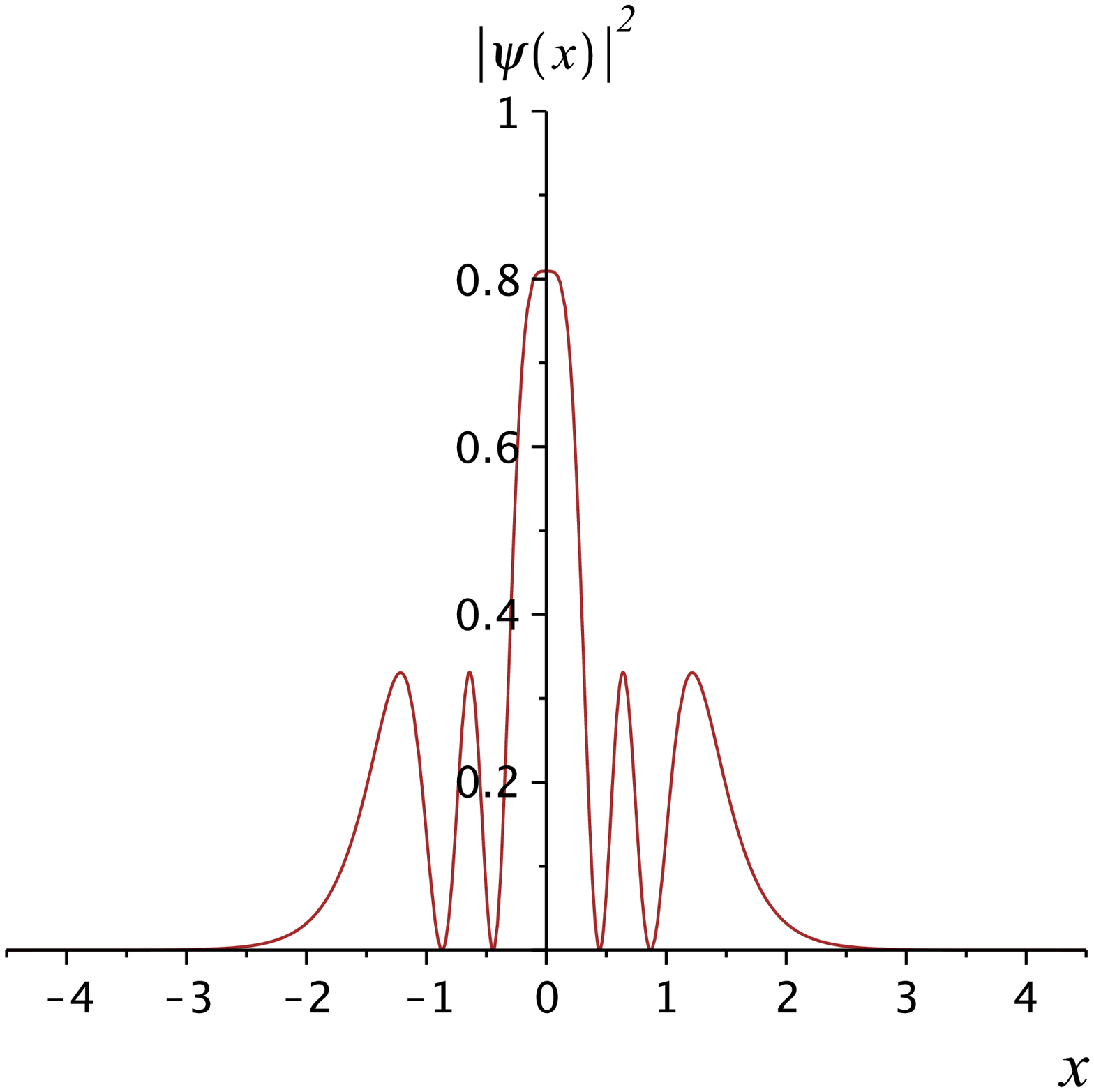}}
\caption{\label{figs_x_C0_Sim}
Plot of symmetric PDM zero-modes $\psi_s(x)$ of $V(x)=  A (\sech^6(x) -  \sech^4(x))$   [up],
and the corresponding probability densities $|\psi_s(x)|^2$ [down],
for  ${A} = -25.125695463186$ (left);
${A} = -209.2999338840$ (center); and  ${A}= -571.605964500$ (right).}
\end{figure}

\begin{figure}[h]
\center
{\includegraphics[width=4.5cm,height=4cm]{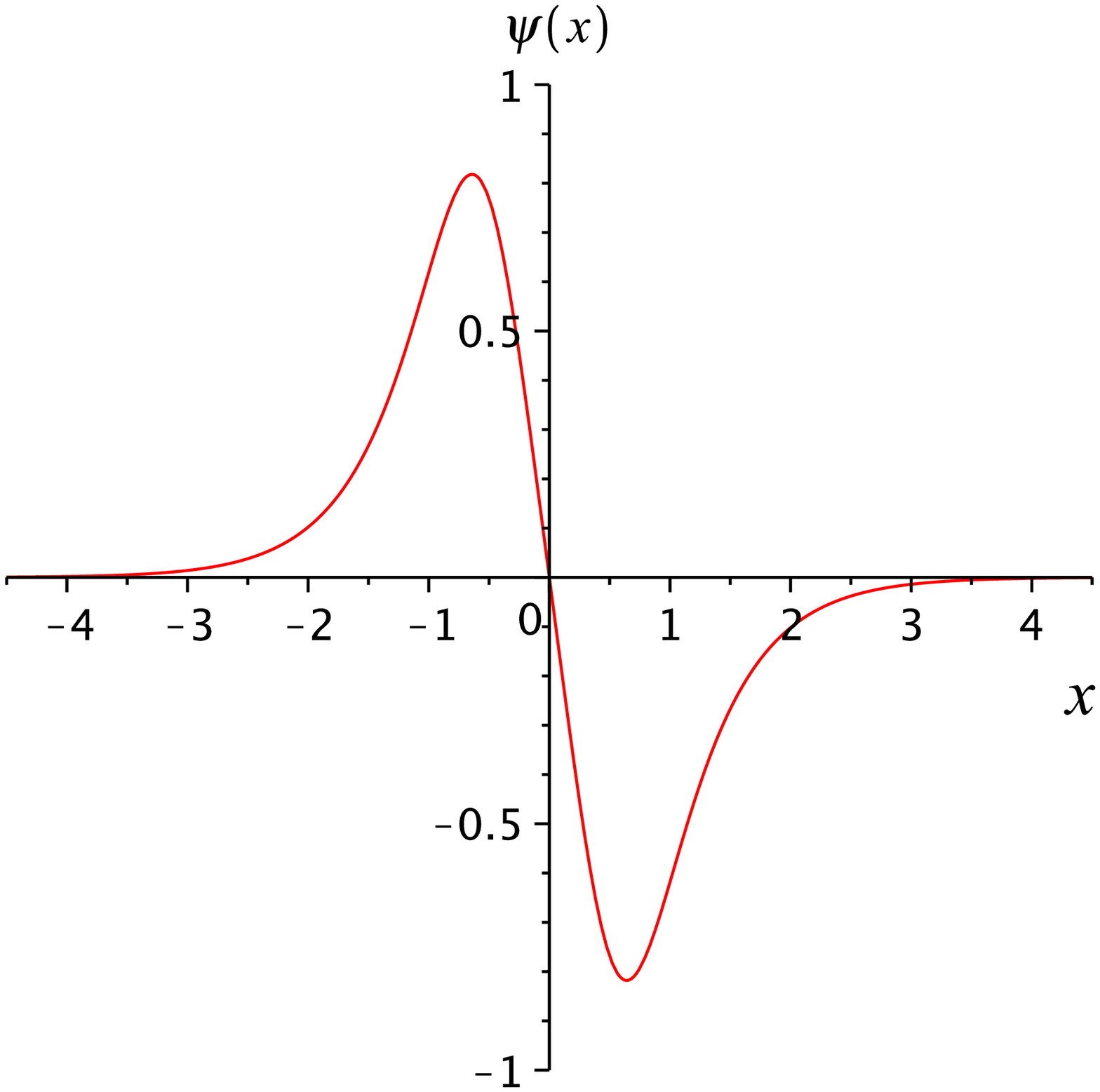}}
{\includegraphics[width=4.5cm,height=4cm]{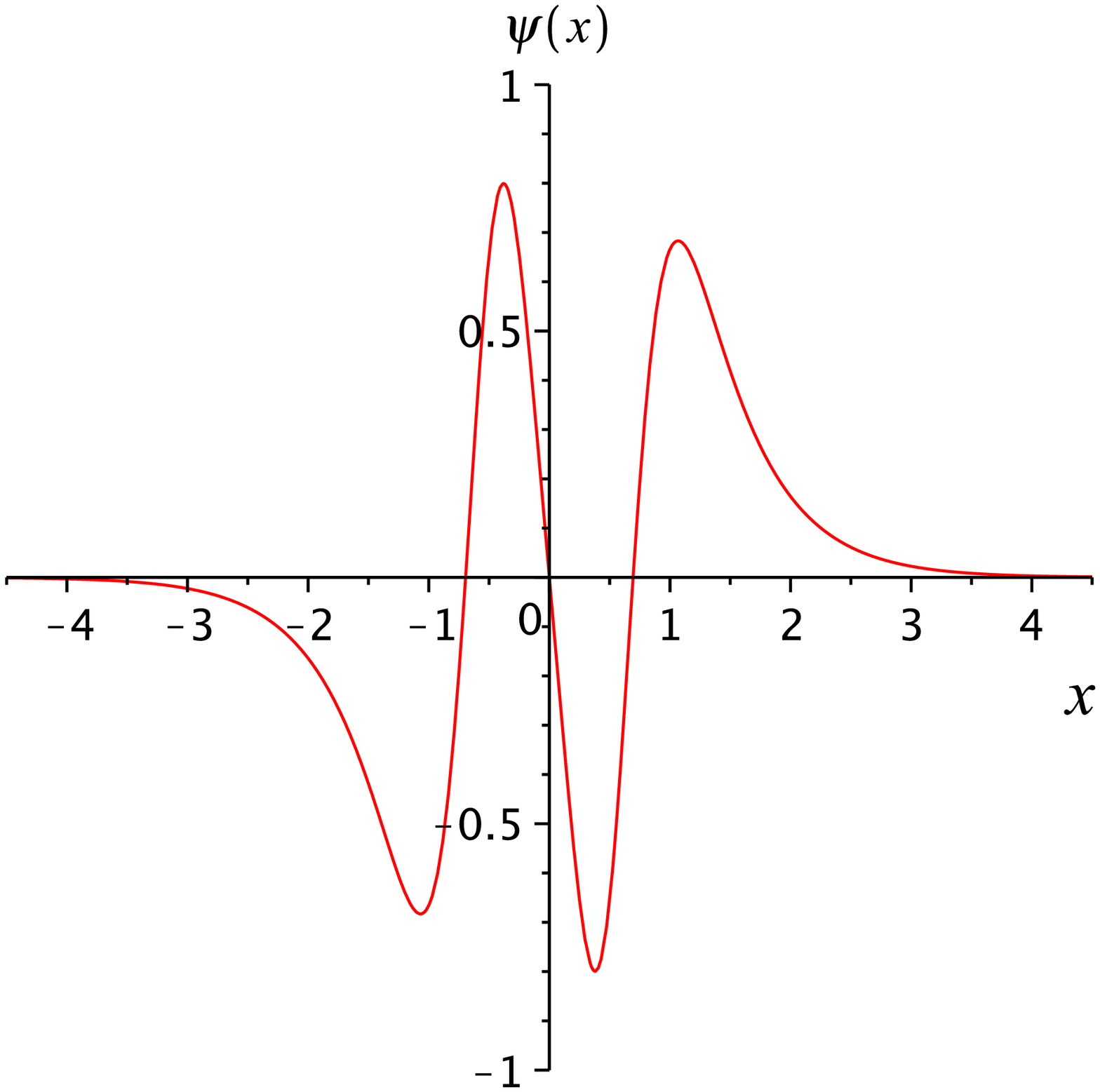}}
{\includegraphics[width=4.5cm,height=4cm]{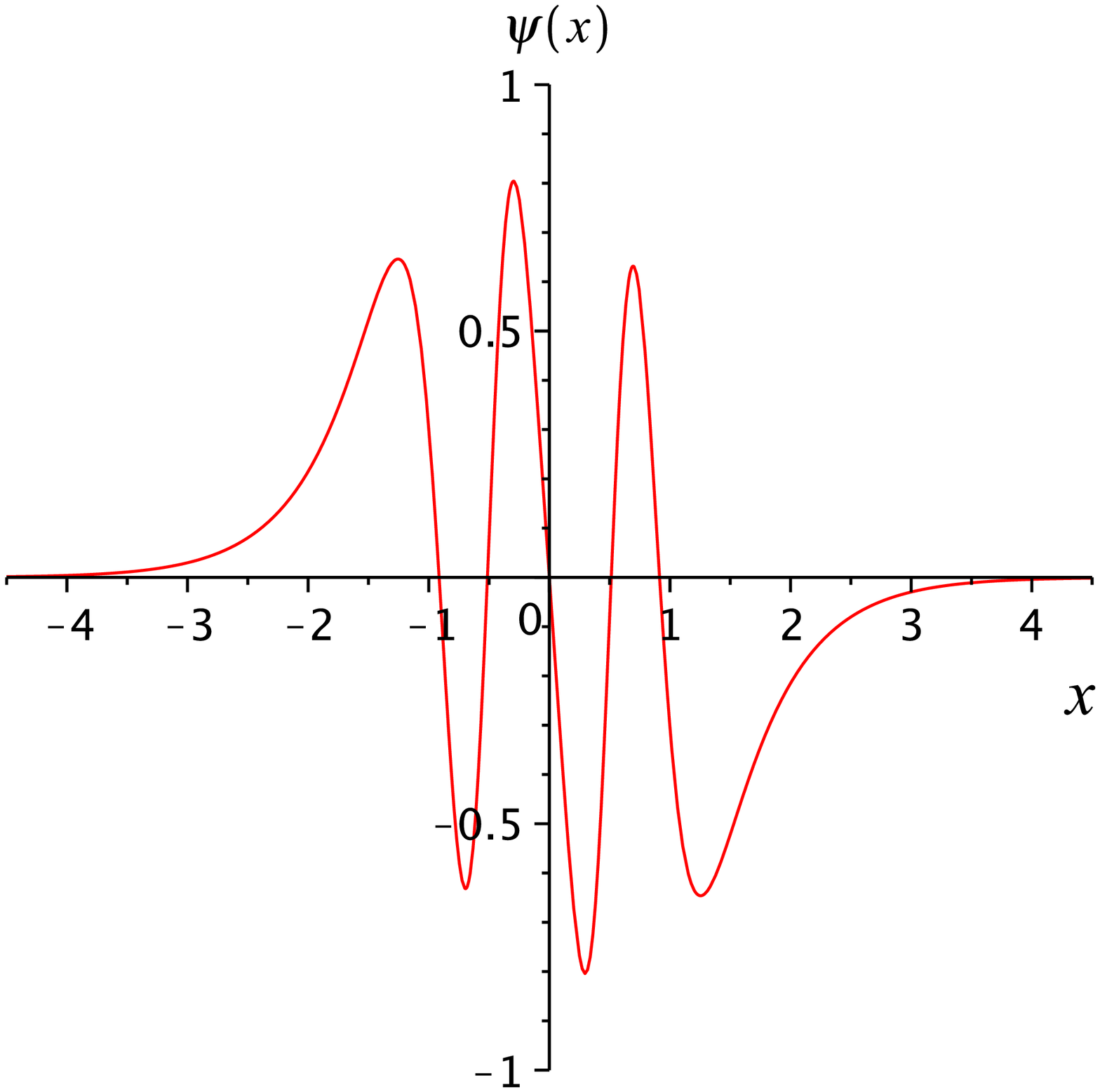}}\\
{\includegraphics[width=4.5cm,height=4cm]{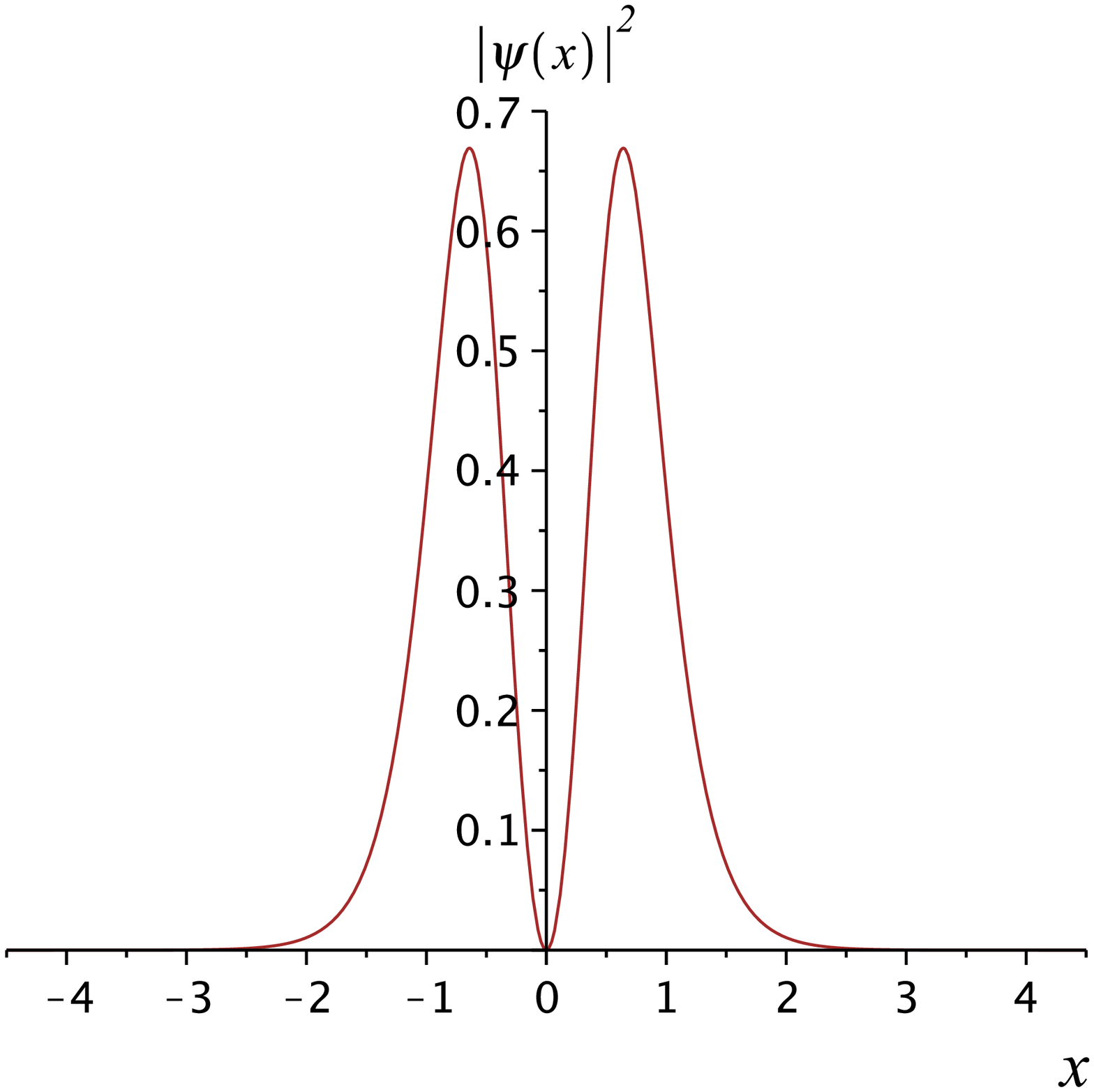}}
{\includegraphics[width=4.5cm,height=4cm]{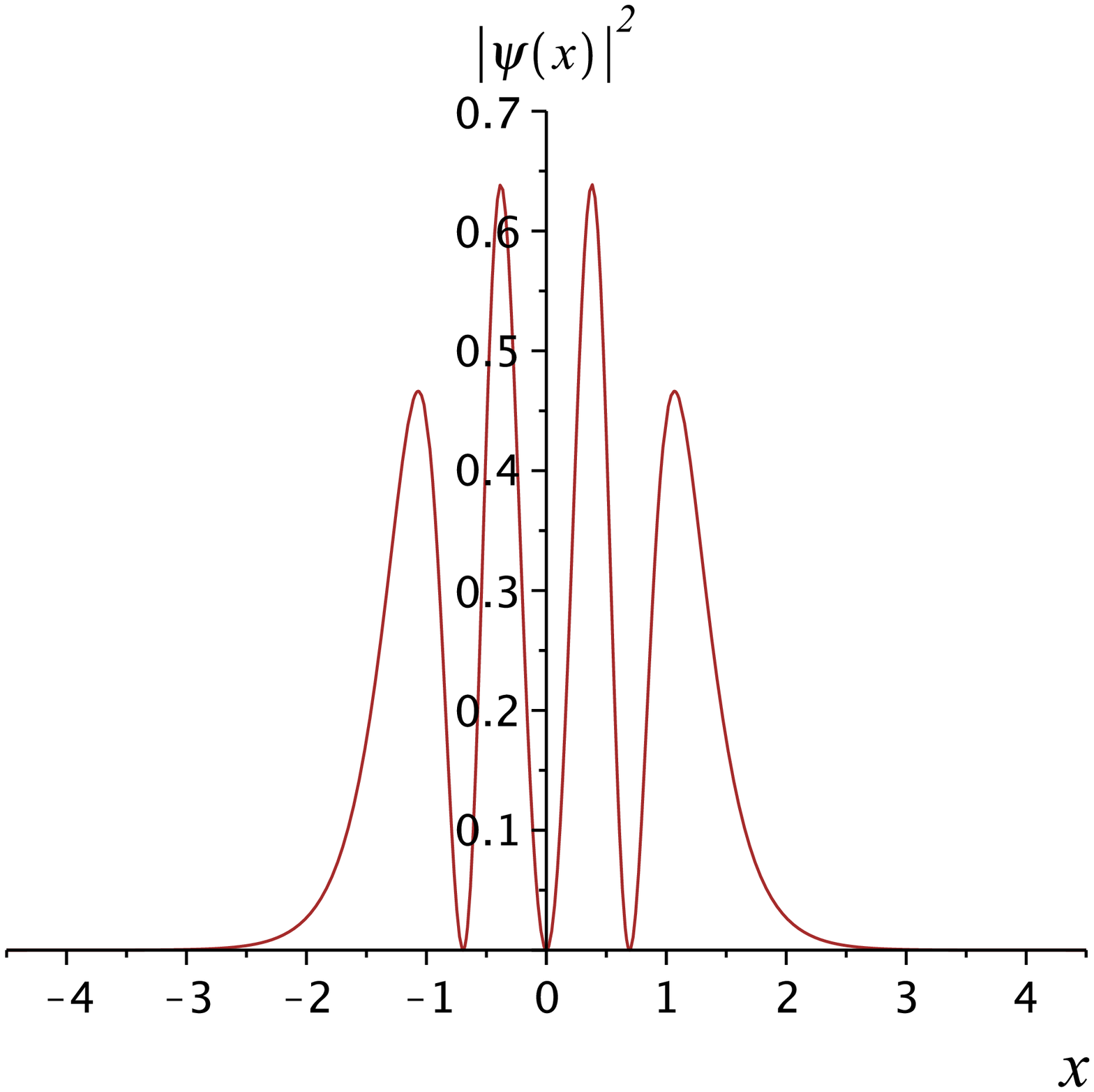}}
{\includegraphics[width=4.5cm,height=4cm]{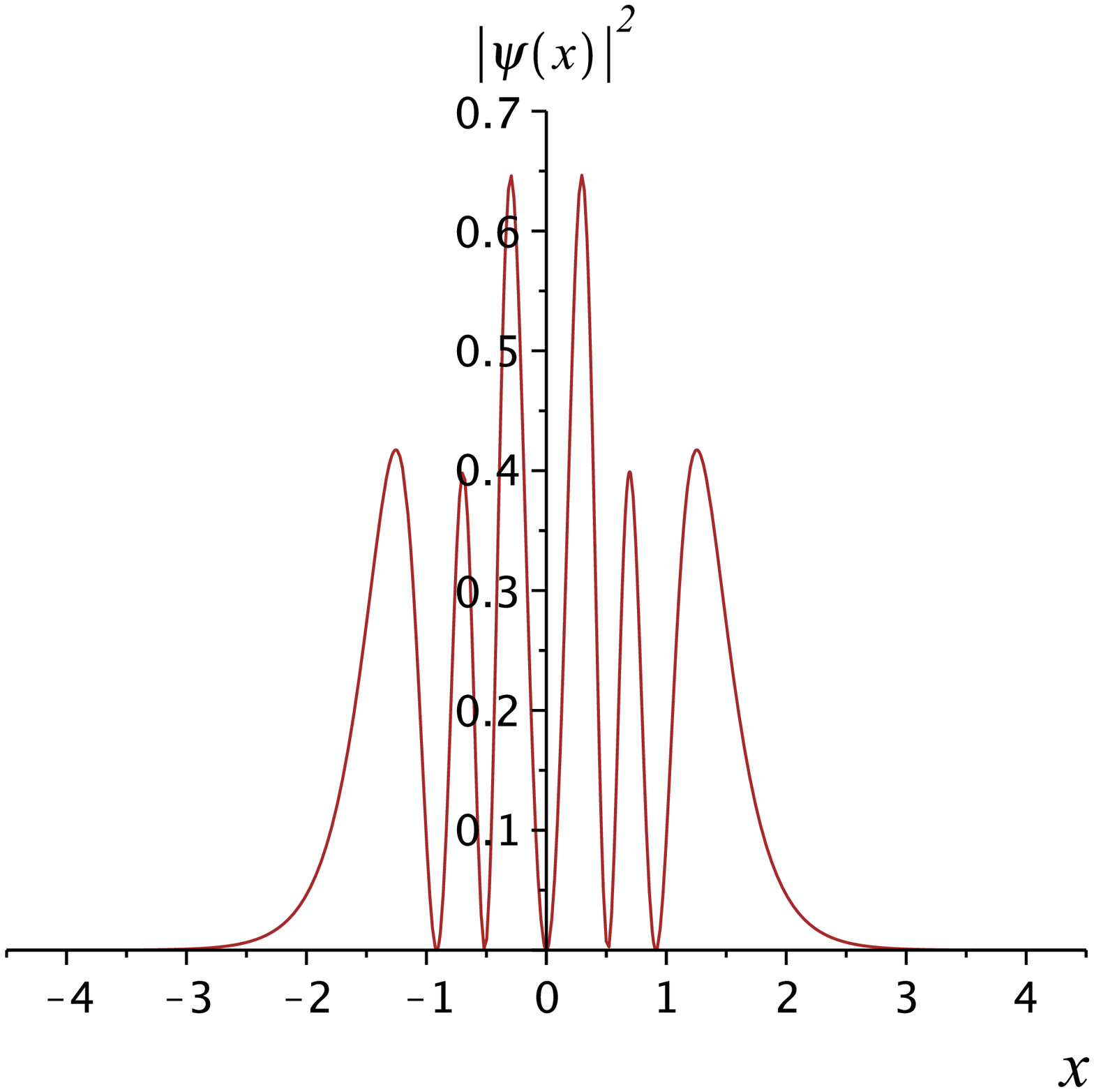}}
\caption{\label{figs_x_C0_Asim}
Plot of antisymmetric PDM zero-modes $\psi_a(x)$ of $V(x)=  A (\sech^6(x) -  \sech^4(x))$
[up] and the corresponding probability densities $|\psi_a(x)|^2$ [down],
for ${A}=-56.05506043241$ (left); ${A} = -284.9369967664$ (center);
${A} = -691.7230772070$ (right).}
\end{figure}

In this case, we found again that just the symmetric and antisymmetric combinations
of Eqs. (\ref{eqsech461}) and (\ref{eqsech462}) are real functions and the only that
fit the boundary conditions (as expected from the parity of the potential).
After a numerical survey of the parameter space
we found a discrete set of values of potential depths compatible with a zero energy eigenstate.
In Figs. \ref{figs_x_C0_Sim} and \ref{figs_x_C0_Asim}
we plot the eigenfunctions in the first six cases,
three being even and the other antisymmetric, as indicated.

%\begin{figure}[h]
%\center
%{\includegraphics[width=5cm,height=5cm]{fig7_x_C0_s.eps}}
%{\includegraphics[width=5cm,height=5cm]{fig8_x_C0_a.eps}}\\
%{\includegraphics[width=5cm,height=5cm]{fig7_prob_x_C0_s.eps}}
%{\includegraphics[width=5cm,height=5cm]{fig8_prob_x_C0_a.eps}}
%\caption{\label{figs_x_C0}  Plot of PDM zero modes of $V(x)=  A (\sech^6(x) -  \sech^4(x))$.
%For $\mathcal{A}=-25.125695463186$, it is a symmetric eigenfunction $\psi_s(x)$ (left) and for $\mathcal{A}=-56.05506$,
%an antisymmetric eigenfunction $\psi_a(x)$ (right).
%Below are the corresponding probability densities $|\psi(x)|^2$.}
%\end{figure}

For a constant mass, the general solution for this potential is \cite{sech246}
\bea
&&\chi^{(1)}(x) =  {\rm e}^{\meio\sqrt{A} \tanh^2x}(\sech x)^{\sqrt{-E}}\,
Hc\! \left(\sqrt{A},-\frac{1}{2}, \sqrt{-E},\ 0,\frac{1-E}{4};\, \tanh^2\!x \right)  \label{ordC01}\\
%= \\ && C_1\sech(x)^{\sqrt{-E}}\, \text{Hc}\left(0, \sqrt{-\mathcal{E}}, -\frac{1}{2}, %-\frac{\mathcal{B}}{4},\, \frac{1}{4}-\frac{\mathcal{E}+\mathcal{C}}{4};\,\sech^2(x)\right)\label{ordManning1}\\
&&\chi^{(2)}(x) = {\rm e}^{\meio\sqrt{A} \tanh^2x}\,(\sech x)^{\sqrt{-E}}\tanh x\,\,
Hc\! \left(\sqrt{A},\frac{1}{2}, \sqrt{-E},\ 0,\frac{1-E}{4};\, \tanh^2\!x \right).\label{ordC02}
\eea

% notar que en el caso PDM el signo en la exponencial no coincide
% notar que las triconfluents tienen apenas 4 entradas contra 6 de las confluentes

It is noteworthy that we found no zero-energy modes in the ordinary constant mass cases
of neither $V(x)=  A \sech^6x$ nor $V(x)= A (\sech^6x -  \sech^4x)$ potentials.

%\newpage

\subsubsection{Three-term potentials}

The three-term potentials given by Eq. (\ref{ipotsech246})
have three possible phases:
hyperbolic single-wells, hyperbolic double-wells and hyperbolic triple-wells.
Since we have already analyzed in detail the first two situations,
among which the PDM Poschl-Teller \cite{JPM2013}
and PDM Manning potentials respectively, we now focus  on
the triple-well case which oblige the three terms. % although we may use just one or two param

\begin{figure}[h]
\center
{\includegraphics[width=9cm,height=7cm]{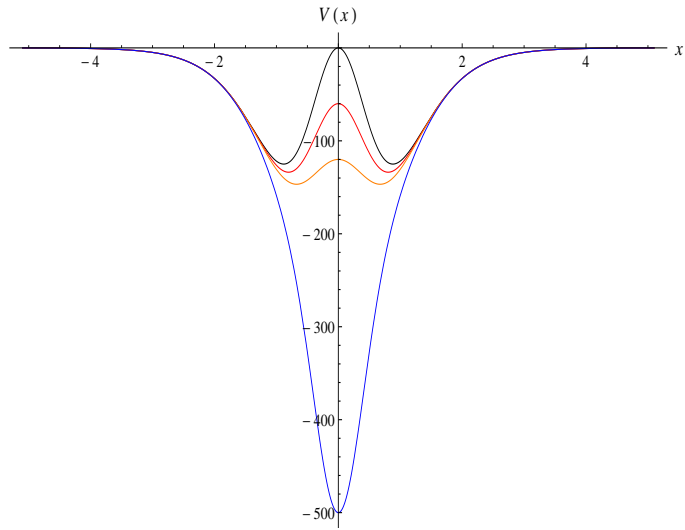}}
\caption{\label{graf_Vx_A0_A60_A120_A500}
Plot of $V(x)= - A \sech^6x - B  \sech^4x - C\sech^2x$
for $ {A}=0$ (black), 60 (red), 120 (orange), 500 (blue)
and $ {B}=- {C}$ =  - 500. The first is the Manning case already
represented at the top of Fig. \ref{graf_VA0} and the following double-wells result
from an $A$ term added to it.}
\end{figure}

\begin{figure}[h]
\center
{\includegraphics[width=4.5cm,height=4cm]{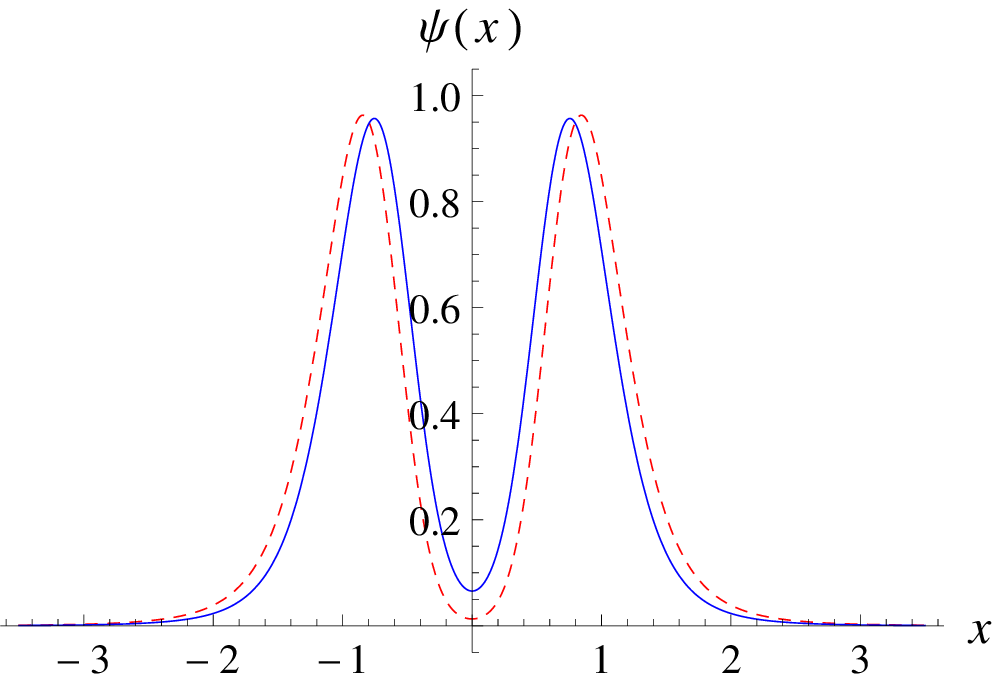}}
{\includegraphics[width=4.5cm,height=4cm]{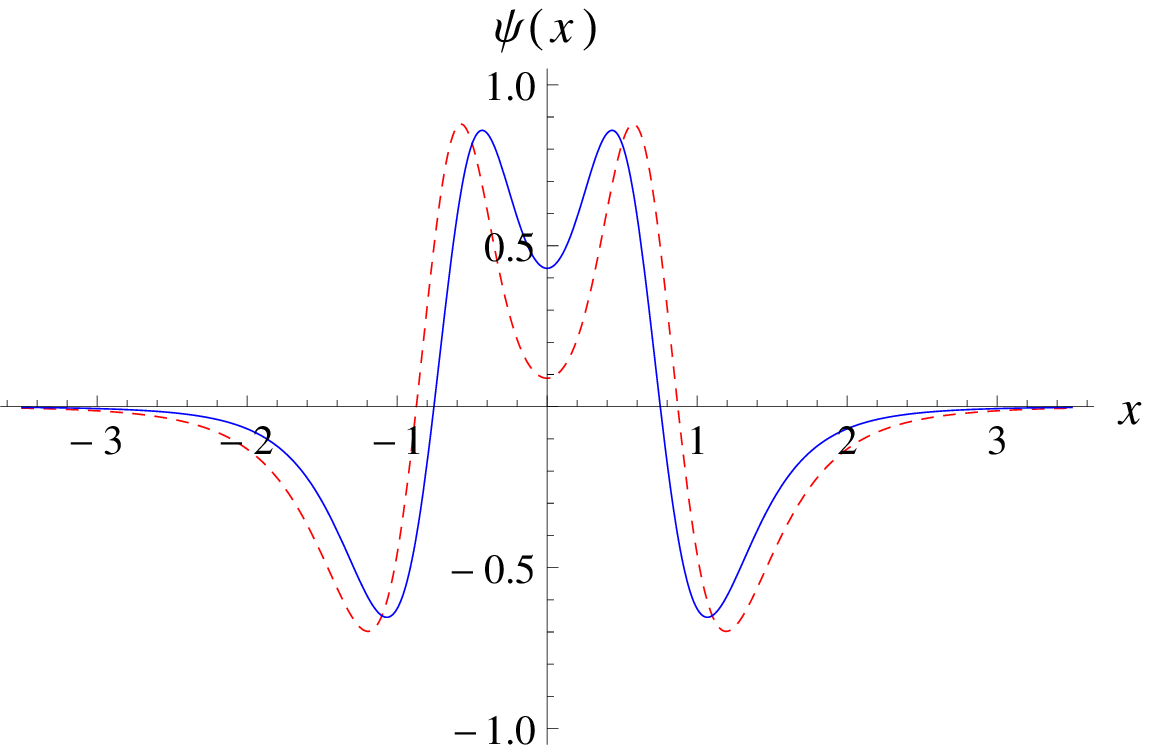}}
{\includegraphics[width=4.5cm,height=4cm]{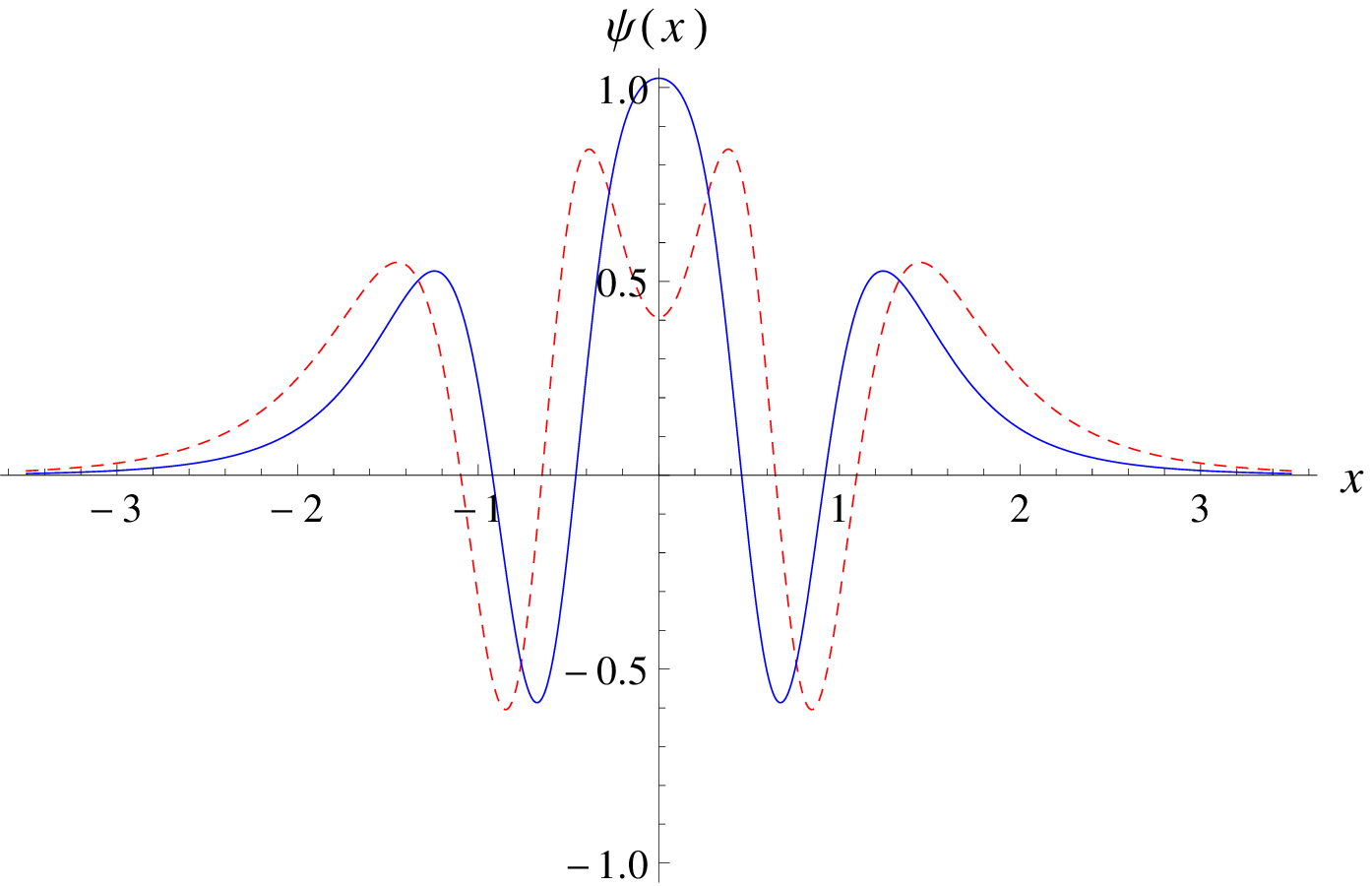}}\\
{\includegraphics[width=4.5cm,height=4cm]{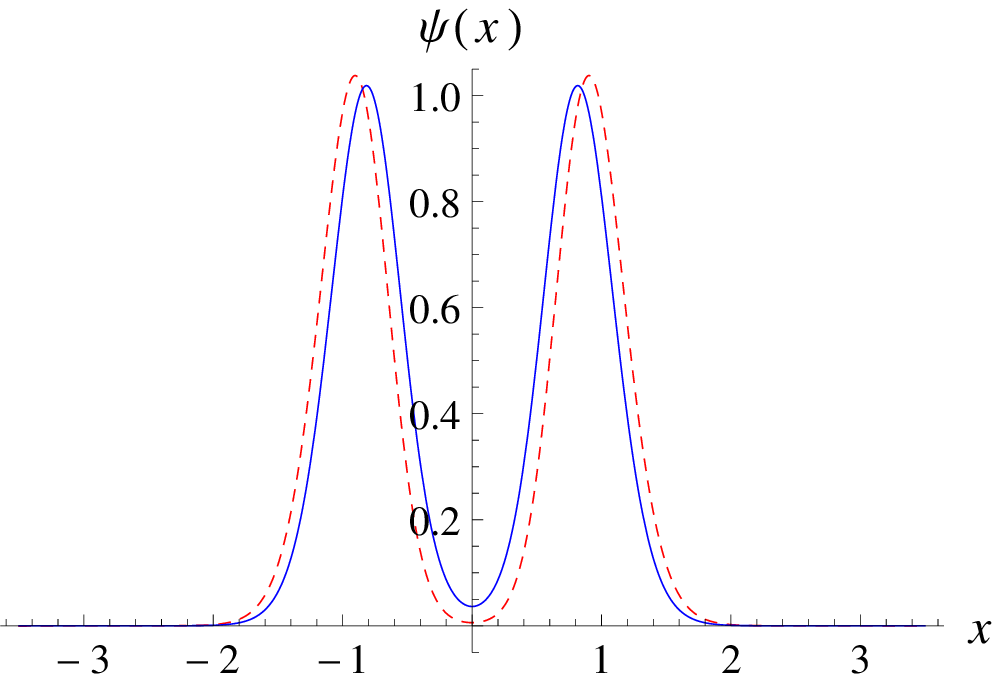}}
{\includegraphics[width=4.5cm,height=4cm]{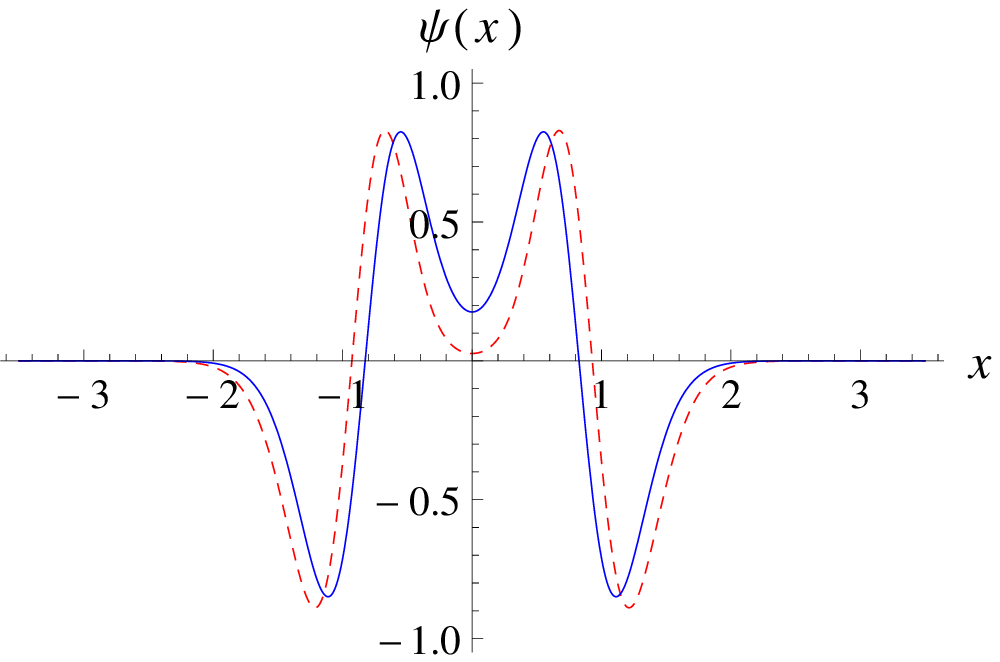}}
{\includegraphics[width=4.5cm,height=4cm]{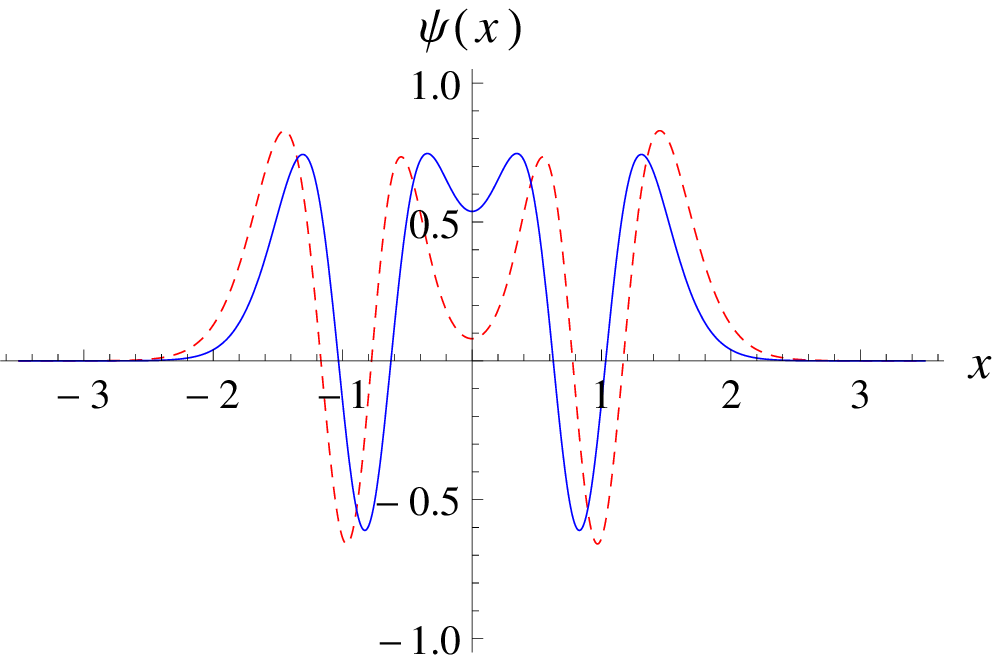}}
\caption{\label{figs_x_comp_A0_A60}
Plot of symmetric eigenstates of  $V_A(x)= V(x)_{Mann}-60 \sech^6(x)$ (solid)
versus $V(x)_{Mann}$ (dashed) for PDM [up] and constant mass [down].}
\end{figure}

In Fig. \ref{graf_Vx_A0_A60_A120_A500} we show a sequence of
triple-wells based on the Manning potential, already represented
at the top of Fig. \ref{graf_VA0}, now with the addition of a ''$\sinh^6x$'' term.
It can be seen that in this case the bigger is $A$ the softer is the barrier.
In Fig. \ref{figs_x_comp_A0_A60} (up) we show the eigenstates of this three-term \textit{PDM}-potential
$A=60$ (solid) together with the $A=0$ (dashed) Manning potential. In Fig. \ref{figs_x_comp_A0_A60} (down)
we show again the $A=60$ and $A=0$  eigenstates  but for an ordinary constant mass.
We put the figures altogether in two lines for a more comprehensive comparison.
In all the eigenstates we see a higher probability density around the origin in the $V_A(x)$ potential
and, remarkably, the PDM particle is always more probably tunneling than the ordinary one.
We have numerically computed the full spectrum of the $A=60$ potential and found that a
for a constant-mass particle there are 14 eigenstates that for PDM merge into eight (see Table \ref{tableA60}).

%%%%%%%%%%%%%%%%%%%%%%%%%%%%%%%%    TABLE  A60   %%%%%%%%%%%%%%%%%%%%%%%%%%%%%%%
\begin{table}[hb]
\caption{\label{tableA60} Full list of the energy eigenvalues of the \textit{PDM} and constant-mass
hamiltonians for $B=-C=-500$ and $A=60$. The $_S$ and $_A$ subindexes indicate
symmetric and antisymmetric states.}
\vskip 0.2cm
\begin{tabular}{c c c }
\hline  \hline
\,  \,               &Constant mass      \, & \, $PDM$\\ \hline
% \,  \, &     \, $--$ 			  \, & \,      $--$           \\
\,  \,               &Constant mass      \, & \, $PDM$\\ \hline
  \, $E_S^1$   \, & \,$-119.74469342961597$\, & \, $-113.818781855$  \\
   \, $E_A^2$   \, & \,$-119.7247052343852$ \, & \, $-113.7572364242$ \\
    \, $E_S^3$   \, & \,$-93.74280924014700$ \, & \, $-79.9396818103$  \\
     \, $E_A^4$   \, & \,$-93.3985291361313$  \, & \, $-78.1858715300$  \\
      \, $E_S^5$   \, & \,$-72.4166803691808$  \, & \,$-55.2994525270$   \\
       \, $E_A^6$   \, & \,$-70.0578863544200$  \, & \,$-44.8009029700$   \\
        \, $E_S^7$   \, & \,$-55.5196963894542$  \, & \,$-26.8437685530$   \\
         \, $E_A^8$   \, & \,$-49.0703204490206$  \, & \,\,\,\,$-9.3961914300$\\
          \, $E_S^9$   \, & \,$-38.5023159579192$  \, & \,       $--$            \\
           \, $E_A^{10}$\, & \,\,$-30.432820320680$ \, & \,      $--$             \\
            \, $E_S^{11}$\, & \,\,$-22.074924478020$ \,\, & \,     $--$             \\
             \, $E_A^{12}$\, & \,\,$-15.125970648791$\,\, & \,      $--$             \\
              \, $E_S^{13}$\, & \,\,\,\,\,$-9.091694203800$\, & \,  $--$              \\
               \, $E_A^{14}$  \,   &  \,$-4.5142592000 $\,\,\,\, & \,     $--$   \\
                 \hline
                  \hline
\end{tabular}
\end{table} %
%%%%%%%%%%%%%%%%%%%%%%%%%%%%%%%%%%   END TABLE A60  %%%%%%%%%%%%%%%%%%

\newpage

{~}

\newpage

\begin{figure}[h]
\center
{\includegraphics[width=4.5cm,height=4cm]{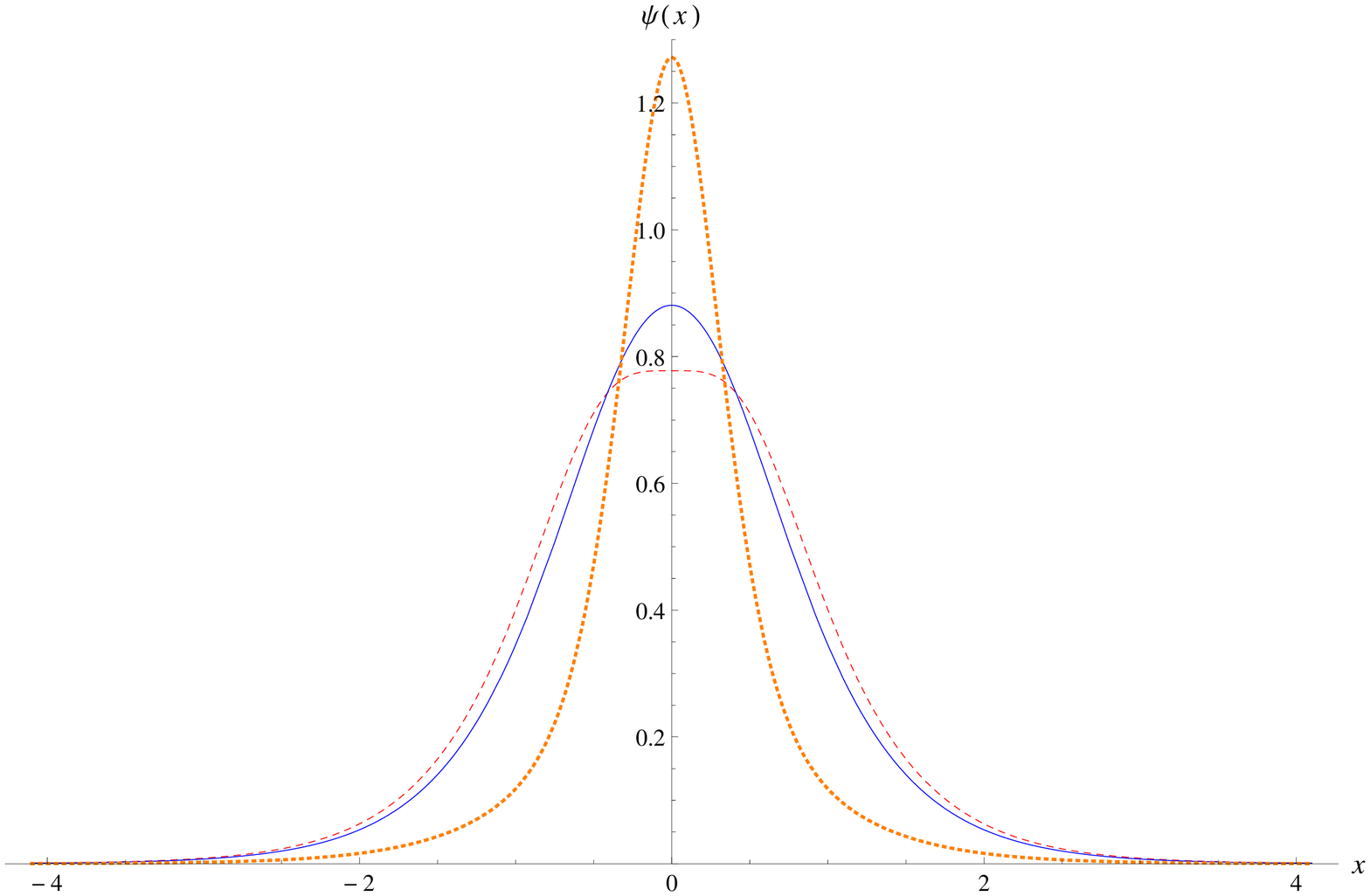}}
{\includegraphics[width=4.5cm,height=4cm]{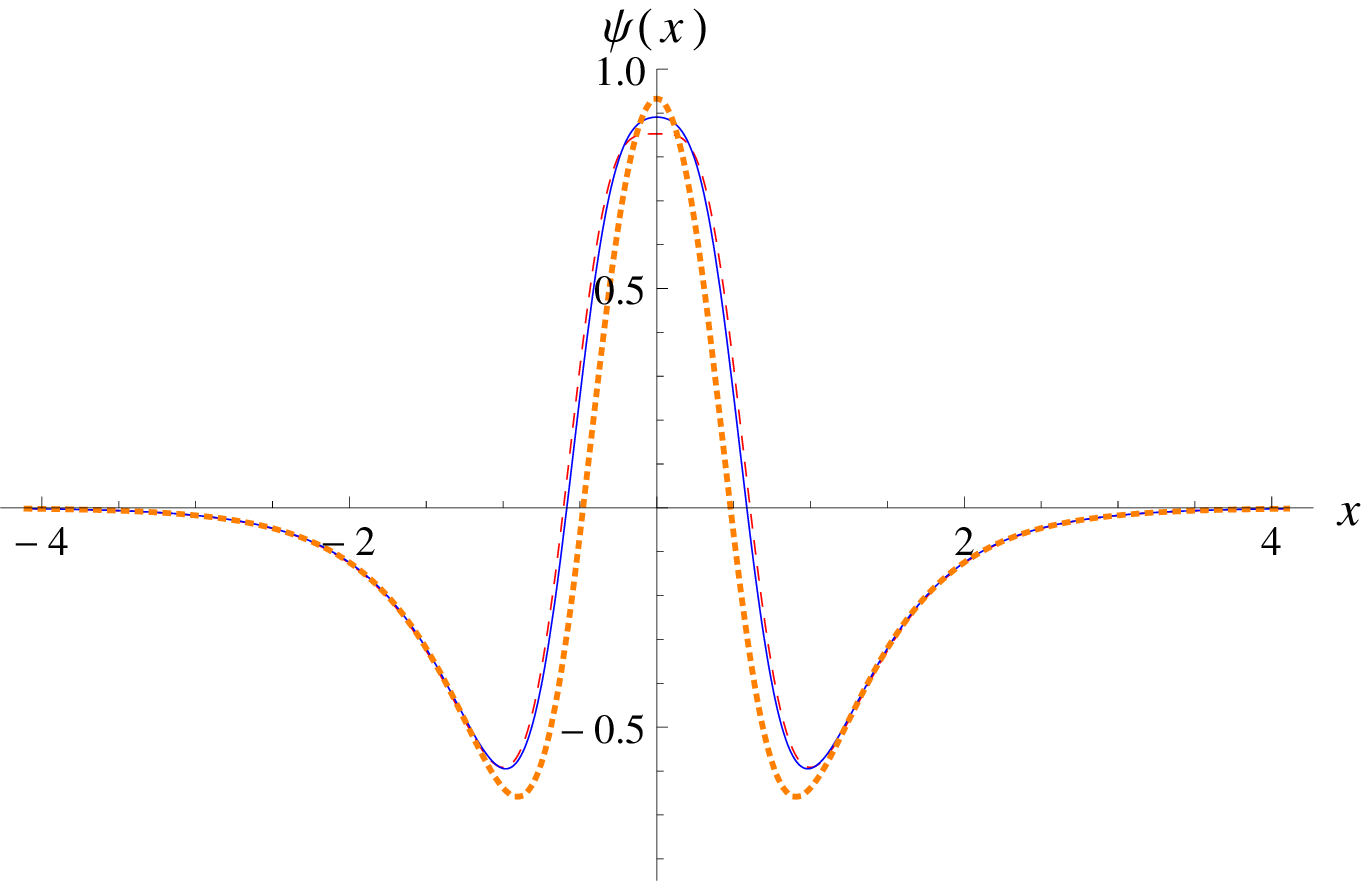}}\\
{\includegraphics[width=4.5cm,height=4cm]{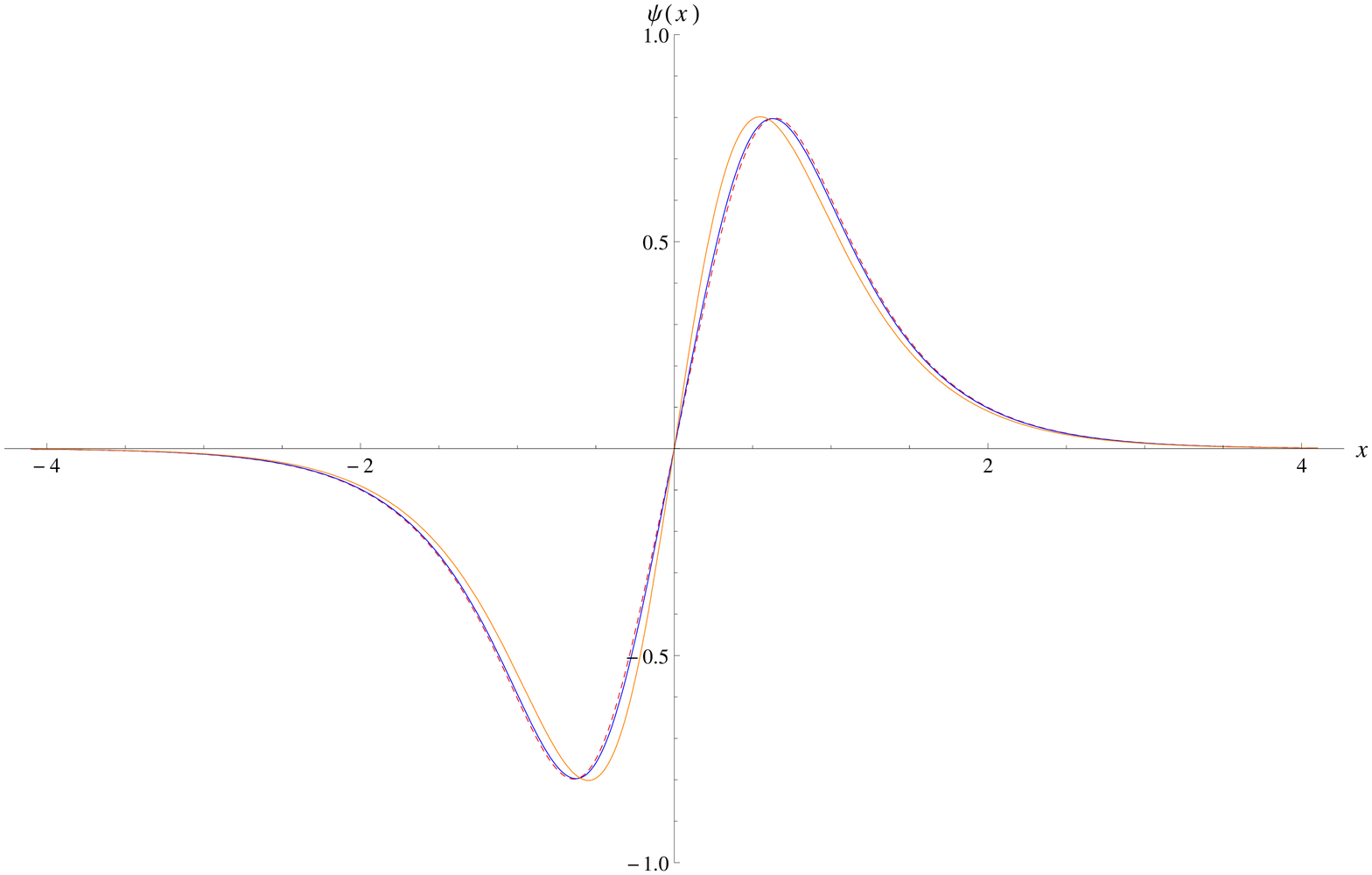}}
{\includegraphics[width=4.5cm,height=4cm]{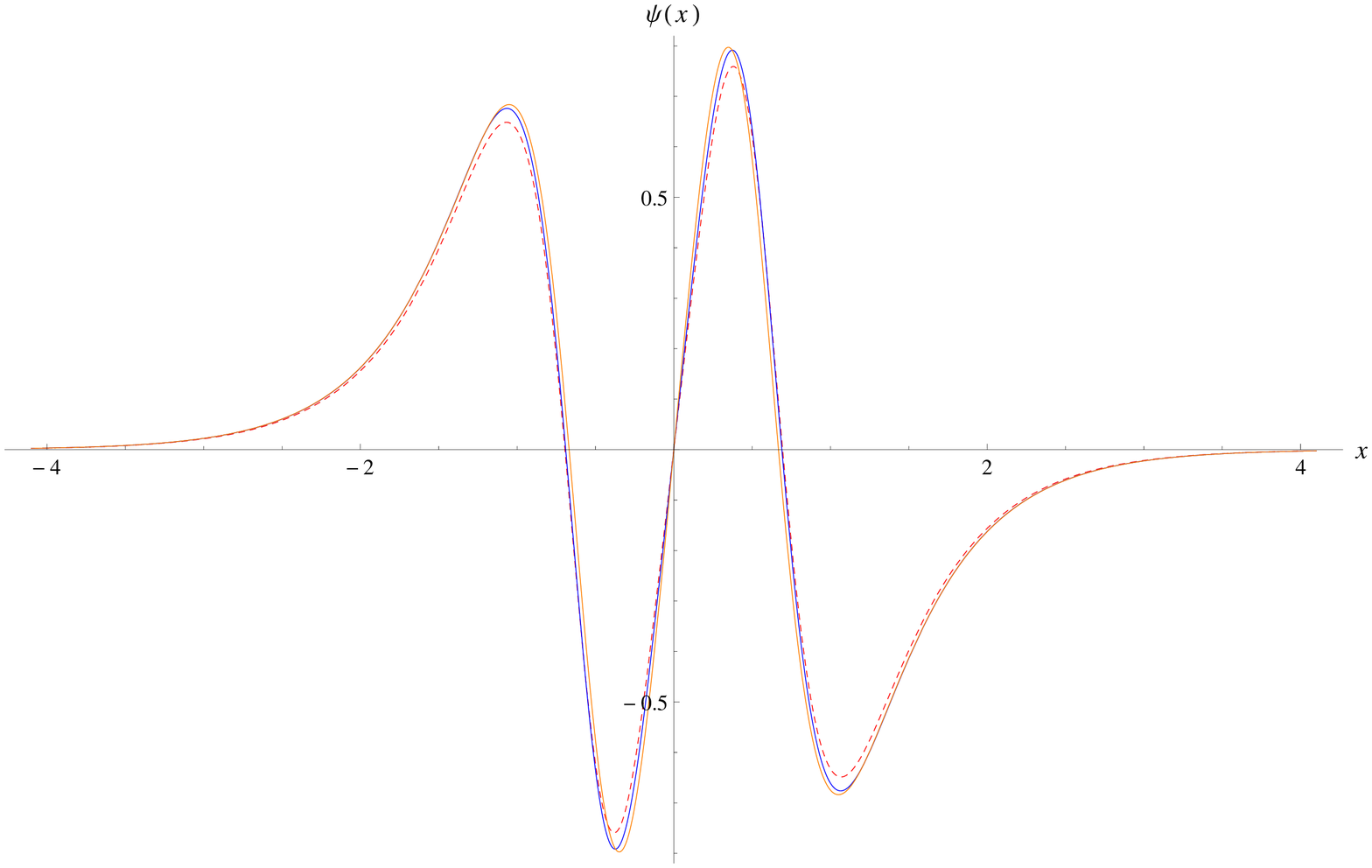}}
\caption{\label{figs_x_c0c2c10}
Comparative plot of $C=10$ (orange solid) vs. $C=2$ (blue dotted) vs. $C=0$ (red dashed)  for
the first PDM zero-modes of potential $V(x)=  A (\sech^6(x) -  \sech^4(x)) + C \sech^2x$; see values of
$A$ in Table \ref{table_A}. }
\end{figure}

Likewise, when we add a  ''$\sech^2x$'' term
the zero-modes found in the previous section \ref{sec:sech6-4}, and
the values of $A$ for them to exist, deviate increasingly as $C$ goes bigger.
In Fig. \ref{figs_x_c0c2c10} we show the first zero-modes for $C=10$ and $C=2$
with respect to $C=0$. These zero-modes take place for $A$ as listed in Table \ref{table_A}.
Adding a $C=10$ term has a stronger effect and the first two ${A}$
values with a zero-mode are in this case positive.
Note that as $C$ increases the zero energy particle tends to stay closer to the origin.
Table \ref{table_A} and the corresponding  figures show that, as $A$ grows,
the values of $A$ and the curves themselves rapidly converge
to a unique one for all the three columns.
In any case, the number of nodes of these eigenstates grows accordingly.
\begin{table}[hb]
\caption{\label{table_A} List of $A$ values for which potential
$V(x)=  A (\sech^6(x) -  \sech^4(x)) + C \sech^2x$ has PDM zero-modes.
The $_S$ and $_A$ subindexes at left indicate symmetric and antisymmetric states.}
\vskip 0.2cm%
\begin{tabular}{c c c c}
\hline  \hline
\,  \,               & $C=0$        \, &  \, $C=2$               &  $C=10$\\ \hline
  \, $A_S^1$   \, & \,$- 25.125695463186$    \, &  \, $- 5.12163713219$ \, &  \,$119.20733625954$\\
   \, $A_A^2$   \, & \,$- 56.05506043240$    \, &  \, $-45.04829005631$ \, &  \, $0.77393328576$\\
    \, $A_S^3$   \, & \,$-209.2999338840$    \, &  \, $-186.17085929030$\, & \, $-96.85687735639$\\
     \, $A_A^4$   \, & \,$-284.9369967664$    \, &  \, $-270.878712213$  \, &  \,$-213.6783204354$\\
      \, $A_S^5$   \, & \,$-571.60596450$    \, &  \,    $-546.7836421587$  \, &  \,$-$447.94257185326\\
       \, $A_A^6$   \, & \,$-691.7230772070$    \, &  \,  $-676.0837563001$  \, &  \,$-$612.9440054961\\
        \, $A_S^7$   \, & \,$-1111.7112347$    \, &  \,   $-1085.76152523664$  , &  \,$-$981.980281775\\
        \, $A_A^{8}$ \, & \,$-1276.268835001$    \, &  \,  $-1259.5595989779$  \, &  \,$-$1192.32726360\\
          \,  \, 			   & \, $\dots$       \, &  \, $\dots$            & \,  $\dots$ \\
            \hline
             \hline
\end{tabular}
\end{table}

\newpage

{~}

{~}

\begin{figure}[h]
\center
{\includegraphics[width=4cm,height=4.cm]{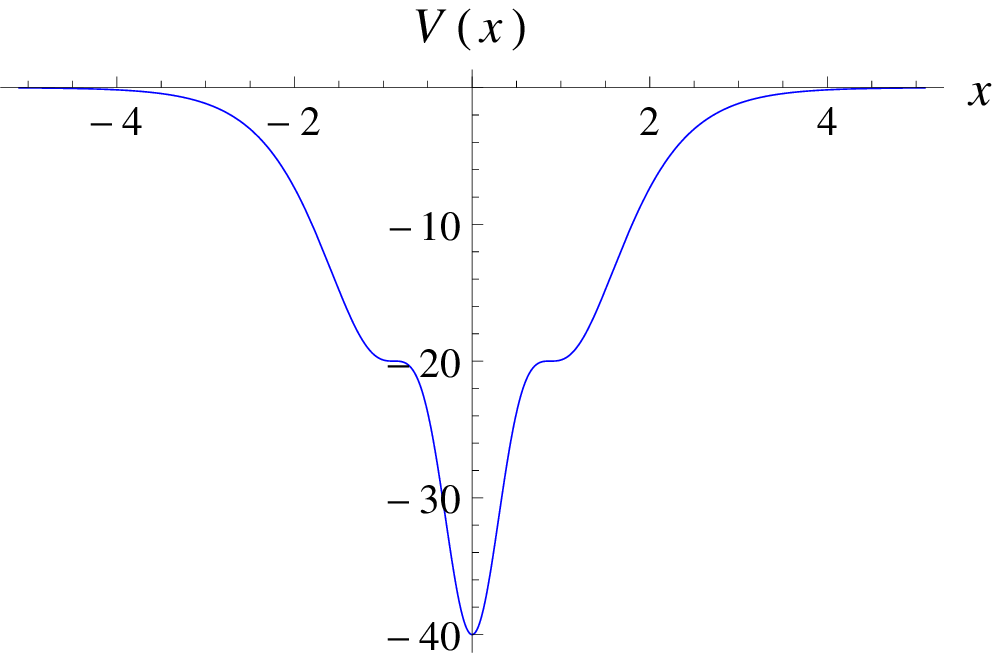}}
{\includegraphics[width=4cm,height=4.cm]{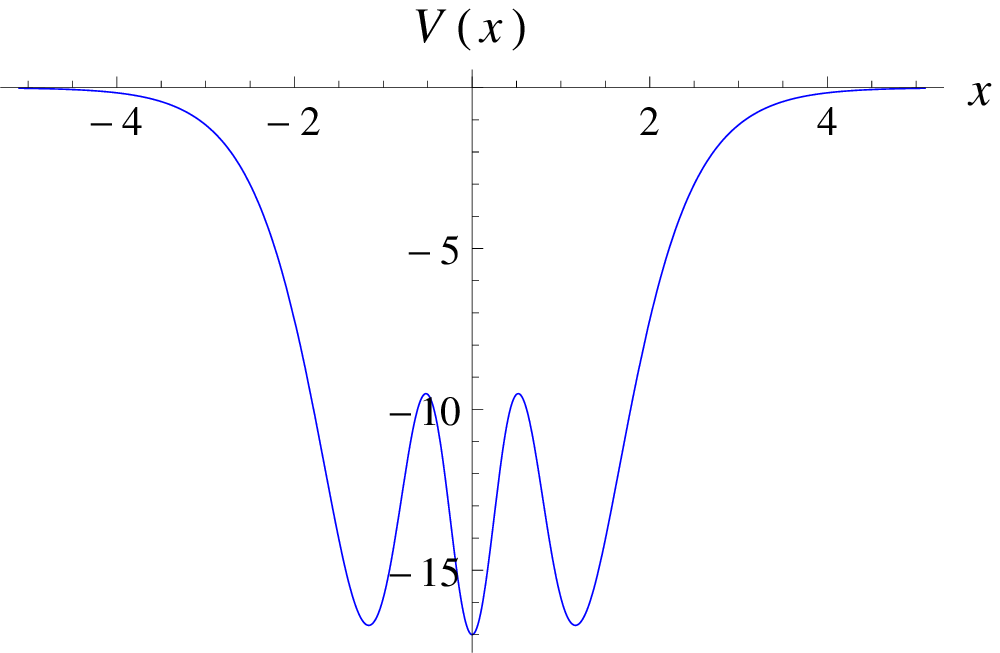}}
{\includegraphics[width=4cm,height=4.cm]{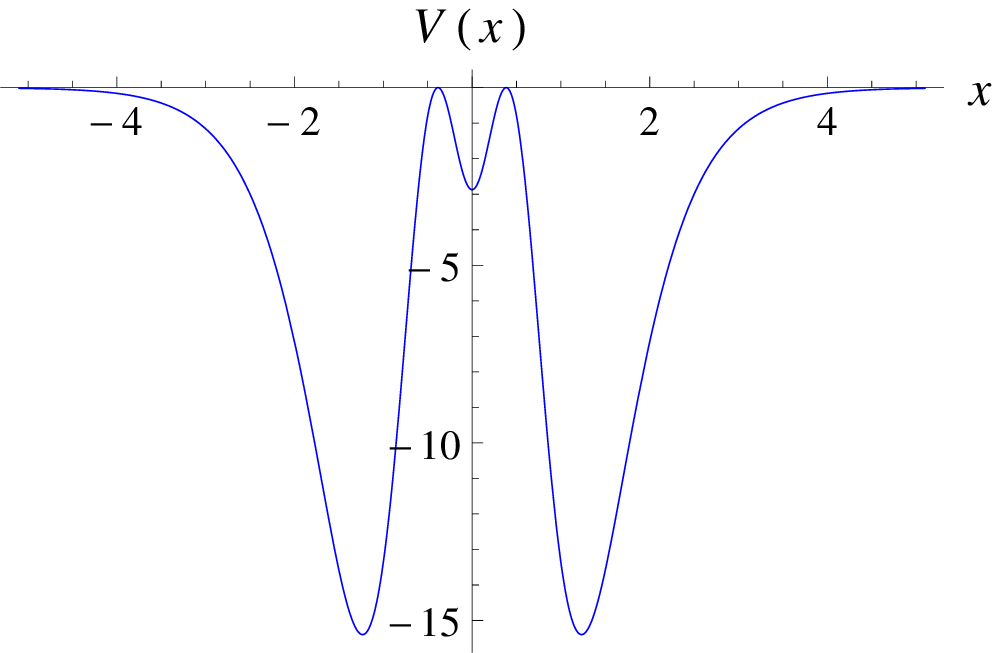}}
{\includegraphics[width=4cm,height=4.cm]{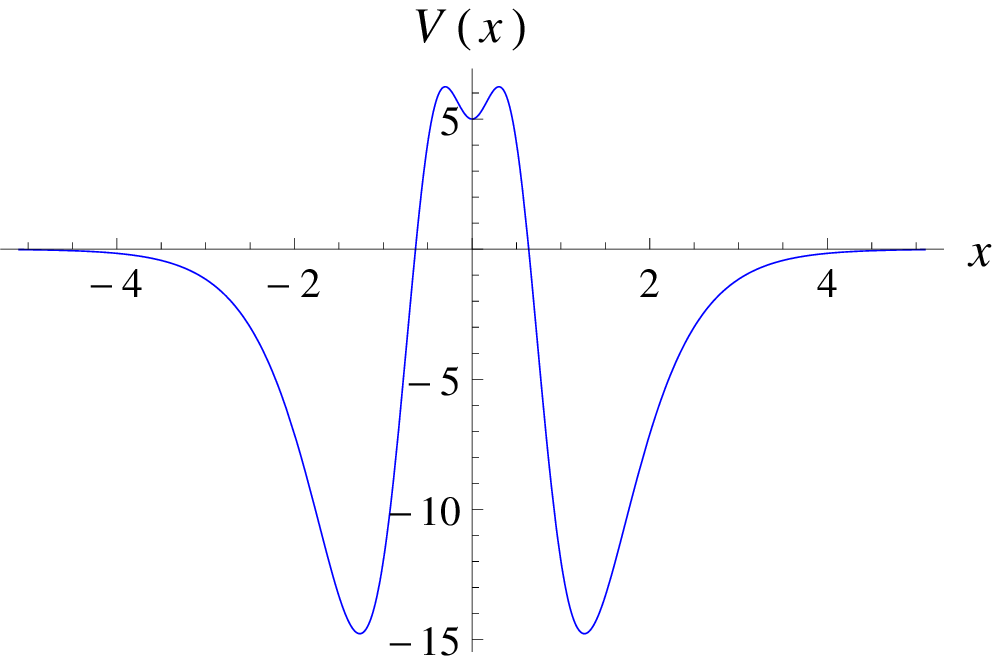}}
%{\includegraphics[width=4cm,height=3.5cm]{graf5_AlessthanC.eps}}
%{\includegraphics[width=4cm,height=3.5cm]{graf6_AlessthanC.eps}}
%{\includegraphics[width=3.5cm,height=3.5cm]{graf7_AlessthanC_B2_4AC.eps}}
\caption{\label{graf_TRIPLOS}  Plot of a sequence of potentials
 $V(x)= - 240\sech^6x - B  \sech^4x - 160\sech^2x$ as described in the text.}
\end{figure}

\newpage
Now, let us finally deal with the triple-well phase of the family.
In Fig. \ref{graf_TRIPLOS} we display a sequence of triple-well potentials constructed
by setting specific relations among the parameters.
In the first (left) figure the potential is about starting the triple-well phase.
In this case, the first and second derivatives are zero at
$x=\pm\arcsech\sqrt{-B/3A}, B/A<0$, for $B^2=3AC$. We adopt $A>0$ so that $C>0$ and $B<0$.
%note that $V'(x)=0$ for $3A\sech^4x+2B\sech^2+C=0$
%namely $\sech^2=(-B\pm\sqrt{B^2-3AC})/3A$ which has only two solutions for $B^2=3AC$.
In the second figure we let $B^2>3AC$ and we get a couple of maxima,
one at each side of the origin, resulting in three cleanly defined wells.
In the third figure we let $B^2=4AC$ when both maxima make simultaneously
$V'(x)=0$ and $V(x)=0$. For $B^2>4AC$ the potential develops a nonnegative barrier (see fourth figure).

\begin{figure}[h]
\center
{\includegraphics[width=8cm,height=7cm]{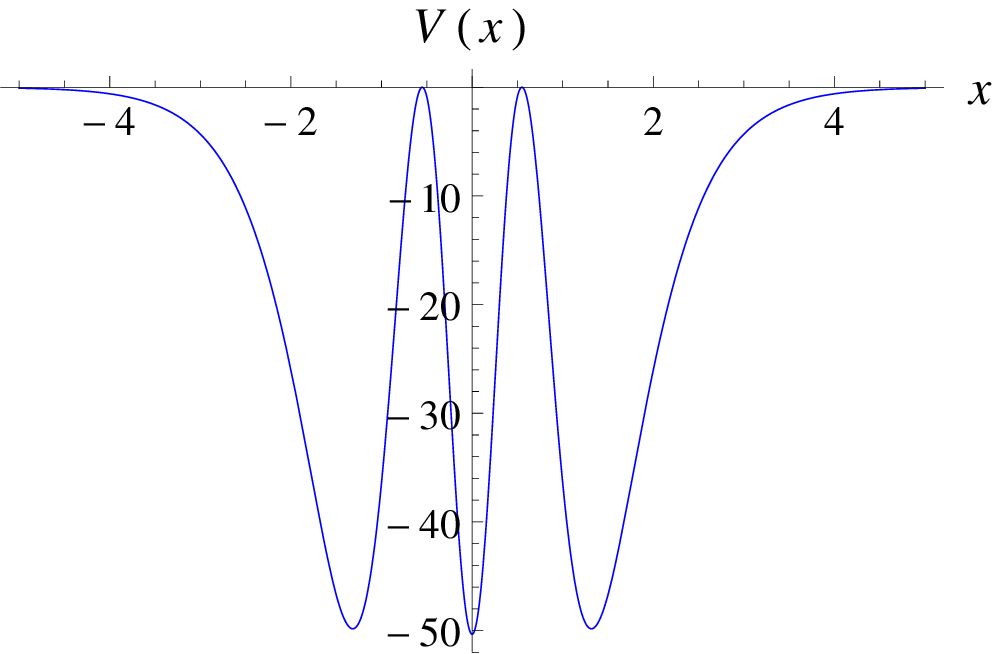}}
\caption{\label{triplo_especial}  Plot of the potential
 $V_T(x)= -800\sech^6x+\sqrt{4AC}\sech^4x-449\sech^2x$.}
\end{figure}

\begin{figure}[h]
\center
{\includegraphics[width=3.5cm,height=4cm]{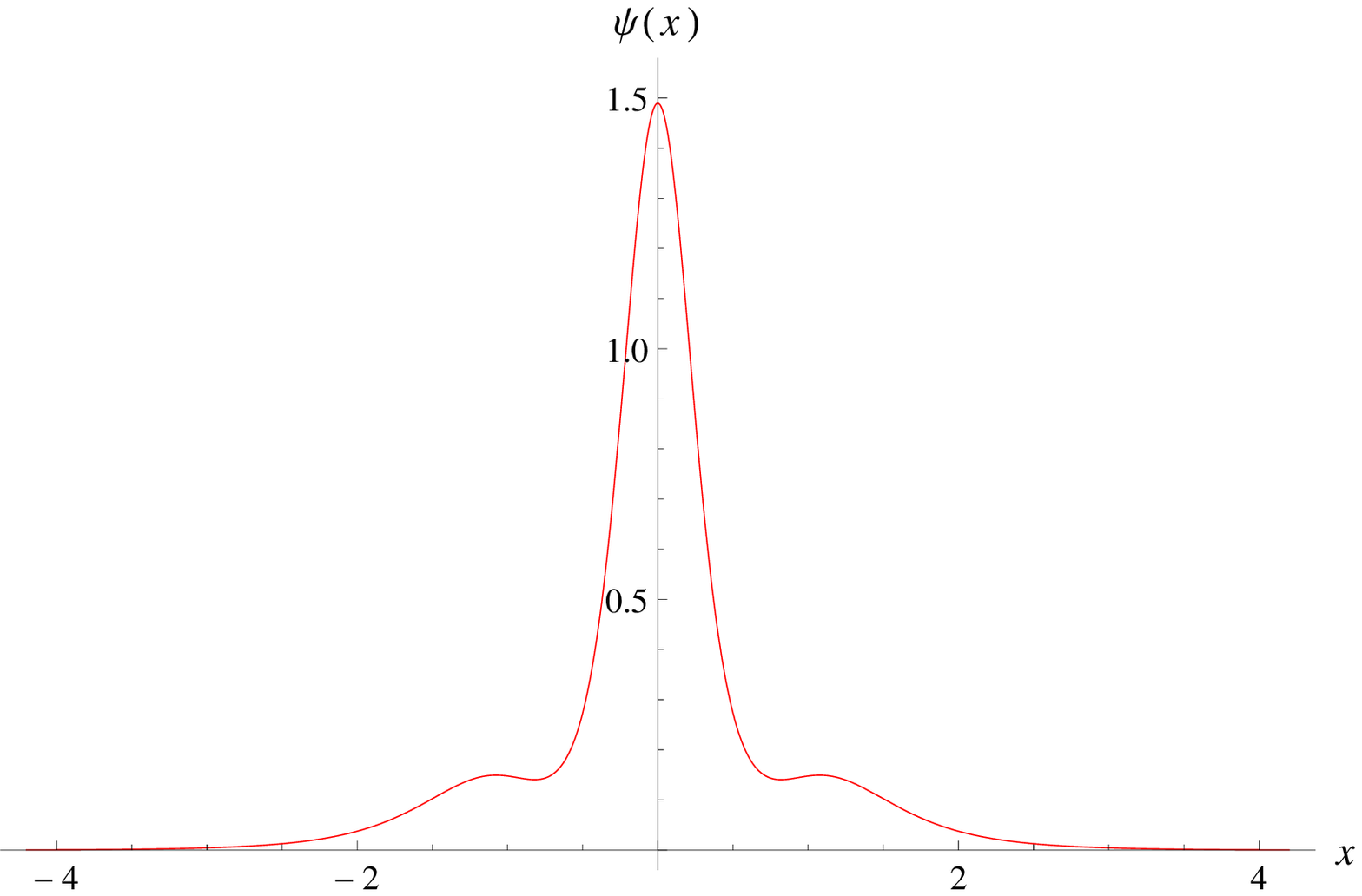}}
{\includegraphics[width=3.5cm,height=4cm]{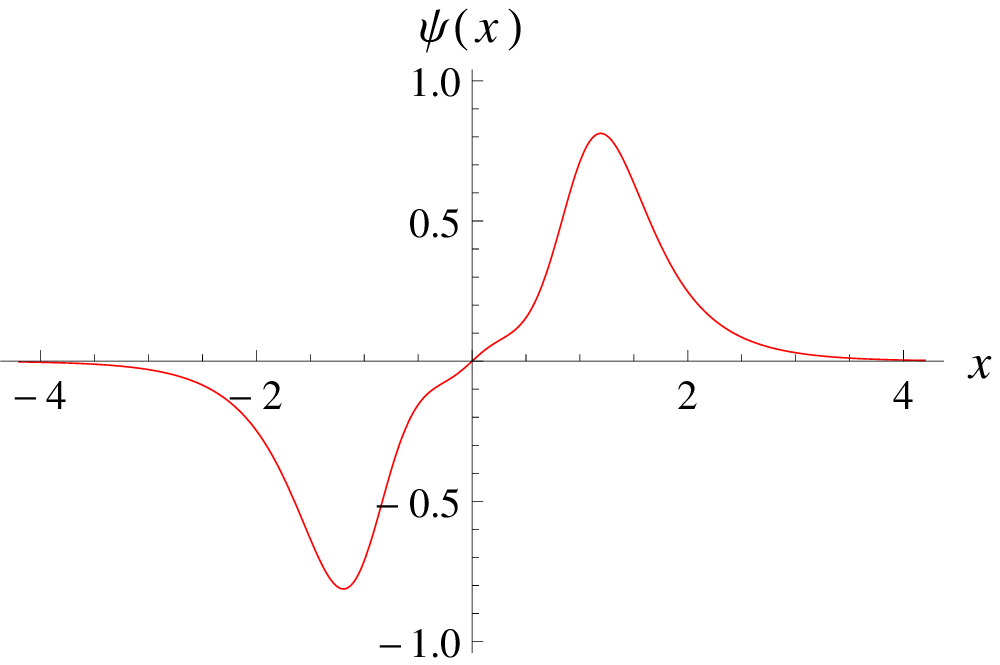}}
{\includegraphics[width=3.5cm,height=4cm]{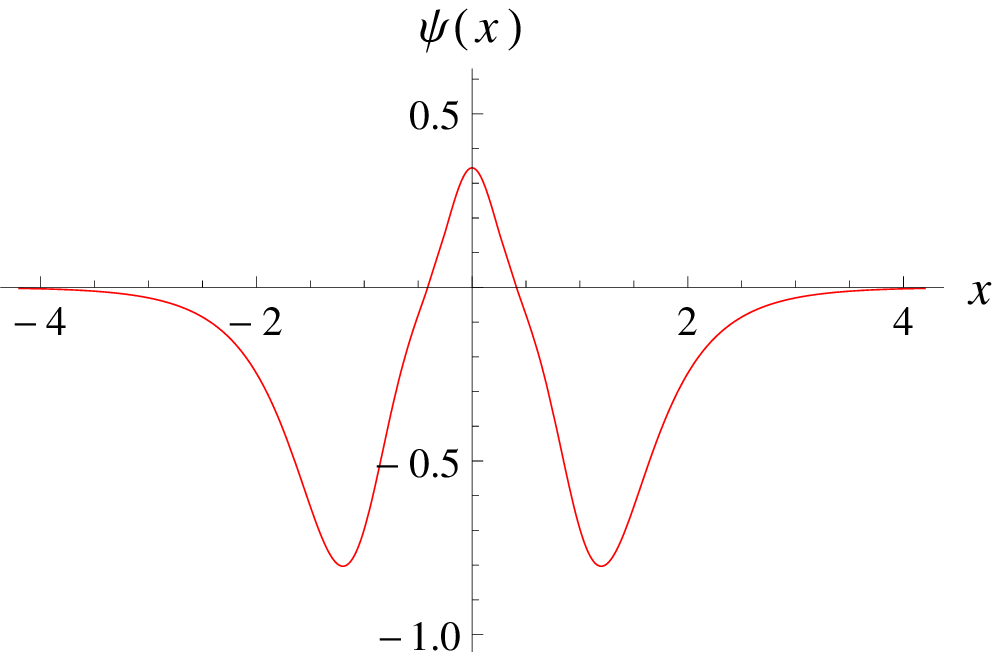}}
{\includegraphics[width=3.5cm,height=4cm]{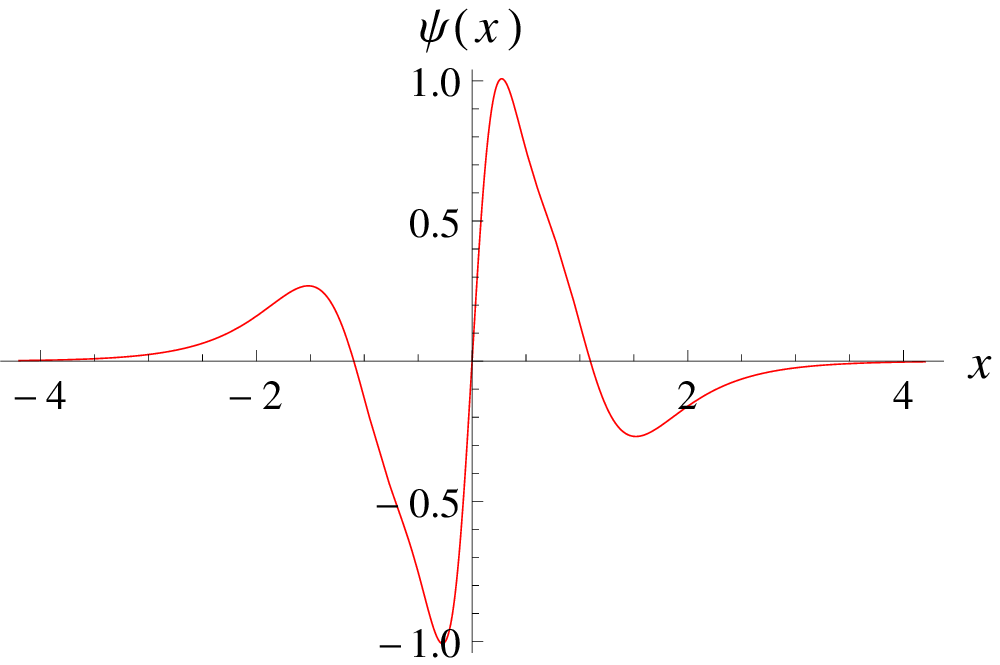}}\\
{\includegraphics[width=3.5cm,height=4cm]{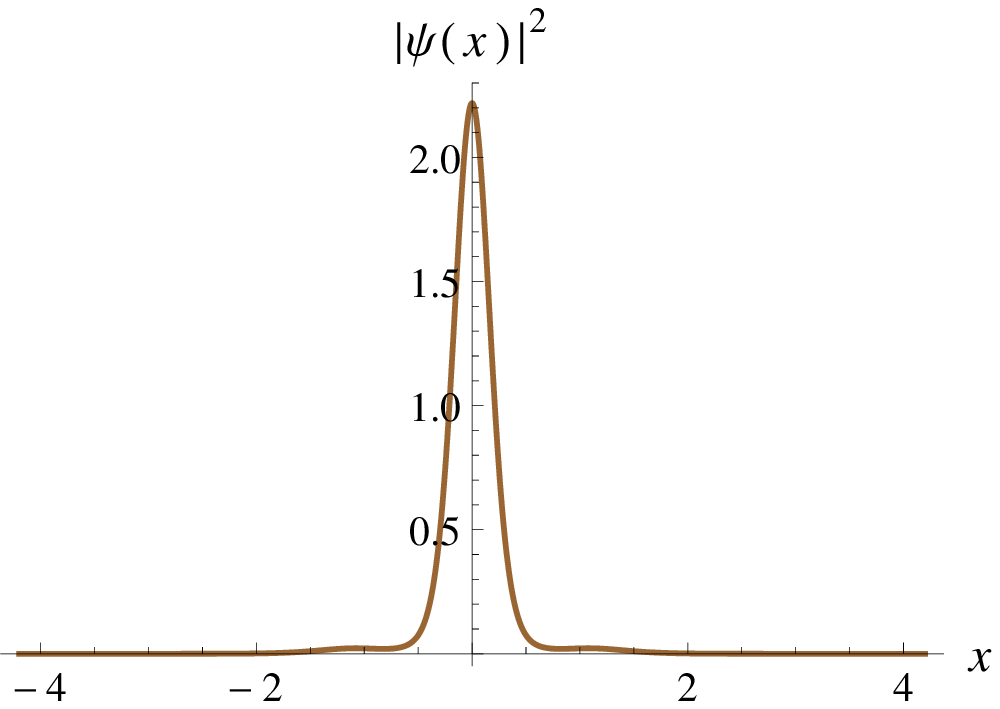}}
{\includegraphics[width=3.5cm,height=4cm]{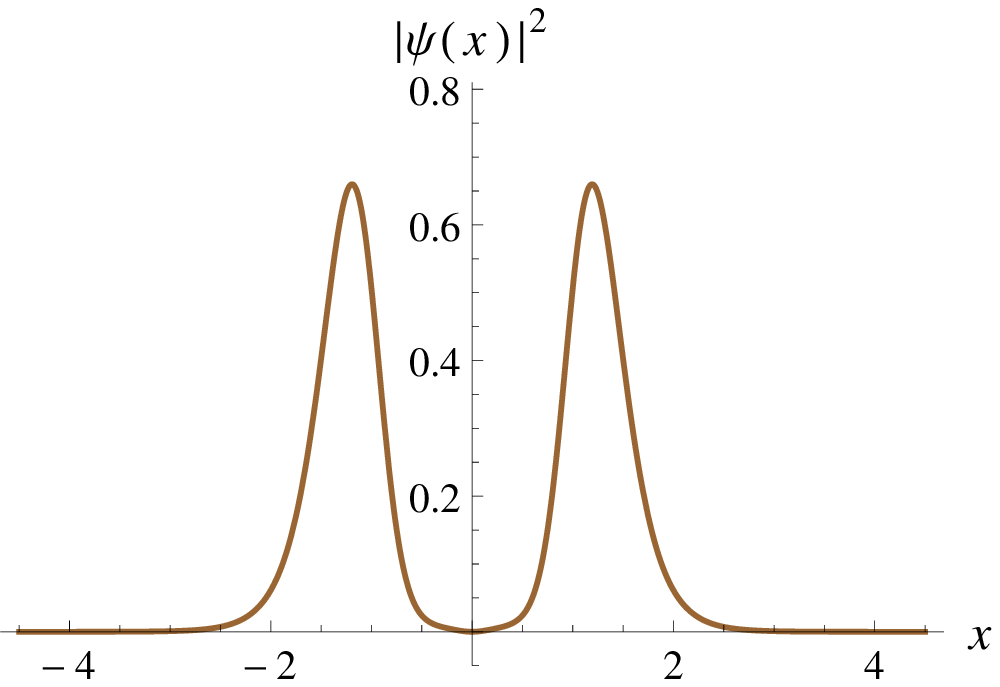}}
{\includegraphics[width=3.5cm,height=4cm]{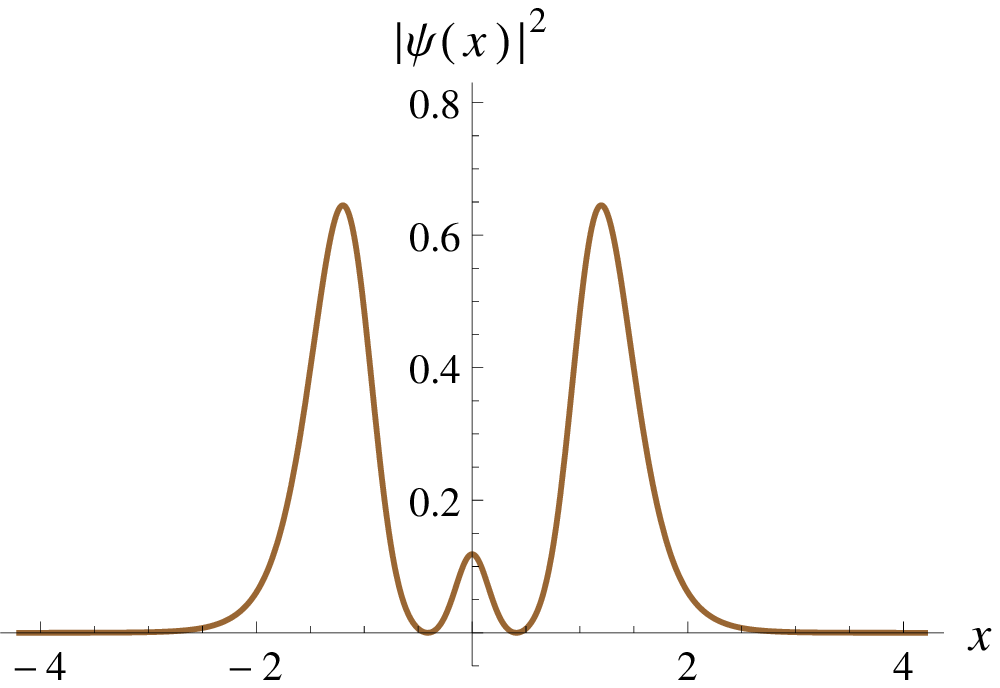}}
{\includegraphics[width=3.5cm,height=4cm]{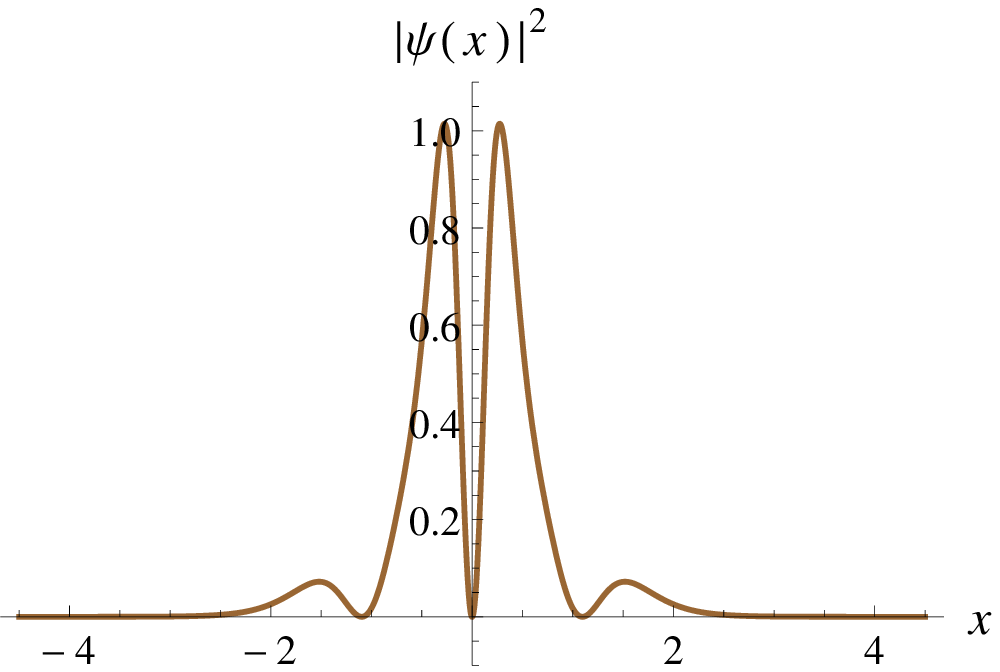}}
\caption{\label{wf_var_triplo}  Plot of PDM eigenfunctions and probabilities for
$V_T(x)= -800\sech^6x+\sqrt{4AC}\sech^4x-449\sech^2x$. The full sequence of eigenergies
has been numerically computed, see Table \ref{table_triplo}.}
\end{figure}

For the triple-well represented in Fig. \ref{triplo_especial},
$V_T(x)= -A\sech^6x+\sqrt{4AC}\sech^4x-C\sech^2x$ ($A=800, C=449$),
we show the full PDM bound spectrum of eigenfunctions and probability
densities in Fig. \ref{wf_var_triplo}.
As shown in Table \ref{table_triplo}, there is once again a reduction
or merging of eigenstates for PDM in which case the ten (constant-mass)
eigenstates result in just four eigenstates when the mass depends on the position.

%%%%%%%%%%%%%%%%%%%%%%%%%%%%%%%%    TABLE  10x4   %%%%%%%%%%%%%%%%%%%%%%%%%%%%%%%
\begin{table}[hb]
\caption{\label{table_triplo} Full list of the energy eigenvalues of the
\textit{PDM} and constant-mass triple-well hamiltonians for $A=800$, $C=449$,
and $B=-\sqrt{4AC}\approx-1198.67$. The $_S$ and $_A$ subindexes
 at left indicate symmetric and antisymmetric states.}
\vskip 0.2cm%
\begin{tabular}{c c c }
\hline  \hline
\,  \,               &Constant mass        \, &  \, $PDM$\\ \hline
 \,  \, & \, $--$ 			    \, &  \, $--$             \\
  \, $E_S^1$   \, & \,$-40.0750771677640$    \, &  \, $-31.3132652539$ \\
   \, $E_A^2$   \, & \,$-40.0731076274700$    \, &  \, $-27.0691086460$ \\
    \, $E_S^3$   \, & \,$-31.8627686815140$    \, &  \, $-26.8175802300$ \\
     \, $E_A^4$   \, & \,$-23.0630939687000$    \, &  \, $-2.05157297020$ \\
      \, $E_S^5$   \, & \,$-23.0065594776400$    \, &  \,                  \\
       \, $E_A^6$   \, & \,$-10.6509913784700$    \, &  \,                  \\
        \, $E_S^7$   \, & \,$-10.3287500184298$    \, &  \,                  \\
         \, $E_A^8$   \, & \,\,\,\,$-3.936337196700$\, &  \,                  \\
          \, $E_S^9$   \, & \,\,\,\,$-2.361296623000$\, &  \,                  \\
          \, $E_A^{10}$\, & \,\,\,\,$-0.782147401000$\, &  \,                  \\
            \hline
             \hline
\end{tabular}
\end{table}

\newpage

\section{Conclusion \label{sec:conclusion}}

In the present paper we have studied the new hyperbolic potential class recently reported in \cite{sech246}.
Here, the mass of the particle has gained a position-dependent status in order to enrich the
phenomenological possibilities of the models involved and to explore its mathematical consequences.
After properly setting up the \textit{PDM} problem we have payed special attention to the celebrated Manning
potential, of great interest in molecular physics, here studied in the PDM case for the first time.
We have analytically obtained the complete set of eigenstates to this hamiltonian and then
shown its full bound-state spectrum and eigenfunctions in a case study.
We have analytically found confluent-Heun expressions in the general case and compared the
PDM eigenfunctions to the constant-mass ones. Heun functions have recently been receiving
increasing attention and have been found in a wide variety of contexts 
\cite{christiansencunha2011,christiansencunha2012,rumania,bulgaria,cvetic2011,herzog}.
PDM particles tend to be more likely tunneling than ordinary ones.
Next, we have addressed the PDM version of the sixth power hyperbolic potential
$V(x) = -{A}~{\sech^6x}-B~{\sech^4x}$
and obtained exact expressions for the zero-modes in the $A$ and $A=-B$ cases belonging to a
discrete set of parameters. All such eigenstates have been found proportional to triconfluent
forms of the Heun functions. Interestingly, we have met with no zero-modes for any value of the
parameter in the ordinary  constant-mass counterpart of this potential.
In both the Manning and sixth-order hyperbolic potentials we have also
analyzed the consequences of considering a complementary three-term potential by comparing
their eigenfunctions. The analysis of these and others three-terms cases has been
performed for constant-mass and PDM hamiltonians and has shown interesting differences between their spectra,
particularly a reduction of eigenstates in the PDM circumstance.
Finally, we have discussed the triple-well phase of the potential class and focused especially
in a $V_T(x)= -A\sech^6x+\sqrt{4AC}\sech^4x-C\sech^2x$ case
showing the full set of PDM eigenfunctions and probability densities.
This triple-well also exposes the phenomenon of merging of the ordinary spectrum when the mass
turns nonuniform. It seems to be a general property of this class of hamiltonians.

\end{document}